\documentclass[11pt]{article}

\usepackage{epsfig,graphics}
\usepackage{amssymb,amsmath,amsthm,bm}

\newtheorem{theo}{Theorem}
\newtheorem{lem}{Lemma}
\newtheorem{pro}{Proposition}

\def\R{{\mathbb R}}

\def\ga{\alpha}
\def\gga{\gamma}

\def\go{\omega}

\def\gb{\beta}

\def\gs{\sigma}

\def\wt{\widetilde}
\def\n{\noindent}

\def\gD{\Delta}

\def\b0{{\bf 0}}
\def\1{{\bf 1}}

\def\wh{\widehat}

\def\wt{\widetilde}

\begin{document}

\title{Adaptive Synchrosqueezing Transform with a Time-Varying Parameter for Non-stationary Signal Separation\thanks{This work was supported in part by the National Natural Science Foundation of China (Grant No. 61201287) and Simons Foundation (Grant No. 353185)}}

\author{Lin Li${}^{1}$, Haiyan Cai${}^{2}$  and Qingtang Jiang${}^2$ }

\date{}

\maketitle

\vskip -1cm
\centerline{1. School of Electronic Engineering, Xidian University, Xi'an 710071, P.R. China}
\centerline{e-mail: lilin@xidian.edu.cn.}

\centerline{2. Dept. of Math \& CS, University of Missouri-St. Louis, St. Louis,  MO 63121, USA}
\centerline{e-mail: $\{$haiyan${}_-$cai, jiangq$\}$@umsl.edu}

\begin{abstract}
The continuous wavelet transform ({CWT}) is a linear time-frequency representation and a powerful tool for analyzing non-stationary signals.
The synchrosqueezing transform (SST) is a special type of the reassignment method which not only enhances the energy concentration of CWT in the time-frequency plane, but also separates the components of multicomponent signals. The ``bump wavelet'' and Morlet's wavelet are commonly used continuous wavelets for the wavelet-based SST. There is a parameter in these wavelets which controls the widths of the time-frequency localization window.
 In most literature on SST, this parameter is a fixed positive constant.
 In this paper, we consider the CWT with a time-varying parameter (called the adaptive CWT) and the corresponding SST (called the adaptive SST) for instantaneous frequency estimation and multicomponent signal separation.
We also introduce the 2nd-order adaptive SST. We analyze the separation conditions for non-stationary multicomponent signals with the local approximation of linear frequency modulation mode. We derive well-separated conditions of a multicomponent signal based on the adaptive CWT.
We propose methods to select the time-varying parameter so that the corresponding adaptive SSTs of the components of a multicomponent signal have sharp representations and are well-separated, and hence the components can be recovered more accurately. We provide comparison experimental results to demonstrate the efficiency and robustness of the proposed adaptive CWT and adaptive SST in separating components of multicomponent signals with fast varying frequencies.

\end{abstract}

\section{Introduction}
Multicomponent signals are common in nature and in many engineering problems. These signals are usually non-stationary, meaning that their frequencies and/or amplitudes change with the time. It is important to separate the components of such a signal $x(t)$ to extract information, such as the underlying dynamics,  hidden in $x(t)$. However, due to its non-stationary property, this is a challenging problem. Sometimes it is even difficult to distinguish a monocomponent signal from a multicomponent signal. For example,
$$
x(t)=\cos(2\pi \xi_1t)+\cos(2\pi \xi_2t)=2\cos(\pi (\xi_1-\xi_2)t)\cos(\pi (\xi_1+\xi_2)t)
$$
has 	two components, but can be seen as a monocomponent signal if  $\xi_1$ is close to  $\xi_2$, because in this case, the amplitude $2\cos(\pi (\xi_1-\xi_2)t)$  changes slowly compared to the carrier wave  $\cos(\pi (\xi_1+\xi_2)t)$.

The empirical mode decomposition (EMD) algorithm along with the Hilbert spectrum analysis introduced in \cite{Huang98} is a popular method to decompose and analyze non-stationary signals.
The intrinsic mode function (IMF) 
is used to represent a monocomponent signal \cite{Huang98}. An IMF satisfies two conditions: (a) the number of its minimum and maximum must either be equal or differ at most by one; and (b) the value of the mean of its upper envelope and lower envelope is close to zero. EMD decomposes a signal
$x(t)$ into finitely many IMFs plus a trend signal, and then the instantaneous frequency ({\bf IF}) of each IMF  is calculated by the Hilbert spectrum analysis which results in a representation of $x(t)$ as
\begin{equation}
\label{AHMold}
x(t)=A_0(t)+\sum_{k=1}^K x_k(t),  \quad x_k(t)=A_k(t) \cos \big(2\pi \phi_k(t)\big)
\end{equation}
with  $A_k(t), \phi_k'(t)>0$, where  $A_k(t)$ is called the instantaneous amplitude (IA)  and $\phi'_k(t)$ the IF of $x_k(t)$.
There are many articles studying the property of EMD or proposing variants of EMD to improve the performance, see e.g. \cite{Flandrin04}-\cite{HM_Zhou16}. In particular, the separation ability of EMD is discussed in \cite{Rilling08}, which shows that EMD cannot decompose two components when their frequencies are close to each other. The ensemble EMD (EEMD) is proposed to suppress the noise interferences \cite{Wu_Huang09}. A weakness of EMD or EEMD is that it can easily lead to mode mixture or artifacts, namely undesirable or false components \cite{Li_Ji09}.

The time-frequency analysis is another class of methods for non-stationary multicomponent signals. Some non-linear time-frequency analyses, such as the Wigner-Ville distribution and the Choi-Williams distribution \cite{Cohen79}-\cite{Stankovic13}, have cross-term interferences and cannot be used to reconstruct the signal components. On the other hand, some linear time-frequency analysis, such as the continuous wavelet transform ({\bf CWT}) \cite{Dau_book, Mallat99} and the short time Fourier transform ({\bf STFT}) \cite{Sejdic09}, have the inverse transforms. The uncertainty principle (see e.g. \cite{Boashash15}) imposes an unavoidable tradeoff between temporal and spectral resolutions.
In addition, the time and frequency reassignments were introduced and studied in \cite{A_Flandrin_reassignment95} and \cite{CM_A_Flandrin_reassignment03}
to enhance the energy concentration in the time-frequency plane.

The synchrosqueezing transform ({\bf SST}), also called the synchrosqueezed wavelet transform, was introduced in \cite{Daub_Maes96} and further developed in the seminal article \cite{Daub_Lu_Wu11}. It is a special type of reassignment method on the CWT which not only sharpens the time-frequency representation of a signal, but also recovers the components of a multicomponent signal.  SST provides an alternative to the EMD method and its variants, and it overcomes some limitations of the EMD and EEMD schemes such as mode-mixing. Many works on SST have been carried out since the publication of  \cite{Daub_Lu_Wu11}.
For example, \cite{Wu_Flandrin_Daub11}-\cite{Flandrin_Wu_etal_review13} studied a comparison between EMD and SST. The stability of SST was studied in \cite{Thakur_etal_Wu13}.
A hybrid EMD-SST computational scheme by applying the modified SST to the IMFs of the EMD was proposed in \cite{Chui_Walt15}. The synchrosqueezed wave packet transform was introduced in \cite{Yang15}.
The SST with vanishing moment wavelets was introduced in \cite{Chui_Lin_Wu15}.
A multitapered SST was introduced in \cite{Daub_Wang_Wu15} to enhance the concentration in the time-frequency plane by averaging over random projections with synchrosqueezing.
The STFT-based SST was introduced and studied in \cite{Thakur_Wu11, Wu_thesis} and also studied in \cite{MOM14} with different conditions on $A'_k(t), \phi''_k(t)$.
The 2nd-order SST was proposed and studied in \cite{MOM15, OM17, BMO18}. \cite{Wang_etal14}
introduced the demodulation-transform based SST with STFT, and \cite{Jiang_Suter17}  studied CWT-SST with the demodulation-transform. The linear and synchrosqueezed time-frequency representations were reviewed in \cite{Iatsenko15}, which also discussed the choice of window and wavelet parameters, the advantages and drawbacks of synchrosqueezing, etc.  A STFT-based signal separation operator was proposed and studied in \cite{Chui_Mhaskar15} for signal separation. The statistical analysis of synchrosqueezed transforms has been studied in \cite{Yang18}.  An empirical signal separation algorithm was presented in \cite{LCJJ17}.

SST has been used in engineering and medical data analysis applications including machine fault diagnosis \cite{Li_Liang_fault12, WCSGTZ18}, anesthesia evaluation \cite{Wu_anaesthesia14, Chui_Lin_Wu15}, breathing dynamics discovery \cite{Wu_breathing14}, sleep stage assessment \cite{Wu_sleep15} and heart beat classification \cite{Wu_heartbeat17}.

The ``bump wavelet'' $\psi_{\rm bump}(x)$ defined by
\begin{equation}
\label{def_bump}
\wh \psi_{\rm bump}(\xi)=
 e^{1-\frac 1{1-\gs^2(\xi-\mu)^2}} \chi_{(\mu-\frac 1\gs, \mu+\frac1\gs)}(\xi),
\end{equation}
where $\gs>0, \mu>0$  with $\gs \mu>1$,
and the (scaled) Morlet wavelet $\psi_{\rm Mor}(x)$ defined by
\begin{equation}
\label{def_Morlet0}
\wh \psi_{\rm Mor}(\xi)=e^{-2\gs^2 \pi^2(\xi-\mu)^2}-e^{-2\gs^2 \pi^2 (\xi^2+\mu^2)},
\end{equation}
where $\gs>0, \mu>0$, are the commonly used continuous wavelets.
For example, the ``bump wavelet'' $\psi_{\rm bump}(x)$  is used in  \cite{Daub_Lu_Wu11} to derive the conditions for
IF estimation and the recovery of the components from the SST of a  multicomponent signal.  In practice, Morlet's wavelet can be more desirable due to its nice localization property in both the time and frequency domains.

The parameter $\gs$ in \eqref{def_bump} and \eqref{def_Morlet0} controls the window widths of the time-frequency localization of the wavelets and has effects on both CWT and SST of a signal. In the literature, the parameter $\gs$ of the wavelets is usually treated as a fixed constant. In this paper, we consider a time-varying $\sigma$, namely $\gs=\gs(t)$ is a positive function of the time variable $t$.   As pointed out in \cite{Iatsenko15}, for a multicomponent signal $x(t)$,
if the CWTs of two components are mixed, the SST will not be able to separate these two components. Thus to separate $x(t)$ with the SST approach, we need to, first of all, separate the CWTs of the components of $x(t)$ in the time-scale plane, that is, the CWTs of the components lie  in non-overlapping regions of the time-scale plane. On the other hand, the error bounds derived in \cite{Daub_Lu_Wu11} imply that the synchrosqueezed representation of a signal is sharper when the width of the continuous wavelet's window in the time domain,
which is $\gs$ (up to a constant),  is smaller.
The main goal of this paper is (i) to study for a given multicomponent signal $x(t)$ as given in \eqref{AHMold} with $A_0(t)=0$, the conditions (called well-separated conditions) under which a suitable time-varying $\gs=\gs(t)$ can be selected such that  the corresponding CWTs (called the adaptive CWTs) of $x_k(t), 1\le k\le K$ do not overlap in the time-scale plane, and (ii)  to provide a formula and an algorithm to select as small as possible
$\gs(t)$ such that the associated SST (called the adaptive SST) of $x_k(t)$ will have a sharper representation which results in a better IF estimation and a more accurate recovery of $x_k(t)$.
In this paper, we will consider the linear chirp model, namely, we consider the case where the CWT of $x_k(t)$ is well-approximated by that of a linear chirp signal.

The adaptive SST with a time-varying window width was recently proposed in \cite{Wu17} and the width of the window is selected through minimizing the R${\rm \acute e}$nyi  entropy of the SST.
The authors of \cite{Guo14} considered the SST based on the STFT with a changing window width $\gs(t)=\frac
{0.7}{\sqrt{2\pi \phi''_{k_0}(t)}}$, where $\phi_{k_0}(t)$ is the phase function of a component of the multicomponent signal. Compared with the approach in \cite{Wu17} and \cite{Guo14},
our work focuses on establishing well-separated conditions for multicomponent signals based on the adaptive CWT and a study on how to select $\gs(t)$ such that the CWTs of the components lie in non-overlapping regions of the time-scale plane based on our well-separated condition.
Here we also remark that the window width of the signal-separation-operator algorithm in \cite{Chui_Mhaskar15} is also time-varying. After we completed our work, we were aware of the very recent work \cite{Saito17} on the adaptive STFT-based SST in which the window function has not only the time-varying parameter but also frequency-varying parameter.

The remainder of this paper is organized as follows.  First we briefly review SST in \S2.  Then we propose the adaptive CWT and SST with a time-varying parameter in \S3. In \S3, we also introduce the 2nd-order adaptive SST. We consider the support zone of a CWT of a signal with a non-bandlimited wavelet in \S4. After that, in \S5 we derive the well-separated conditions for multicomponent signals based on the adaptive CWT.
We propose a method and an algorithm to select the parameter for blind source signal separation in \S6. We provide the experimental results in \S7. Finally we give the conclusion in \S8.

\section{Synchrosqueezing transform (SST)}

A function $\psi(t) \in L_2(\R)$ is called a continuous wavelet (or an admissible wavelet) if it satisfies (see e.g. \cite{Meyer_book, Dau_book}) the admissible condition:
\begin{equation}
\label{def_C_psi}
0<C_\psi=\int_{-\infty}^\infty |\wh \psi(\xi)|^2\frac {d\xi}{|\xi|}<\infty,
\end{equation}
where $\wh \psi$ is  the Fourier transform of $\psi(t)$, defined by
\begin{equation*}
\wh \psi(\xi)=\int_{-\infty}^\infty \psi(t) e^{-i2\pi \xi t}dt.
\end{equation*}
Denote
$
\psi_{a, b}(t)= \frac 1{a}\psi \big(\frac{t-b}a\big).
$
The continuous wavelet transform (CWT) of a signal $x(t)\in L_2(\R)$ with a continuous wavelet $\psi$ is defined by
\begin{equation}
\label{def_CWT}
W_x(a, b)=\langle x, \psi_{a, b}\rangle=\int_{-\infty}^\infty x(t) \frac 1{a}\overline{\psi \big(\frac{t-b}a\big)} dt.
\end{equation}
The variables $a$ and $b$ are called the scale and time variables respectively. The signal $x(t)$ can be recovered by the inverse wavelet transform (see e.g. \cite{Chui_book,Meyer_book, Dau_book, Chui_Jiang_book})
$$
x(t)=\frac 1{C_\psi}\int_{-\infty}^\infty\int_{-\infty}^\infty W_x(a, b) \psi_{a, b}(t)db\;  \frac {da}{|a|}.
$$

A function $x(t)$ is called an analytic signal if it satisfies $\wh x(\xi)=0$ for $\xi<0$.
In this paper, we consider analytic continuous wavelets.
In addition, we assume $\psi$ also satisfies
\begin{equation}
\label{def_c_psi}
0\not= c_\psi=\int_0^\infty \overline{\wh \psi(\xi)} \frac {d\xi}{\xi}<\infty.
\end{equation}

For an analytic signal $x(t)\in L_2(\R)$, it can be recovered by 
(refer to \cite{Daub_Maes96, Daub_Lu_Wu11}):
\begin{equation}
\label{CWT_recover_analytic_a_only}
x(b)=\frac 1{c_\psi} \int_0^\infty W_x(a, b) \frac {d a}a,
\end{equation}
where $c_\psi$ is defined by \eqref{def_c_psi}. In addition,
a real signal $x(t)\in L_2(\R)$ can be recovered by the following formula (see \cite{Daub_Lu_Wu11}):
\begin{equation}
\label{CWT_recover_real_a_only}
x(b)={\rm Re }\Big(\frac 2{c_\psi} \int_0^\infty W_x(a, b) \frac {d a}a \Big).
\end{equation}

The Fourier transform and the CWT given above can be applied to a slowly growing $x(t)$ 
if the wavelet function $\psi$ has certain decay order as $|t|\to \infty$. In addition, the above two formulas still hold for such a $x(t)$.
Recall that a function $x(t)$ is called a slowly  growing function if there is a nonnegative integer $L$ such that
 $x(t)/(1+|t|^L)$ is bounded on $(-\infty, \infty)$. We will assume components of $x(t)$ in \eqref{AHMold} are all slowly growing.

\begin{figure}[th]
\centering
\begin{tabular}{ccc}
\resizebox{2.1in}{1.3in}{\includegraphics{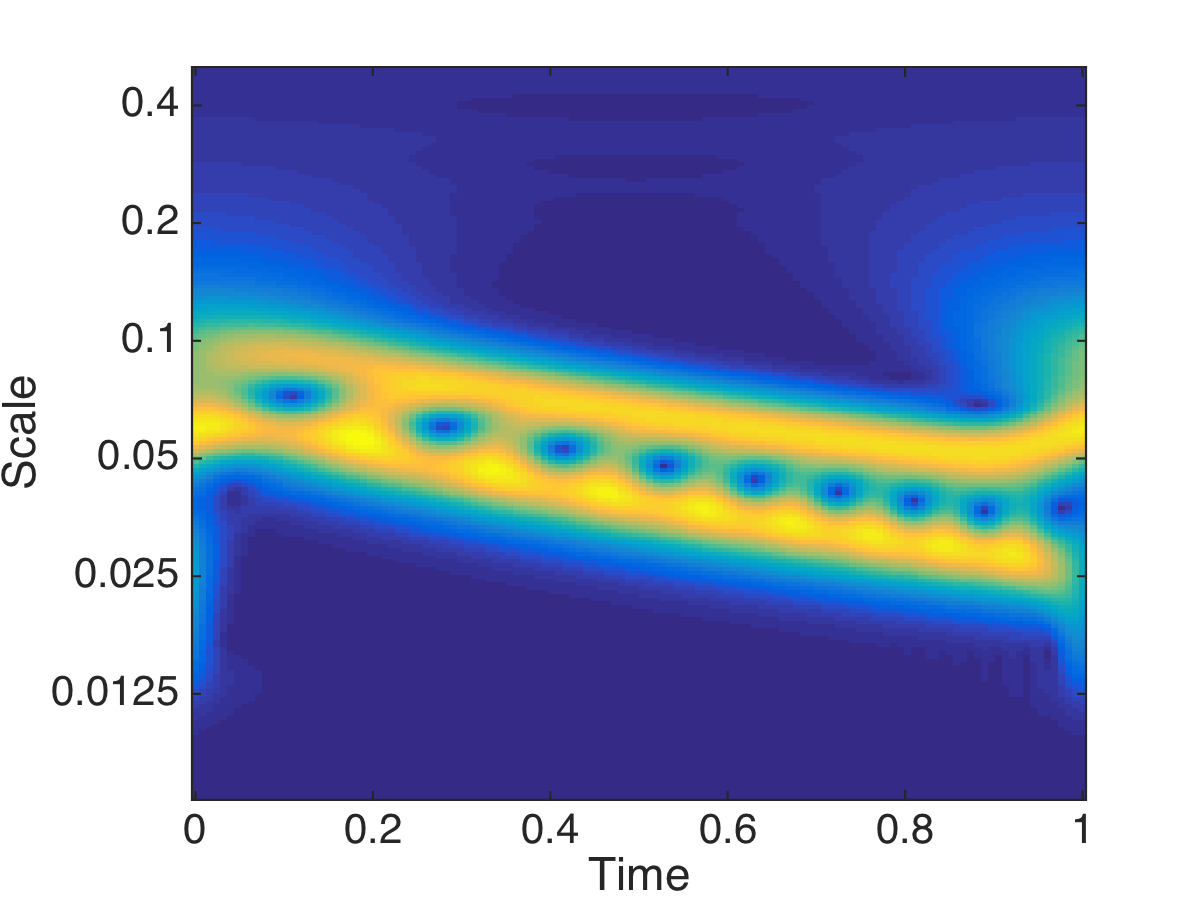}} &
\resizebox{2.1in}{1.3in}{\includegraphics{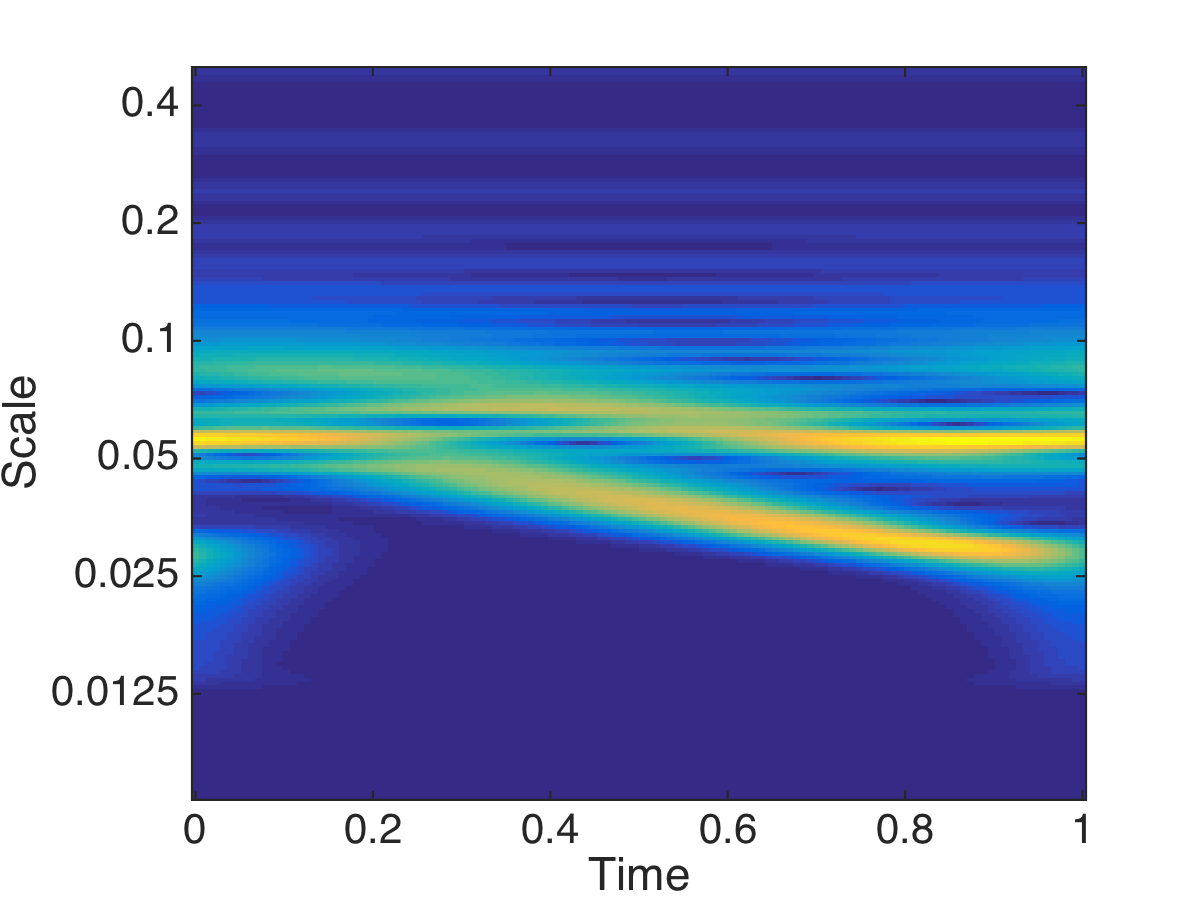}}
\resizebox{2.1in}{1.3in}{\includegraphics{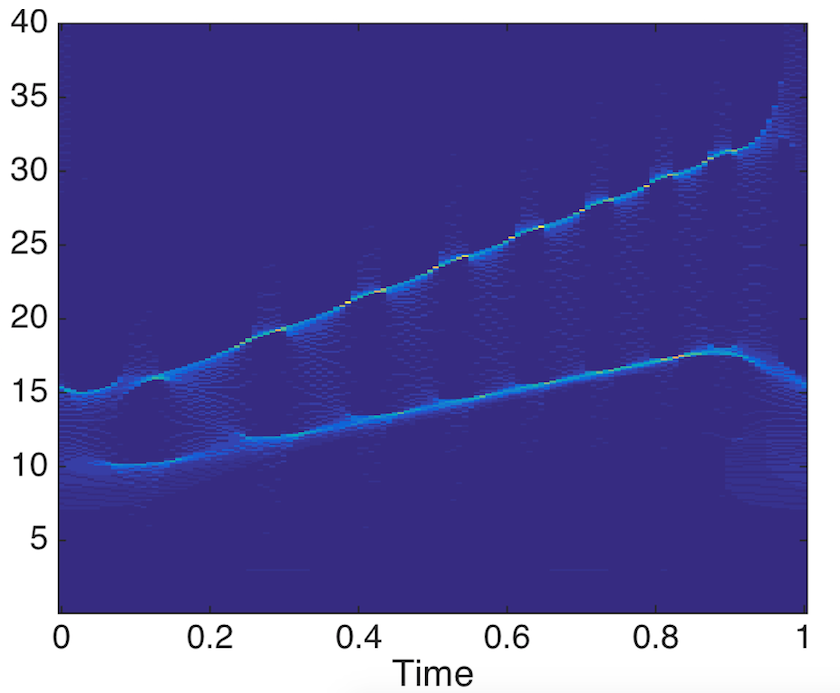}}
\end{tabular}
\caption{\small The CWT $|W_x(a, b)|$ of $x(t)$ in \eqref{def_y}
by using Morlet's wavelet $\psi$ with $\gs=1, \mu=1$ (left panel) and with
$\gs=2, \mu=1$ (middle panel).  The right panel shows $|T_{x}(\xi, b)|$, the SST of $x(t)$ with $\gs=1, \mu=1$.
}
\label{figure:CWT_twochirps_13t_9t}
\end{figure}

As mentioned earlier, the parameter $\gs$ of the ``bump wavelet" in \eqref{def_bump} or Morlet's wavelet in \eqref{def_Morlet0} controls the shape of $\psi$ and has effects on the CWT of a signal. For a simple multicomponent signal
\begin{equation}
\label{const_AHM}
x(t)=\sum_{k=1}^K A_k \cos (2\pi c_k t \big)
\end{equation}
with positive $A_k, c_k$ and $c_k\not=c_{k+1}$, if $\gs$ is large then
the CWTs of the components  $A_k \cos \big(2\pi c_k t \big)$ in $|W_x(a, b)|$ with the ``bump wavelet" will not overlap. On the other hand, for a superposition \eqref{AHMold} of AHMs with $\phi'_k(t)$$\not=$constant,
a larger $\gs$ does not necessarily provide a better separation of AHMs,
as can be illustrated by the following example with Morlet's wavelet.
Let
\begin{equation}
\label{def_y}
x(t)=e^{i 2\pi (9t+5t^2)}+e^{i 2\pi (13t+10t^2)}, \; 0\le t\le 1,
\end{equation}
which is sampled uniformly with 128 sample points. The CWT of $x(t)$ with Morlet's wavelet with $\gs=1, \mu=1$ and $\gs=2, \mu=1$ are shown
in the left and middle panels of Fig.\ref{figure:CWT_twochirps_13t_9t} respectively. Observe that the wavelet with a larger $\gs$ results in a more blurred representation of $x(t)$ in the time-scale plane.

\bigskip
To achieve a sharper time-frequency representation of a signal, the synchrosqueezed wavelet transform (SST) reassigns the scale variable $a$ to a frequency variable. For a given signal $x(t)$, let $\go_x(a, b)$ be the {phase transformation} \cite{Daub_Lu_Wu11}
(also called the {reference IF function} in \cite{Chui_Walt15}) defined by
\begin{equation}
\label{def_phase}
\go_x(a, b) =\frac{1}{i2\pi}\frac{\partial}{\partial b}\log |W_x (a,b)|= \frac{\frac {\partial}{\partial b} W_x(a, b)}{i2\pi W_x(a, b)}, \quad \hbox{for $W_x(a, b)\not=0$}.
\end{equation}
SST is to transform the CWT $W_x(a, b)$ of $x(t)$ to a quantity, denoted by $T_x(\xi, b)$, on the time-frequency plane as defined by
\begin{equation}
\label{def_SST_simple}
{  T_x(\xi, b)}=\int_{\{a \in \R_+: \; W_x(a, b)\not=0\}} W_x(a, b) \delta\big(\go_x(a, b)-\xi\big) \frac {d a}a,
\end{equation}
where $\xi$ is the frequency variable. The reader is referred to \cite{Daub_Lu_Wu11} for more details. As an example, the right panel  in Fig.\ref{figure:CWT_twochirps_13t_9t} shows the SST of $x(t)$ given in \eqref{def_y}. It displays a sharp contrast of SST against CWT in terms of the power in estimating the IFs of the components of the signal $x(t)$.

The input signal $x(t)$ can be recovered from its SST in a similar way. For an analytic $x(t)\in L_2(\R)$, by \eqref{CWT_recover_analytic_a_only}, we have
\begin{equation}
\label{reconst_SST_complex}
x(b)= \frac 1{c_\psi} \int_0^\infty T_x(\xi, b)d\xi;
\end{equation}
and for a  real-valued $x(t)\in L_2(\R)$, by \eqref{CWT_recover_real_a_only}
\begin{equation}
\label{reconst_SST}
x(b)= {\rm Re} \Big(\frac 2{c_\psi}\int_0^\infty T_x(\xi, b)d\xi\Big),
\end{equation}
where $c_\psi$ is the constant defined by \eqref{def_c_psi}.

For a multicomponent signal $x(t)$ in \eqref{AHMold} with $A_0(t)=0$, when $A_k(t), \phi_k(t)$ satisfy certain conditions (see \cite{Daub_Lu_Wu11}), each component $x_k(b)$ can be recovered from SST:
\begin{equation}
\label{reconst_SST_component}
x_k(b)\approx {\rm Re} \Big(\frac 2{c_\psi}\int_{|\xi-\phi'_k(b)|< \Gamma_1} T_x(\xi, b)d\xi\Big),
\end{equation}
for certain $\Gamma_1>0$.

\begin{figure}[th]
\centering
\begin{tabular}{ccc}
\resizebox{2.1in}{1.5in}{\includegraphics{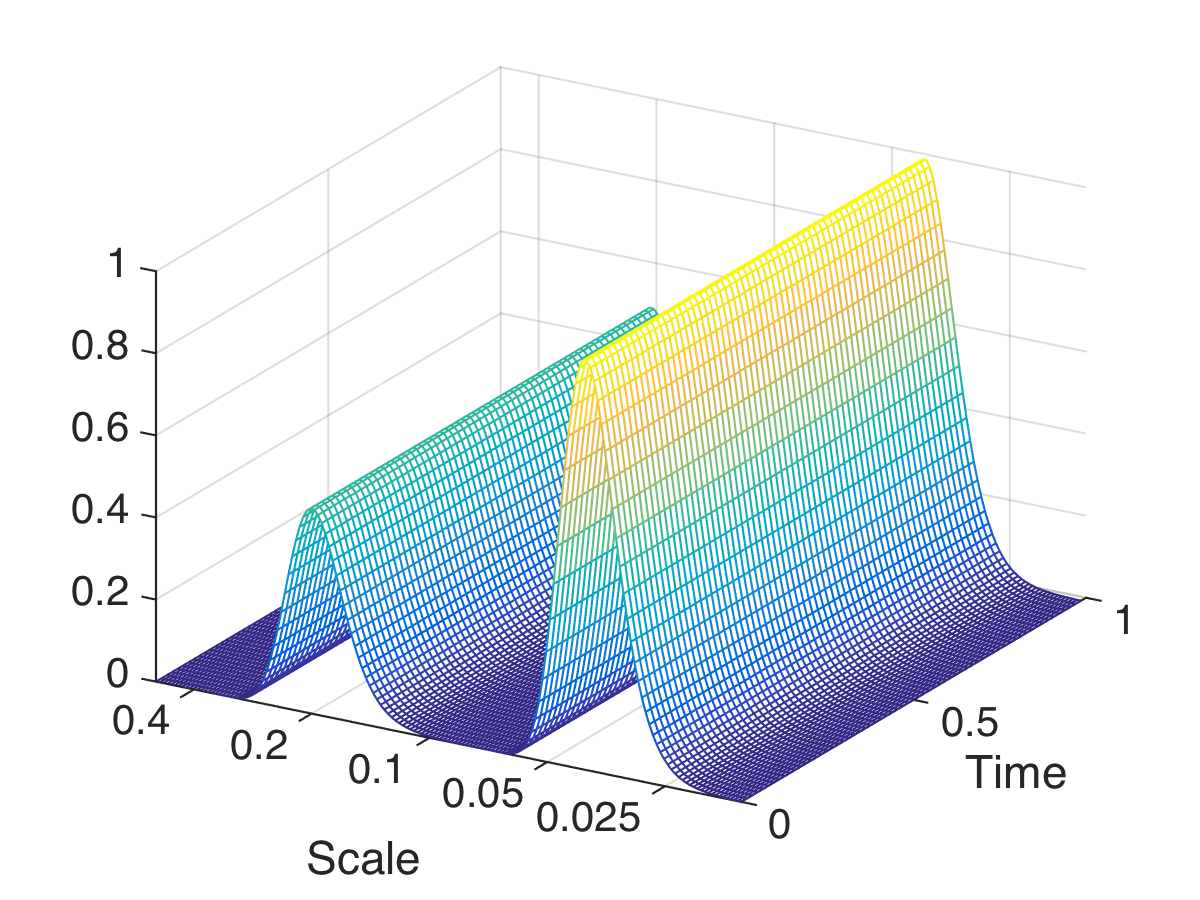}}
&
\resizebox{2.1in}{1.5in}{\includegraphics{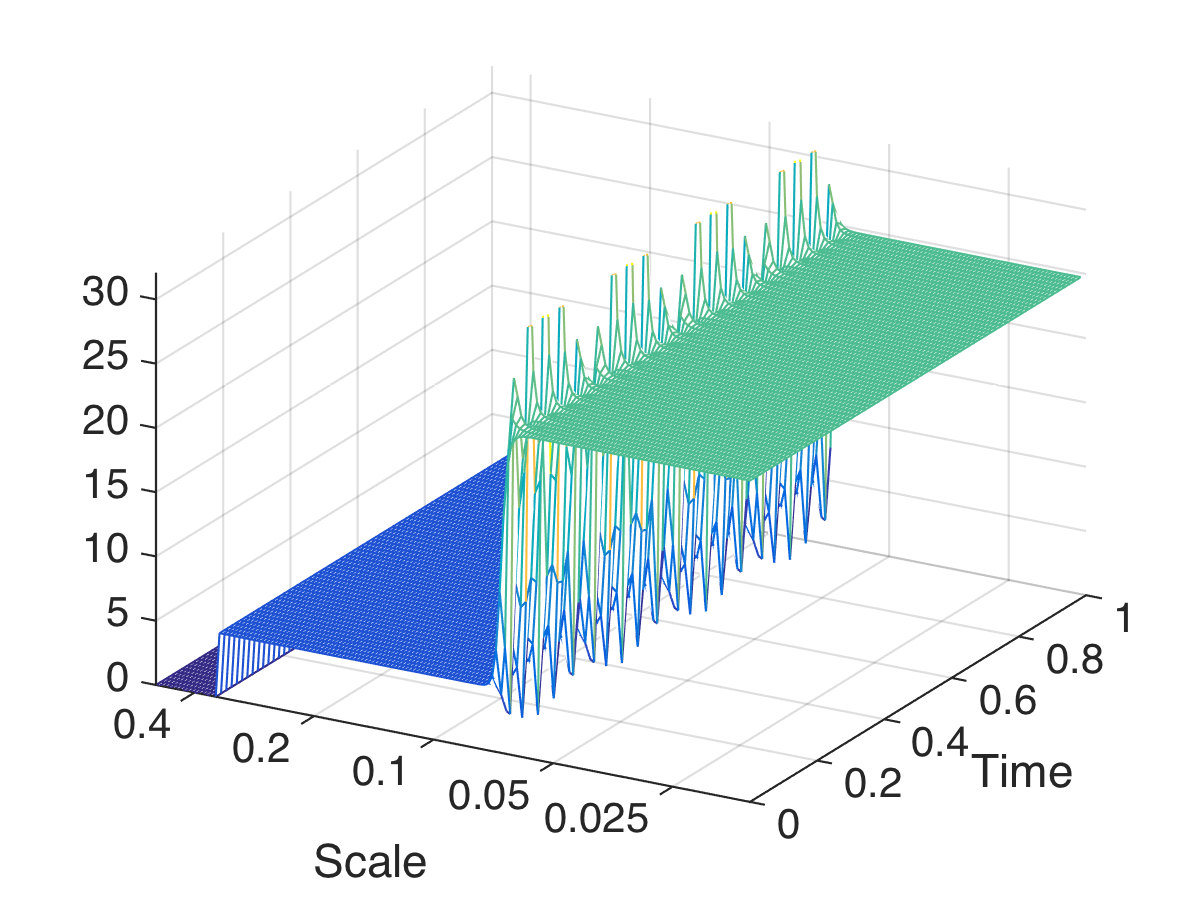}}
\resizebox{2.1in}{1.5in}{\includegraphics{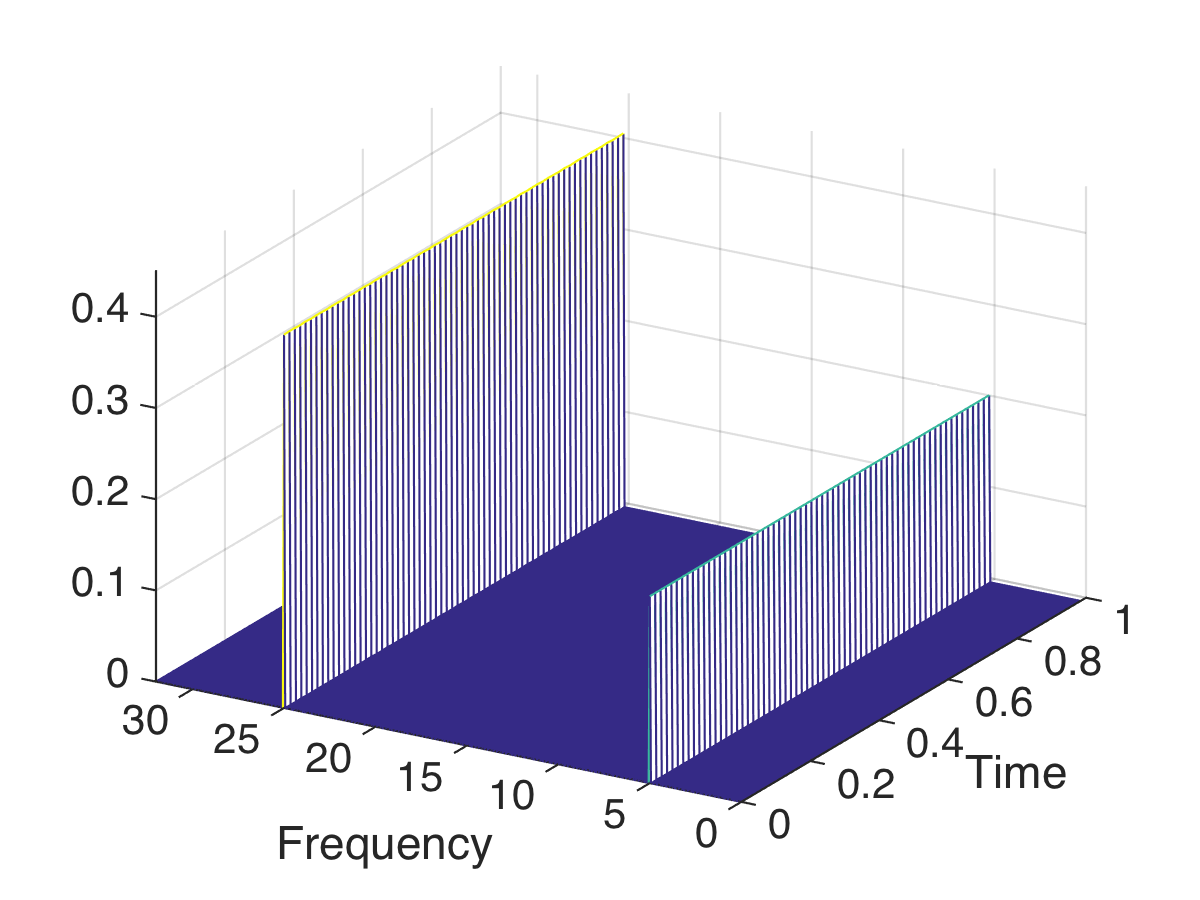}}
&
\end{tabular}
\caption{\small  Left:  $|W_r(a, b)|$,  CWT of $r(t)=r_1(t)+r_2(t)=\cos\big(2\pi(5t)\big)+2\cos\big(2\pi(25t)\big)$; Middle:
 $\go_r(a, b)$ with $\gga=10^{-5}$;
Right: $|T_{r}(\xi, b)|$, the SST of $r(t)$.
}
\label{fig:SST_5_25t}
\end{figure}

Here is an example of SST. Let $r(t)=r_1(t)+r_2(t)$ with
$r_1(t)=\cos\big(2\pi(5t)\big)$ and $r_2(t)=2\cos\big(2\pi(25t)\big)$. The sample points are $t_n=\frac n{64}$, $0\le n\le 63$.
Fig.\ref{fig:SST_5_25t} shows the CWT of $r(t)$ with Morlet's wavelet of $\gs=1$ and $\mu=1$, the phase transformation $\go_{r}(a, b)$ with $\gga=10^{-5}$, and
the SST of $r(t)$. The two bumps in the left panel of Fig.\ref{fig:SST_5_25t} correspond to the CWTs of $r_1(t)$ and $r_2(t)$ respectively. The phase transformation $\go_{r}(a, b)$ in the middle panel of Fig.\ref{fig:SST_5_25t} takes constant values $5$ and $25$ respectively for $a$ in two intervals and for all $b$. These are the IFs of the two components of $r(t)$. Note also from this panel that at the boundary  $a_0$ between the two intervals of the scale variable, $\go_r(a_0, b)$ have large values for some $b$. However, since the corresponding $W_r(a_0, b)$ is small, $W_r(a_0, b) a_0^{-1}$ is also small.
Thus we still have two sharp representations of the IFs of $r_1(t)$ and $r_2(t)$ through SST, as shown in the right panel of Fig.\ref{fig:SST_5_25t}.

\section{CWT and SST with a time-varying parameter}

\subsection{CWT with a time-varying parameter}

We consider continuous wavelets of the form
\begin{equation}
\label{wavelet_general0}
\psi_\gs(t)=\frac 1\gs \overline{g(\frac t\gs)} e^{i2\pi \mu t}-\frac 1\gs  \overline{g(\frac t\gs)} c_\gs(\mu),
\end{equation}
or, in the frequency domain,
\begin{equation}
\label{wavelet_general0_freq}
\wh \psi_\gs(\xi)=\overline{\wh g\big(\gs(\mu-\xi)\big)}- c_\gs(\mu) \overline{\wh g(-\gs \xi)},
\end{equation}
where $\mu>0$,  $g$ is a function in $L_2(\R)$ with certain decaying order as $t\rightarrow \infty$,
and $c_\gs(\mu)$ is a constant such that $\wh \psi_\gs(0)=0$. If $\wh g\big(\gs \mu)=0$, then we just set $c_\gs(\mu)=0$; otherwise, if in addition $\wh g(0)\not=0$, we let $c_\gs(\mu)=\overline{\wh g(\gs \mu)}/\overline{\wh g(0)}$. For example, if $g(t)$ is given by $\wh g(\xi)= e^{1-\frac 1{1-\xi^2}} \chi_{(-1, 1)}(\xi)$, then $\psi_\gs(t)=\frac 1\gs g(\frac t\gs) e^{i2\pi \mu t}$ is the ``bump wavelet" defined in \eqref{def_bump}, and if
\begin{equation}
\label{gaussian}
g(t)=\frac 1{\sqrt {2\pi}} e^{-\frac {t^2}2},
\end{equation}
then $\psi_\gs$ is Morlet's wavelet in \eqref{def_Morlet0}.

In the following, we will assume our signal $x(t)$ to be a slowly increasing function.
The CWT of such an $x(t)$ with the $\psi_\gs$ considered above is well-defined as long as $g(t)$ decays to $0$ fast enough as $t\rightarrow \infty$.

As observed from Fig.\ref{figure:CWT_twochirps_13t_9t}, the choice of the parameter $\gs$ for the wavelet $\psi_\gs$ affects the representation of the CWT.
In this paper, we introduce a CWT with time-varying $\gs$. More specifically, let $\psi_\gs$ be a continuous wavelet defined by \eqref{wavelet_general0} and $x(t)$ be a given signal (a slowly increasing function). The CWT of $x(t)$ with a time-varying parameter is defined by
\begin{equation}
\label{def_CWT_para_old}
\wt W_x(a, b)=
\int_{-\infty}^\infty x(t) \frac 1{a}\overline{\psi_{\gs(b)} \big(\frac{t-b}a\big)} dt.
\end{equation}
where $\gs$ is a positive function of $b$.
We call $\wt W_x(a, b)$ the adaptive CWT of $x(t)$ with $\psi_\gs$.
One can easily obtain
\begin{equation*}
{ \wt W_x(a, b)}
=\int_{-\infty}^\infty \wh x(\xi) \overline{\wh \psi_{\gs(b)} \big(a \xi\big)} e^{i2\pi b \xi} d\xi.
\end{equation*}
Thus, if $\psi_\gs$ or $x(t)$ is analytic, then we have for $a>0$,
\begin{equation}
\label{CWT_xreqdomain_analytic}
{ \wt W_x(a, b)}
=\int_0^\infty \wh x(\xi) \overline{\wh \psi_{\gs(b)} \big(a \xi\big)} e^{i2\pi b \xi} d\xi.
\end{equation}

As shown in the following proposition, the original signal $x(b)$ can be recovered from $\wt W_x(a, b)$ by formulas similar to \eqref{CWT_recover_analytic_a_only} and  \eqref{CWT_recover_real_a_only}.
\begin{pro}
\label{pro:recover_CWT_para}
Let $\wt W_x(a, b)$  be the time-varying CWT of a signal $x(t)$ defined by \eqref{def_CWT_para}. 
Then the following holds.

\begin{itemize}
\item[{\rm (1).}] If $x(t)$ is analytic, then it can be recovered by
\begin{equation}
\label{CWT_para_recover_analytic_a_only}
x(b)=\frac 1{c_\psi(b)} \int_0^\infty \wt W_x(a, b) \frac {d a}a,
\end{equation}
where $c_\psi(b)$ is defined by
\begin{equation}
\label{def_c_psi_para}
c_\psi(b)
=\int_0^\infty \overline{\wh \psi_{\gs(b)}(\xi)}
\frac {d\xi}{\xi}
\end{equation}
\item[{\rm (2).}] In addition, if $\psi_{\sigma}$ is analytic, then for real-valued $x(t)$ we have
\begin{equation}
\label{CWT_para_recover_real_a_only}
x(b)={\rm Re }\Big(\frac 2{c_\psi(b)} \int_0^\infty \wt W_x(a, b) \frac {d a}a \Big).
\end{equation}
\end{itemize}
\end{pro}

The proof  Proposition \ref{pro:recover_CWT_para}  is straightforward in the sense that it can be followed directly from that in \cite{Daub_Lu_Wu11} for the conventional CWT. For self-containedness, it is provided in Appendix.

\bigskip

We remark that, in practice $\wh g$ is usually chosen to be a fast decaying function, and thus
numerically, the second term in  \eqref{wavelet_general0_freq},
$c_\gs(\mu) \overline{\wh g(-\gs \xi)}$,  is very small.
For example, if $\psi_\gs$ is Morlet's wavelet, the second term
in \eqref{wavelet_general0_freq}  equals $e^{-2\gs^2 \pi^2 (\xi^2+\mu^2)}$.  
When $\mu=1$ and $\gs=1$,
$e^{-2\gs^2 \pi^2 (\xi^2+\mu^2)}\le  \exp(-2 \pi^2)=2.6753 \times 10^{-9}$, a negligible quantity. 
Thus for the simplicity of presentation, we will assume \begin{equation}
\label{wavelet_general}
\psi_\gs(t)= \frac 1\gs \overline{g(\frac t\gs)} e^{i2\pi \mu t}
\end{equation}
or equivalently
\begin{equation}
\label{wavelet_general_freq}
\wh {\psi}_\gs(\xi)=\overline{\wh {g}(\gs(\mu-\xi))},
\end{equation}
and the associated adaptive CWT $\wt W_x(a, b)$ is defined by
\begin{equation}
\label{def_CWT_para}
\wt W_x(a, b)=
\int_{-\infty}^\infty x(b+a t) \frac 1{\gs(b)} g\big( \frac t{\gs(b)} \big) e^{-i2\pi \mu t} dt.
\end{equation}
In particular, when $g$ is the Gaussian function given in \eqref{gaussian}, then
\begin{equation}
\label{def_Morlet0_time}
\psi_\gs(t)=\frac 1{\gs \sqrt {2\pi}}e^{-\frac 12 (\frac t \gs)^2}e^{i2\pi \mu t},
\end{equation}
or equivalently
\begin{equation}
\label{def_Morlet}
\wh \psi_\gs(\xi)=e^{-2\pi^2 \gs^2  (\xi-\mu)^2},
\end{equation}
is the simplified version of Morlet's wavelet.

We note that the improper integrals in Proposition \ref{pro:recover_CWT_para} will converge with this simpler $\psi_\gs$ if we exclude a small neighborhood of the origin in integrations and the numerical results are close approximations of original integrals.

\subsection{SST with a time-varying parameter}

We now define the phase transformation $\go^{adp}_x(a, b)$ associated with the adaptive CWT. To this regard, denote
$g_2(t)=t g'(t)$, and we use
$\wt W^{g_2}_x(a, b)$ to denote the CWT defined by \eqref{def_CWT_para} with $g$ replaced by 
$g_2$, namely,
$$
\wt W^{g_2}_x(a, b)=
\int_{-\infty}^\infty x(b+at)\frac t{\gs^2(b)} g'(\frac t{\gs(b)})e^{-i2\pi \mu t} dt.
$$

To motivate the definition of the phase transformation $\go^{adp}_x(a, b)$ to be given below, let us look at a simple example  $x(t)=s(t)=A e^{i 2\pi c t}$.
From
$$
\wt W_s(a, b)
=A\int_{-\infty}^\infty e^{i2\pi c (b+at)} \frac 1{\gs(b)}{g(\frac t{\gs(b)})}e^{-i2\pi \mu t} dt,
$$
we have
\begin{eqnarray*}
&&\frac {\partial} {\partial b} \wt W_s(a, b)=A\int_{-\infty}^\infty (i2\pi c)e^{i2\pi c (b+at)}\;\frac 1{\gs(b)}{g(\frac t{\gs(b)})}e^{-i2\pi \mu t} dt
\\
&&\qquad + A \int_{-\infty}^\infty e^{i2\pi c (b+at)}(-\frac {\gs'(b)}{\gs(b)^2}){g(\frac t{\gs(b)})}e^{-i2\pi \mu t} dt
 + A\int_{-\infty}^\infty e^{i2\pi c (b+at)} (-\frac {\gs'(b) t}{\gs(b)^3}){g'(\frac t{\gs(b)})}e^{-i2\pi \mu t} dt\\
&&=i2\pi c\; \wt W_s(a, b)- \frac {\gs'(b)}{\gs(b)}\; \wt W_s(a, b) -\frac {\gs'(b)}{\gs(b)}\; \wt W^{g_2}_s(a, b).
\end{eqnarray*}
Thus, if $\wt W_s(a, b)\not=0$, we have
\begin{equation*}
\frac {\frac {\partial}{\partial b} \wt W_s(a, b)}{i 2\pi \wt W_s(a, b)}=
c- \frac {\gs'(b)}{i2\pi \gs(b)}- \frac {\gs'(b)}{\gs(b)} \frac {\wt W^{g_2}_s(a, b)}{i2\pi \wt W_s(a, b)}.
\end{equation*}
Therefore,  the IF of $s(t)$, which is $c$,  can be obtained by
\begin{equation}
\label{para_derivation}
c=\frac {\frac{\partial}{\partial b} \wt W_s(a, b)}{i2\pi \wt W_s(a, b)}
+ \frac {\gs'(b)}{\gs(b)} \frac {\wt W^{g_2}_s(a, b)}{i2\pi \wt W_s(a, b)} +\frac {\gs'(b)}{i2\pi \gs(b)}.
\end{equation}

Following this example, we define, for a general $x(t)$ and at $(a, b)$ for which $\wt W_x(a, b)\not=0$,  the {\it phase transformation} or the {\it reference IF function} to be the real part of the quantity on the right-hand side of \eqref{para_derivation}:
\begin{equation}
\label{def_phase_para}
\go^{adp}_x(a, b)={\rm Re}\Big\{\frac{\partial_b\big(\wt W_x(a, b)\big)}{i2\pi \wt W_x(a, b)}\Big\}+ \frac {\gs'(b)}{\gs(b)} {\rm Re}\Big\{ \frac {\wt W^{g_2}_x(a, b)}{i2\pi \wt W_x(a, b)} \Big\},  \quad \hbox{for $\wt W_x(a, b)\not=0$}.
\end{equation}

The SST with a time-varying parameter (also called the adaptive SST of $x(t)$) is defined by
\begin{equation}
\label{def_SST_para_simple}
T^{adp}_x(\xi, b)=\int_{\{a \in \R_+: \; \wt W_x(a, b)\not=0\}} \wt W_x(a, b) \delta\big(\go^{adp}_x(a, b)-\xi\big) \frac {d a}a,
\end{equation}
where $\xi$ is the frequency variable. For an analytic $x(t)\in L_2(\R)$, by \eqref{CWT_para_recover_analytic_a_only}, we have
\begin{equation}
\label{reconst_SST_para_complex}
x(b)= \frac 1{c_\psi(b)} \int_0^\infty T^{adp}_x(\xi, b)d\xi;
\end{equation}
and for a  real-valued $x(t)\in L_2(\R)$, by \eqref{CWT_para_recover_real_a_only}
\begin{equation}
\label{reconst_SST_para}
x(b)= {\rm Re} \Big(\frac 2{c_\psi(b)}\int_0^\infty T^{adp}_x(\xi, b)d\xi\Big),
\end{equation}
where $c_\psi(b)$ is defined by \eqref{def_c_psi_para}. In addition, we can use the following formula to recover the $k$th component $x_k(b)$ of a multicomponent signal (satisfying certain conditions) from the adaptive SST:
\begin{equation}
\label{reconst_SST_para_component}
x_k(b)\approx {\rm Re} \Big(\frac 2{c_\psi(b)}\int_{|\xi-\phi'_k(b)|< \Gamma_2} T^{adp}_x(\xi, b)d\xi\Big),
\end{equation}
for certain $\Gamma_2>0$.

Here we remark that if $\psi_\gs$ is a simplified version of Morlet's wavelet
given by \eqref{def_Morlet},  then
\begin{equation*}
\wt W^{g_2}_s(a, b)=(4\pi^2\gs^2(b)(a c -\mu)^2-1)\wt W_s(a, b).
\end{equation*}
Thus
$$
\frac {\wt W^{g_2}_s(a, b)}{i2\pi \wt W_s(a, b)} =\frac 1{i2\pi} (4\pi^2\gs^2(a c-\mu)^2-1).
$$
Therefore, the second term on the right-hand side of \eqref{def_phase_para} is zero, and hence, one may define
\begin{equation*}
\wt \go_x(a, b)={\rm Re}\Big\{\frac{\partial_b\big(\wt W_x(a, b)\big)}{i2\pi \wt W_x(a, b)}\Big\},  \quad \hbox{for $\wt W_x(a, b)\not=0$}.
\end{equation*}
as the phase transformation.

\subsection{Second-order SST with a time-varying parameter}
The 2nd-order SST was introduced in \cite{MOM15}.
The main idea is to define a new phase transformation $\go_x^{2nd}$ which is associated with the 2nd order partial derivatives of the CWT of $x(t)$ such that when $x(t)$ is a linear frequency modulation (LFM) signal (linear chirp), then $\go_x^{2nd}$ is exactly the IF of $x(t)$. We say $s(t)$ is an LFM signal if
\begin{equation}
\label{def_chip_At}
s(t)=A(t) e^{i2\pi \phi(t)}=Ae^{pt+\frac q2 t^2} e^{i2\pi (ct +\frac 12 r t^2)}
\end{equation}
with phase function $\phi(t)=ct +\frac 12 r t^2$, the IF  $\phi'(t)=c +r t$, chirp rate  $\phi''(t)=r$, the instantaneous amplitude (IA)
$A(t)=Ae^{pt+\frac q2 t^2}$, where $p, q$ are real numbers and $|p|$ and $|q|$ are much smaller than $c$, which is positive.

Now we show how to derive the phase transformation $\go_s^{2nd}$. Note that our derivation is slightly different from that in \cite{MOM15} and \cite{OM17}, where it was based on reassignment operators. The formulation for $\go_s^{2nd}$ provided here is also slightly different from that in \cite{OM17}. Our derivation can easily be generalized to the case of adaptive CWT and SST.

For a given wavelet $\psi$, let $W_s(a, b)$ be the CWT of a signal $s(t)$ with $\psi$ as defined in \eqref{def_CWT}. For $\psi_1(t)=t\psi(t)$, let  $W_s^{\psi_1}(a, b)$ denote the CWT of $s(t)$ with $\psi_1(t)$, namely, the integral on the right-hand side of \eqref{def_CWT} with $x(t)$ and $\psi(t)$ replaced by $s(t)$ and $\psi_1(t)$ respectively.

Observe that for $s(t)$ given by \eqref{def_chip_At}
\begin{equation*}
s'(t)=\big(p+qt+i2\pi (c+rt )\big) s(t).
\end{equation*}
Thus from
$$
W_s(a, b)=\int_{-\infty}^\infty s(b+at)\; \overline{\psi(t)}dt,
$$
we have
\begin{eqnarray*}
&&
\frac{\partial}{\partial b} W_s(a, b)
=\int_{-\infty}^\infty s'(b+at)\; \overline{\psi(t)}dt\\
&&=\int_{-\infty}^\infty\big(p+q(b+at)+i2\pi (c+rb +ra t )\big) s(b+at)\; \overline{\psi(t)}dt\\
&&=(p+qb+i2\pi (c+rb)\big)W_s(a, b)+(q+i 2\pi r) a \; W^{\psi_1}_s(a, b).
\end{eqnarray*}
Thus at $(a, b)$ on which $W_s(a, b)\not=0$, we have
\begin{equation}
\label{2nd_derivation}
\frac {\frac{\partial}{\partial b} W_s(a, b)}{W_s(a, b)}=
p+q b+i2\pi (c+rb) +(q+i 2\pi r) a \; \frac {W^{\psi_1}_s(a, b)}{W_s(a, b)}.
\end{equation}
Taking partial derivative $\frac {\partial }{\partial a}$ to both sides of \eqref{2nd_derivation}, we have
$$
\frac {\partial }{\partial a}
\Big(
\frac {\frac{\partial}{\partial b} W_s(a, b)}{W_s(a, b)}
\Big)=
(q+i 2\pi r) U(a, b),
$$
where we use $U(a, b)$ to denote
$$
U(a, b)= \frac {\partial }{\partial a}\Big(\frac{a W^{\psi_1}_s(a, b)}{W_s(a, b)}\Big)
=\frac{W^{\psi_1}_s(a, b)}{W_s(a, b)}+a \frac {\partial }{\partial a}\Big(\frac{W^{\psi_1}_s(a, b)}{W_s(a, b)}\Big).
$$
Thus if $U(a, b)\not=0$, then
$$
q+i 2\pi r =\frac 1{U(a, b)} \frac {\partial }{\partial a}
\Big(\frac {\frac{\partial}{\partial b} W_s(a, b)}{W_s(a, b)}\Big).
$$
Back to \eqref{2nd_derivation}, we have
$$
\frac {\frac{\partial}{\partial b} W_s(a, b)}{W_s(a, b)}=
p+qb+i2\pi (c+rb) + a \; \frac{W^{\psi_1}_s(a, b)}{W_s(a, b)U(a, b)} \; \frac {\partial }{\partial a}
\Big(\frac{ \frac {\partial} {\partial b} W_s(a, b)}{W_s(a, b)}\Big).
$$
Therefore,
$$
\phi'(b)=c+rb ={\rm Re}\Big\{\frac {\frac{\partial}{\partial b} W_s(a, b)}{i2\pi W_s(a, b)}\Big\}
- a \;{\rm Re}\Big\{ \frac{W^{\psi_1}_s(a, b)}{W_s(a, b)U(a, b)} \; \frac {\partial }{\partial a}
\Big(\frac{ \frac {\partial} {\partial b} W_s(a, b)}{i2\pi W_s(a, b)}\Big)\Big\}.
$$
Hence, one may define the phase transformation as
\begin{equation}
\label{2nd_phase}
\go^{2nd}_s(a, b)=\left\{
\begin{array}{ll}
{\rm Re}\Big\{\frac {\frac{\partial}{\partial b} W_s(a, b)}{i2\pi W_s(a, b)}\Big\}
- a \;{\rm Re}\Big\{ \frac{W^{\psi_1}_s(a, b)}{W_s(a, b)U(a, b)} \; \frac {\partial }{\partial a}
\Big(\frac{ \frac {\partial} {\partial b} W_s(a, b) }{i2\pi W_s(a, b)}\Big)\Big\}, &\hbox{if $U(a, b)\not=0, W_s(a, b)\not=0,$}\\
{\rm Re}\Big\{\frac {\frac{\partial}{\partial b} W_s(a, b)}{i2\pi W_s(a, b)}\Big\}, &\hbox{if $U(a, b)=0, W_s(a, b)\not=0.$}
\end{array}
\right.
\end{equation}
From the above derivation, we know $\go^{2nd}_s(a, b)$ is exactly the IF $\phi'(t)$ of $s(t)$ if $s(t)$ is an LFM signal  given by \eqref{def_chip_At}. For a signal $x(t)$, with the phase transformation $\go^{2nd}_x(a, b)$ in \eqref{2nd_phase}, the 2nd-order SST of a signal $x(t)$ is defined by
\begin{equation}
\label{def_2ndSST_simple}
T^{2nd}_x(\xi, b)=\int_{\{a \in \R_+: \; W_x(a, b)\not=0\}} W_x(a, b) \delta\big( \go^{2nd}_x(a, b)-\xi\big) \frac {d a}a,
\end{equation}
where $\xi$ is the frequency variable.

\bigskip

Next we consider the CWT with a time-varying parameter.
Recall that $\wt W^{g_2}_s(a, b)$ denotes the adaptive CWT defined by \eqref{def_CWT_para} with $g$ replaced by $g_2(t)=tg'(t)$. Now we define $g_1(t)=t g(t)$ and
use $\wt W^{g_1}_s(a, b)$ to denote the CWT defined by \eqref{def_CWT_para} with $g$ replaced by $g_1$,
namely,
\begin{equation*}
\wt W^{g_1}_s(a, b)
=\int_{-\infty}^\infty s(b+at) \frac t{\gs^2(b)}{g(\frac t{\gs(b)})}e^{-i2\pi \mu t} dt.
\end{equation*}

For a signal $x(t)$, in the following we define the phase transformation as
\begin{equation}
\label{2nd_phase_para}
\go^{2adp}_x(a, b)=\left\{
\begin{array}{l}
{\rm Re}\Big\{\frac {\frac{\partial}{\partial b} \wt W_x(a, b)}{i2\pi \wt W_x(a, b)}\Big\}
+ \frac {\gs'(b)}{\gs(b)} {\rm Re}\Big\{ \frac {\wt W^{g_2}_x(a, b)}{i2\pi \wt W_x(a, b)} \Big\}
- a \;{\rm Re}\Big\{ \frac{\wt W^{g_1}_x(a, b)}{i2\pi \wt W_x(a, b)} R_0(a, b)\Big\},\\
\hskip 4cm \hbox{if $\frac {\partial}{\partial a}\Big(a \frac {\wt W^{g_1}_x(a, b)}{\wt W_x(a, b)}\Big)\not=0$ and $\wt W_x(a, b)\not=0;$}\\
\\
{\rm Re}\Big\{\frac {\frac{\partial}{\partial b} \wt W_x(a, b)}{i2\pi \wt W_x(a, b)}\Big\}+ \frac {\gs'(b)}{\gs(b)} {\rm Re}\Big\{ \frac {\wt W^{g_2}_x(a, b)}{i2\pi \wt W_x(a, b)} \Big\},
\hbox{if $\frac {\partial}{\partial a}\Big(a \frac {\wt W^{g_1}_x(a, b)}{\wt W_x(a, b)}\Big)=0,  \wt W_x(a, b)\not=0$}
\end{array}
\right.
\end{equation}
where
\begin{equation}
\label{def_R0}
R_0(a, b)=\frac 1{\frac {\partial}{\partial a}\Big(a \frac {\wt W^{g_1}_x(a, b)}{\wt W_x(a, b)}\Big) }\Big\{\frac {\partial}{\partial a}\Big(\frac {\frac {\partial}{\partial b} \wt W_x(a, b)}{\wt W_x(a, b)}\Big)+ \frac {\gs'(b)}{\gs(b)}
\frac {\partial}{\partial a}\Big(\frac {\wt W^{g_2}_x(a, b)}{\wt W_x(a, b)}\Big)\Big\}.
\end{equation}
We have the following theorem with its proof given in Appendix.
\begin{theo}
\label{theo:2nd_phase_para}
If $x(t)$ is an LFM signal given by \eqref{def_chip_At}, then at $(a, b)$ where $\frac {\partial}{\partial a}\Big(a \frac {\wt W^{g_1}_x(a, b)}{\wt W_x(a, b)}\Big)\not=0$ and $\wt W_x(a, b)\not=0$,
$\go^{2adp}_x(a, b)$ defined by \eqref{2nd_phase_para} is the IF of $x(t)$, namely $\go^{2adp}_x(a, b)=c+r b$.
\end{theo}

With the phase transformation $\go^{2adp}_x(a, b)$ in \eqref{2nd_phase_para}, we define the 2nd-order SST with a time-varying parameter (also called the 2nd-order adaptive SST) of a signal $x(t)$ as in  \eqref{def_SST_para_simple}:
\begin{equation}
\label{def_2ndSST_para_simple}
T^{2adp}_x(\xi, b)=\int_{\{a \in \R_+: \; \wt W_x(a, b)\not=0\}} \wt W_x(a, b) \delta\big(\go^{2adp}_x(a, b)-\xi\big) \frac {d a}a,
\end{equation}
where $\xi$ is the frequency variable. We also have the reconstruction formulas for $x(t)$ and $x_k(t)$ similar to \eqref{reconst_SST_para_complex},  \eqref{reconst_SST_para} and \eqref{reconst_SST_para_component} with $T^{adp}_x(\xi, b)$ replaced by $T^{2adp}_x(\xi, b)$.

\section{Support zones of CWTs of linear frequency modulation signals}

In this section we consider the support zone of CWT in the time-scale plane.
The ``bump wavelet'' $\psi_{\rm bump}$ is bandlimited (namely, $\wh \psi_{\rm bump}$ is compactly supported), and hence it has a better frequency localization than Morlet's wavelet.
On the other hand,
Morlet's wavelet as given in \eqref{def_Morlet0} or its simplified version given by \eqref{def_Morlet0_time}
enjoys a nice localization property in both the time and frequency domains. We will focus on  simplified  Morlet's wavelet below.

Now let $x(t)$ be a multicomponent signal as given in \eqref{AHMold} with $A_0(t)=0$. Recall the fact (see the discussion in  \cite{Iatsenko15}) that
if the CWTs $W_{x_{k-1}}(a, b)$ and $W_{x_{k}}(a, b)$ of two components $x_{k-1}(t)$ and $x_k(t)$ are mixed, then the SST approach is unable to separate these components.
In addition, as observed from Fig.\ref{figure:CWT_twochirps_13t_9t} that the choice of the parameter $\gs$ for the wavelet affects the representation of the CWT. Our goal is to formulate the conditions (called well-separated conditions) such that we can find (if possible) a suitable positive function $\gs(b)$ of $b$ with which the corresponding adaptive CWTs of different components $x_k(t)$ defined in \eqref{def_CWT_para} are well separated, and hence,  the associated adaptive SST can separate all components $x_k(t)$ of $x(t)$.

To study the separability of CWTs (including CWTs with a time-varying parameter) of different components $x_k(t)$ of $x(t)$, we need to consider the support zone of $W_{x_k}(a, b)$ in the time-scale plane, the region outside which $W_{x_k}(a, b)\approx 0$ . For $s(t)=A e^{i2\pi ct}$, for example, its CWT  $W_{s}(a, b)$ with an analytic wavelet $\psi$ is  given by
\begin{equation*}
\label{CWT_constant}
W_s(a, b)=\frac 12 { A  \; \overline{\wh \psi \big(a c \big)} e^{i 2\pi b c}}.
\end{equation*}
Thus the support zone of $W_s(a, b)$ in the time-scale plane is determined by the region outside which $\wh \psi(\xi)\approx 0$.
Therefore, first of all, we need to define the ``support'' of  $\wh \psi$.
For the ``bump wavelet" $\psi_{\rm bump}$, it is bandlimited, and the support of $\wh \psi_{\rm bump}$
is $[-\frac 1\gs, \frac 1\gs]$.  If $\psi$ is non-bandlimited, the corresponding CWTs  $W_{x_{k-1}}(a, b)$ and  $W_{x_k}(a, b)$  overlap theoretically even for the case when $x_{k-1}(t)$ and $x_k(t)$ are sinusoidal signals. For example, the CWTs $W_{r_1}(a, b)$ and  $W_{r_2}(a, b)$ of $r_1(t)$ and $r_2(t)$ with Morlet's wavelet in Fig.\ref{fig:SST_5_25t} overlap. However, the values of these CWTs are very small over the overlapping region and are hardly noticeable. Instead, what we can see in Fig.\ref{fig:SST_5_25t} are two bumps lying in two separated zones of the time-scale plane.
In such a case we can treat $W_{x_k}(a, b)$ as zero whenever its value is small. We describe this mathematically. Given threshold $0<\tau_0<1$, if a function $h(\xi)$ satisfies $|h(\xi)|/\max_{\xi}{h(\xi)}<\tau_0$ for $|\xi|\ge \xi_0$, then we say $h(\xi)$ is ``supported'' in $[-\xi_0, \xi_0]$. In particular, for the Gaussian function $g(t)$ defined by \eqref{gaussian} with $\wh g(\xi)=e^{-2\pi^2\xi^2}$, if $\wh g(\ga)=\tau_0$ then we have
\begin{equation}
\label{def_ga}
\ga=\frac 1{2\pi}\sqrt{2\ln (1/\tau_0)}.
\end{equation}
Thus we regard that $\wh g$ vanishes outside  $[-\ga, \ga ]$ and hence $\wh g$ is ``supported`` in $[-\ga, \ga ]$. We use $L_{\wh g}$ to denote the length of the ``support'' of $\wh g$, i.e.
$$
L_{\wh g}=2\ga.
$$
We also call $L_ {\wh g}$ the duration of $\wh g$.  For $\wh \psi_\gs$ defined by \eqref{def_Morlet}, $\wh \psi_\gs$ is ``supported'' in $[\mu-\frac \ga{\gs}, \mu+\frac \ga{\gs}]$ and hence, $L_{\wh \psi_\gs}=\frac{2\ga}\gs$.
Since we hope that $\psi_\gs$ is ``analytic", it is desirable that $\mu-\frac \ga \gs\ge 0$. Thus, in the following, we always assume that
$$
\gs\ge \frac\ga\mu.
$$

Recall that for  $s(t)=A e^{i2\pi c t}$, its CWT with $\psi_\gs(t)$ defined by \eqref{def_Morlet} is
$$
W_s(a, b)={ A  \; \overline{\wh \psi_\gs \big(a c \big)} e^{i 2\pi b c}}.
$$
Since $\wh \psi_\gs \big(a c \big)$ is ``supported''  in $\mu-\frac \ga\gs\le ac \le \mu+\frac \ga\gs$,
$W_s(a, b)$ concentrates  around $a=\frac \mu c$ and lies within the zone (a strip) of the time-scale plane of $(a, b)$:
\begin{equation}
\label{zone_constant}
\frac{\mu-\ga/\gs}c\le a \le \frac{\mu+\ga/\gs}c
\end{equation}
for all $b$.

Next we consider LFM signals (linear chirps). For simplicity of presentation, we consider the case that $A(t)$ in \eqref{def_chip_At} is a constant. Namely, we consider
\begin{equation}
\label{def_chip}
s(t)=A e^{i2\pi (ct +\frac 12 r t^2)}.
\end{equation}
First we find the CWT of  $s(t)$. To this regard, we need the following formula.

\begin{lem}
{\rm (\cite{Leon_Cohen, Gibson06})}
\label{lem:FT_for_LinearChirp}
 For real $\ga$, $\beta$ and $\go$ with $\ga>0$,
$$
\int_{-\infty}^\infty e^{-(\ga+i\beta)t^2+i \go t}dt =\frac{\sqrt \pi}{\sqrt{\ga+i\beta}} e^{-\frac{\go^2}{4(\ga+i\beta)}}.
$$
\end{lem}

Next proposition gives the CWT of LMF signal $s(t)$ with $\psi_\gs$.
\begin{pro}
\label{pro:CWT_linear_chirp}
Let $s(t)$ be the LFM signal defined by \eqref{def_chip}. Then the CWT of $s(t)$ with $\psi_\gs$ given by \eqref{def_Morlet}
is
\begin{equation}
\label{CWT_LinearChip}
W_s(a, b)=\frac {A}{\sqrt{1-i2\pi \gs^2 a^2 r}}\;  e^{i2\pi \big(c b+\frac r2 b^2\big)} \; h(c+rb),
\end{equation}
where
\begin{equation*}
h(\xi)= e^{-\frac{2\pi^2 (a\gs)^2}{1+(2\pi r a^2\gs^2)^2}(\xi-\frac \mu a)^2(1+i2\pi a^2 \gs^2  r)}.
\end{equation*}
\end{pro}

The proof of Proposition \ref{pro:CWT_linear_chirp} is presented in Appendix.

Observe that $|h(\xi)|$ is a Gaussian function with duration
$$
L_{|h|}=2\ga \sqrt{\frac{1+(2\pi r a^2\gs^2)^2}{(a\gs)^2}}=2\ga \sqrt{\frac1{(a\gs)^2}+(2\pi r a\gs)^2}.
$$
Thus the ridge of $W_s(a, b)$ concentrates around  $c+rb=\frac \mu a$ in the time-scale plane of $(a, b)$,  and  $W_s(a, b)$ lies within the zone of time-scale plane:
$$
-\frac 12 L_{|h|}\le c+rb-\frac \mu a\le \frac 12 L_{|h|},
$$
or equivalently
\begin{equation}
\label{def_zone}
 c+rb-\ga  \sqrt{\frac1{(a\gs)^2}+(2\pi r a\gs)^2} \le \frac \mu a\le  c+rb+ \ga \sqrt{\frac1{(a\gs)^2}+(2\pi r a\gs)^2}.
\end{equation}
We call the region in the time-scale plane given by \eqref{def_zone} the time-scale zone of $W_s(a, b)$.

$L_{|h|}$ reaches its minimum when $ \frac1{(a\gs)^2}=(2\pi r a\gs)^2$, namely,
\begin{equation}
\label{opt_gs}
\gs=\frac1{a \sqrt{2\pi |r|}}=\frac1{a \sqrt{2\pi |\phi''(b)|}}.
\end{equation}
In this case $L_{|h|}=4\ga \sqrt {\pi |r|}$, and the time-scale zone of $W_s(a, b)$ is
 $$
 c+rb-2\ga \sqrt {\pi |r|}\le \frac \mu a\le  c+rb+2\ga \sqrt {\pi |r|}.
$$

Observe that $\gs$ in \eqref{opt_gs} depends on both $a$ and $b$. Our goal is to design a method to select the parameter $\gs=\gs(b)$ depending on $b$ only so that (i) the corresponding time-varying CWTs of the components of a multicomponent signal can be separated in the time-scale plane and the
adaptive SST defined by \eqref{def_SST_para_simple} with this $\gs(b)$ has a sharp representation and (ii) the components  can be recovered accurately from the SST with $\gs(b)$ by \eqref{reconst_SST_para_component}.
The obtained time-scale zone in \eqref{def_zone} for a linear chirp helps us to formulate the well-separated conditions and develop the method to find suitable $\gs(b)$, which are the problems we will focus on in the next two sections.

\section{Well-separated conditions for multicomponent signals}

In this section we derive the well-separated conditions for multicomponent signals based on the adaptive CWT.
First, we consider the sinusoidal signal model.
Recall from \S4 that the CWT of $x(t)=A e^{i2\pi c t}$ with $\psi_\gs(t)$ defined by \eqref{def_Morlet0_time} 
is supported in the zone of the time-scale plane given by
\eqref{zone_constant}.  For $x(t)=\sum_{k=1}^K x_k(t)=\sum_{k=1}^K A_k e^{i2\pi c_k t}$ with $c_{k-1}<c_k$, 
its CWT is
$$
W_x(a, b)=\sum_{k=1}^KA_k \overline{\wh \psi_\gs \big(a c_k \big)} e^{i 2\pi b c_k}.
$$
Since the CWT of the $k$-component lies within 
the zone $\frac{\mu-\ga/\gs}{c_k}\le a \le \frac{\mu+\ga/\gs}{c_k}$ of the time-scale plane,
 the components of $x(t)$ will be well-separated in the time-scale plane if
  $$
  \frac{\mu+\ga/\gs}{c_k}\le \frac{\mu-\ga/\gs}{c_{k-1}},
  $$
or equivalently
$$
  \frac{c_k-c_{k-1}}{c_k+c_{k-1}}\ge  \frac{\ga}{\mu\gs}, \quad \hbox{for $k=2, 3, \cdots, K$.}
  $$
Hence, we can separate the components of $x(t)$ in the time-scale plane if we choose $\gs$ such that
$$
  \gs\ge \frac \ga\mu \; \frac{c_k+c_{k-1}}{c_k-c_{k-1}}, \quad  \hbox{for $k=2, 3, \cdots, K$.}
 $$
  More general, for $x(t)$ given by
\begin{equation}
  \label{AHM}
  x(t)=\sum_{k=1}^K x_k(t)=\sum_{k=1}^K A_k(t) e^{i2\pi\phi_k(t)},
  \end{equation}
  if for eack $k$, the (adaptive) CWT of $x_k(t)$ with $\psi_\gs$, which is
  \begin{equation}
  \label{CWT_AHM}
\int_{-\infty}^\infty A_k(b+at) e^{i 2\pi \phi_k(at+b)}\overline{\psi_\gs(t)}dt,
\end{equation}
  can be well-approximated by
  \begin{equation}
  \label{sinsuodal_model}
\int_{-\infty}^\infty A_k(b) e^{i 2\pi  (\phi_k(b)+\phi'_k(b) at )}\overline{\psi_\gs(t)}dt=
A_k(b)\overline{\wh \psi_\gs \big(a \phi'_k(b) \big)} e^{i 2\pi \phi_k(b)},
\end{equation}
then the adaptive CWTs of the components $x_k(t), k=1, \cdots, K$, are separated in the time-scale plane provided that
\begin{equation}
  \label{sinsuodal_sep_condition}
  \gs\ge \frac \ga\mu \; \frac{\phi'_k(b)+\phi'_{k-1}(b)}{\phi'_k(b)-\phi'_{k-1}(b)}, \quad  \hbox{for $k=2, 3, \cdots, K$}
\end{equation}
 for each $b$. The condition in \eqref{sinsuodal_sep_condition} is the well-separated condition based on the sinusoidal signal model.

 The error bounds derived in \cite{Daub_Lu_Wu11} imply that for a signal, its synchrosqueezed representation is sharper when the window width (in time) of the continuous wavelet is smaller. 
This fact was also noticed in our various experiments. 
 The parameter $\gs$ is the window width (in time) of $\psi_\gs$ (up to a constant).
 Thus we choose the smallest $\gs$ satisfying \eqref{sinsuodal_sep_condition}. Hence, we propose the sinusoidal signal-based choice for $\gs$, denoted by $\gs_1(b)$, to be
  \begin{equation}
\label{def_gs1}
  \gs_1(b)=\max_{2\le k\le K}\Big\{\frac \ga\mu\; \frac{\phi'_k(b)+\phi'_{k-1}(b)}{\phi'_k(b)-\phi'_{k-1}(b)}\Big\}.
\end{equation}

\begin{figure}[th] 
\centering
\begin{tabular}{c}
\resizebox{4.2in}{2.6in}{\includegraphics{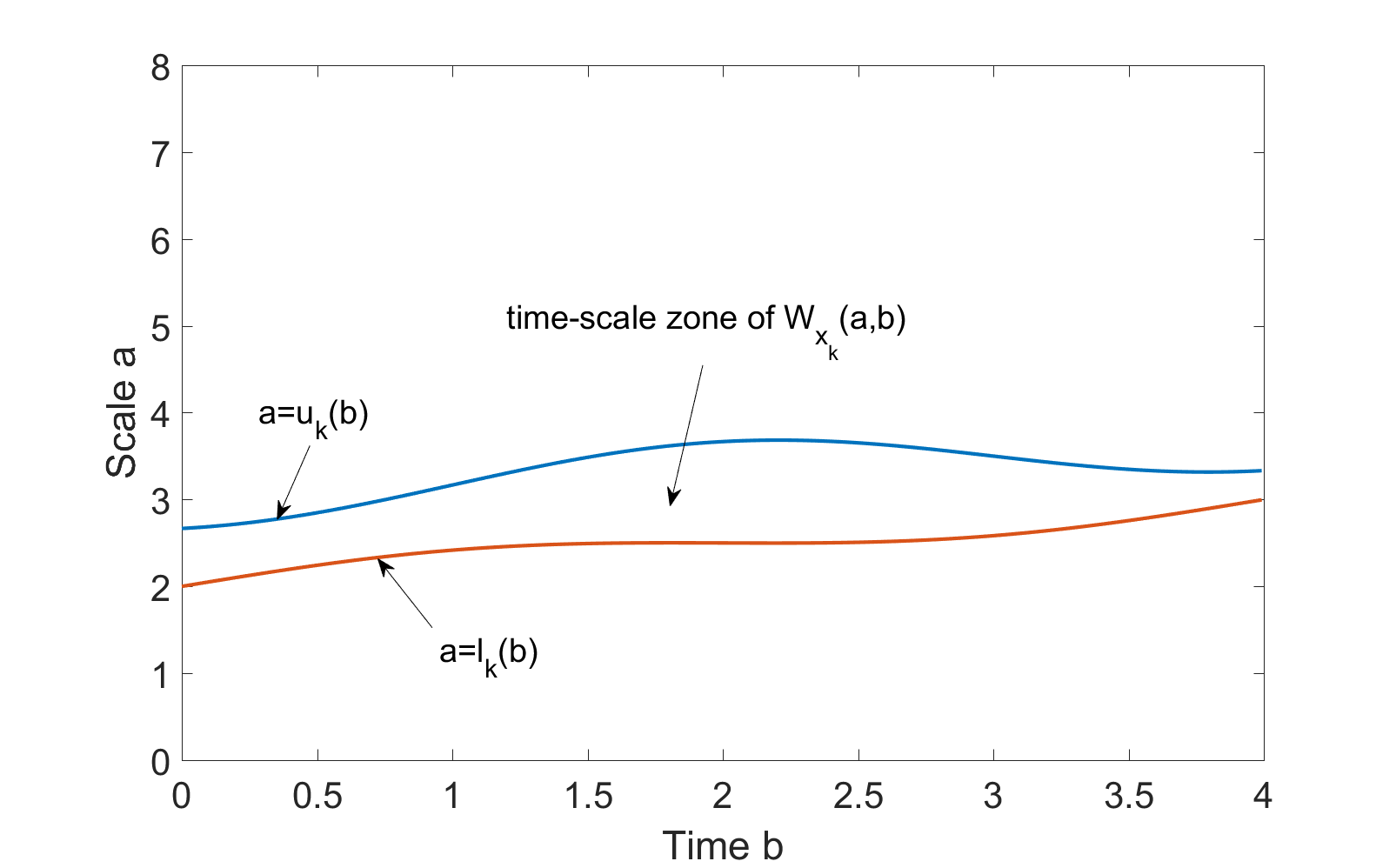}}
\end{tabular}
\caption{\small Time-scale zone of  $W_{x_k}(a, b)$}
\label{fig:wavelet_zones1}
\end{figure}
\begin{figure}[th] 
\centering
\begin{tabular}{c}
\resizebox{4.2in}{2.6in}{\includegraphics{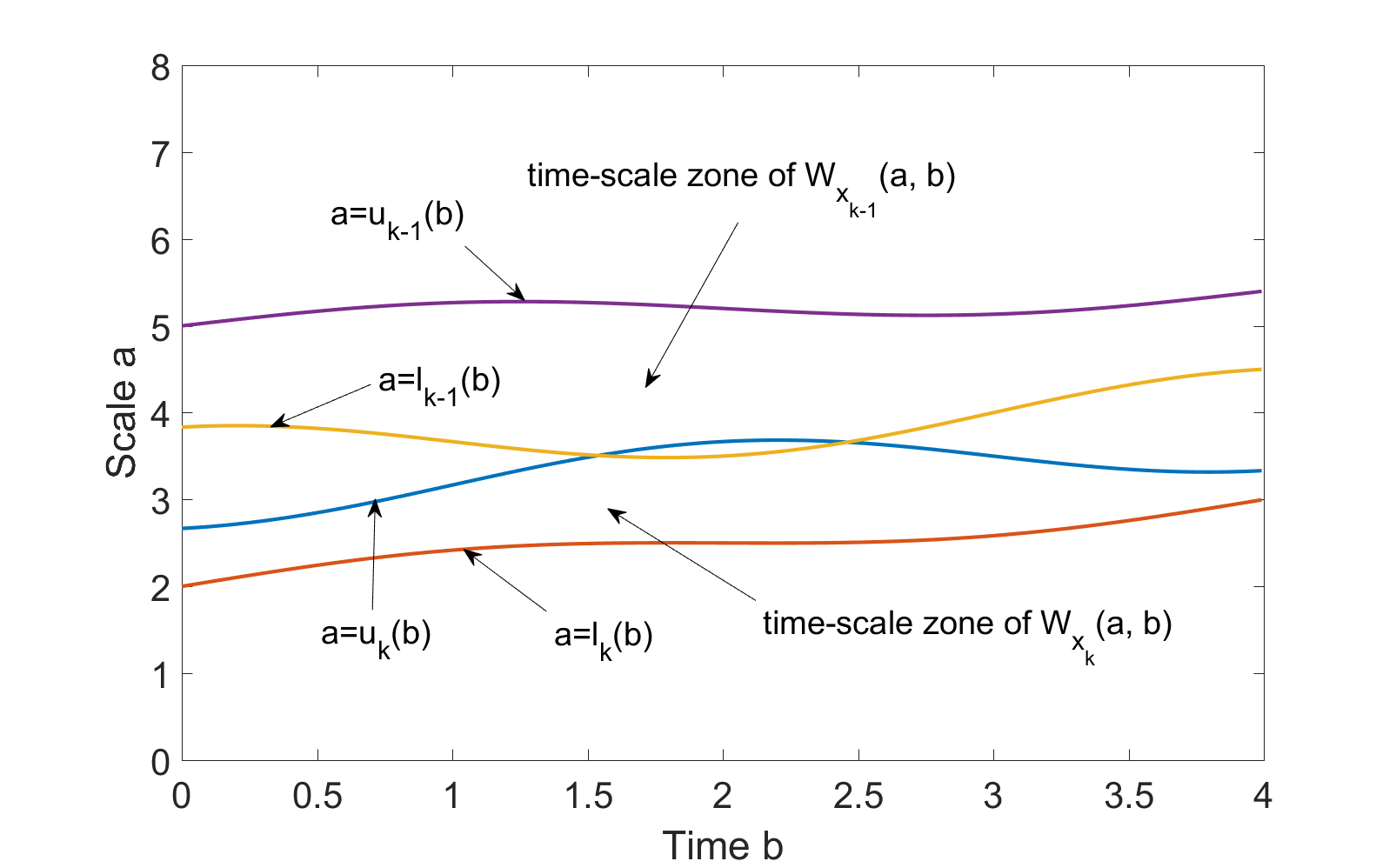}}
\end{tabular}
\caption{\small Time-scale zones of $W_{x_{k-1}}(a, b), W_{x_k}(a, b)$}
\label{fig:wavelet_zones2}
\end{figure}

Next we consider the LFM model (linear chirp model). More precisely,
we consider $x(t)=\sum_{k=1}^K x_k(t)$, where each $x_k(t)$ is a linear chirp, namely,
$$
x_k(t)=A_k e^{i2\pi (c_kt +\frac 12 r_k t^2)}
$$
with the phase $\phi_k(t)=c_kt +\frac 12 r_k t^2$ satisfying $\phi'_{k-1}(t)< \phi'_k(t)$. From \eqref{def_zone},  the CWT $W_{x_k}(a, b)$ of $x_k(t)$ with $\psi_\gs$ lies within the zone of time-scale plane:
 \begin{equation}
\label{def_zone_k}
 c_k+r_k b-\ga  \sqrt{\frac1{(a\gs)^2}+(2\pi r_k a\gs)^2} \le \frac \mu a\le  c_k+r_kb+ \ga  \sqrt{\frac1{(a\gs)^2}+(2\pi r_k a\gs)^2}.
\end{equation}
The two equalities in \eqref{def_zone_k} give the boundaries $u_k(b)$ (upper boundary) and $l_k(b)$ (lower boundary) for the support zone of  $W_{x_k}(a, b)$. More precisely, solving the following two equations in \eqref{def_zone_k_eq} for $a$ gives $u_k(b)$ and $l_k(b)$ respectively:
 \begin{equation}
\label{def_zone_k_eq}
 c_k+r_k b-\ga  \sqrt{\frac1{(a\gs)^2}+(2\pi r_k a\gs)^2}=\frac \mu a, \quad
c_k+r_kb+ \ga  \sqrt{\frac1{(a\gs)^2}+(2\pi r_k a\gs)^2}=\frac \mu a
\end{equation}
See Fig.\ref{fig:wavelet_zones1} for the time-scale zone of $W_{x_k}(a, b)$.
Our goal is to obtain the conditions on $\phi_k$ and $\phi_{k-1}$ under which we can
choose $\gs$, depending on $b$ only, such that the support zones of  $W_{x_k}(a, b)$, $k=1, \cdots, K$ are not overlapped in the time-scale plane, namely, $u_k(b)$ and $l_k(b)$ satisfy
\begin{equation}
\label{boundary_ineq}
u_k(b)\le l_{k-1}(b), \quad  k=2, 3, \cdots, K.
\end{equation}
The case shown in Fig.\ref{fig:wavelet_zones2} is not what we pursue because the condition \eqref{boundary_ineq} is invalid with $u_k(b), l_{k-1}(b)$ entangled.

Observe that
 $$
\sqrt{\frac1{(a\gs)^2}+(2\pi |r_k| a\gs)^2}\le
\frac1{a\gs}+2\pi |r_k| a\gs\le \sqrt 2  \sqrt{\frac1{(a\gs)^2}+(2\pi |r_k| a\gs)^2}.
$$
In the following we use $\frac1{a\gs}+2\pi |r_k| a\gs$ in the place of $\sqrt{\frac1{(a\gs)^2}+(2\pi |r_k| a\gs)^2}$ in equations \eqref{def_zone_k} and  \eqref{def_zone_k_eq}.
More generally, for $x(t)$ given by \eqref{AHM}, if the CWT $W_{x_k}(a, b)$ of $x_k(t)=A_k(t)e^{i2\pi\phi_k(t)}$ can be well approximated by
\begin{equation}
\label{linear_chirp_model}
\int_{-\infty}^\infty A_k(b) e^{i2\pi \pi\big(\phi_k(b)+\phi'_k(b) at +\frac 12 \phi''_k(b)(at)^2\big)}\psi_\gs(t)dt,
\end{equation}
then $W_{x_k}(a, b)$ lies within the following time-scale zone:
 \begin{equation}
\label{def_zone_k_big}
\phi_k'(b)-\ga (\frac1{a\gs }+2\pi |\phi''_k(b)| a\gs) \le \frac \mu a\le  \phi_k'(b)+\ga (\frac1{a\gs }+2\pi |\phi''_k(b)| a\gs).
\end{equation}
Therefore, boundaries $a=u_k$ and $a=l_k$ of the support zone of $W_{x_k}(a, b)$ are the solutions of the following two equations respectively:
\begin{eqnarray}
&&\label{def_zone_k_eq_big_upper}
\frac \mu{u_k}+\ga (\frac1{u_k \gs }+2\pi |\phi''_k(b)| u_k  \gs) =\phi_k'(b), \\
&&\label{def_zone_k_eq_big_lower}
\frac \mu{l_k}-\ga (\frac1{l_k \gs }+2\pi |\phi''_k(b)| l_k\gs ) =\phi_k'(b).
\end{eqnarray}
One can obtain from \eqref{def_zone_k_eq_big_upper} and \eqref{def_zone_k_eq_big_lower} that
\begin{eqnarray}
u_k=u_k(b)\hskip -0.6cm && =\frac{\phi'_k(b)-\sqrt{\phi'_k(b)^2-8\pi \ga (\mu \gs+\ga)|\phi''_k(b)|}}{4\pi \gs \ga |\phi''_k(b)|} \nonumber \\
\label{zone_k_eq_big_upper}
&&=\frac {2(\mu+\frac \ga\gs)}{\phi'_k(b)+\sqrt{\phi'_k(b)^2-8\pi \ga (\ga+\mu \gs)|\phi''_k(b)|}},\\
&&\nonumber \\
l_k=l_k(b)\hskip -0.6cm &&=\frac{-\phi'_k(b)+\sqrt{\phi'_k(b)^2+8\pi \ga (\mu \gs-\ga)|\phi''_k(b)|}}{4\pi \gs \ga |\phi''_k(b)|} \nonumber \\
\label{zone_k_eq_big_low}
&&=\frac {2(\mu-\frac \ga\gs)}{\phi'_k(b)+\sqrt{\phi'_k(b)^2+8\pi \ga (\mu \gs-\ga)|\phi''_k(b)|}}.
\end{eqnarray}
One can verify directly $l_k(b)\le u_k(b)$ for $k=1, 2, \cdots, K$. In order to separate the components of $x_k(t)$, we need to choose $\gs$ such that the support zones of $W_{x_{k-1}}(a, b)$ and  $W_{x_k}(a, b)$ do not overlap, namely \eqref{boundary_ineq} holds. By careful and tedious calculations, one can obtain that inequality 
\eqref{boundary_ineq} with $u_k(b)$ and $l_{k}(b)$ given by \eqref{zone_k_eq_big_upper} and \eqref{zone_k_eq_big_low}
 can be written as
$$
\ga_k(b) \gs ^2 -\gb_k(b) \gs +\gga_k(b)\le 0,
$$
where
\begin{eqnarray*}
\label{gak}
&&\ga_k(b)=2\pi \ga \mu (|\phi''_k(b)|+|\phi''_{k-1}(b)|)^2,\\
\label{gbk}
&&\gb_k(b)=\big(\phi'_k(b)|\phi''_{k-1}(b)|+\phi'_{k-1}(b)|\phi''_{k}(b)|\big) \big(\phi'_k(b)-\phi'_{k-1}(b)\big)+4\pi \ga^2  \big(\phi''_k(b)^2-\phi''_{k-1}(b)^2\big),\\
\label{ggak}
&&\gga_k(b)=\frac{\ga}\mu\Big\{
\big(\phi'_k(b)|\phi''_{k-1}(b)|+\phi'_{k-1}(b)|\phi''_{k}(b)|\big)\big(\phi'_k(b)+\phi'_{k-1}(b)\big)+2\pi \ga^2  \big(|\phi''_k(b)|-|\phi''_{k-1}(b)|\big)^2\Big\}.
\end{eqnarray*}
Thus, if
\begin{eqnarray*}
&&\Upsilon_k(b):=\gb_k(b)^2-4\ga_k(b)\gga_k(b)\\
&&=\big(\phi'_k(b)|\phi''_{k-1}(b)|+\phi'_{k-1}(b)|\phi''_{k}(b)|\big)^2\Big\{\big(\phi'_k(b)-\phi'_{k-1}(b)\big)^2-16\pi \ga^2 \big(|\phi''_k(b)|+|\phi''_{k-1}(b)|\big)
\Big\}\ge 0,
\end{eqnarray*}
  then \eqref{boundary_ineq} holds if and only if $\gs$ satisfies
\begin{equation}
\label{gs_ineq}
\frac{\gb_k(b)-\sqrt{\Upsilon_k(b)}}{2\ga_k(b)}\le \gs\le \frac{\gb_k(b)+\sqrt{\Upsilon_k(b)}}{2\ga_k(b)}.
\end{equation}
Otherwise, if $\Upsilon_k(b)<0$, then there is no suitable solution of the parameter $\gs$ for 
\eqref{boundary_ineq},
which means that components  $x_{k-1}(t)$ and $x_k(t)$ of multicomponent signal $x(t)$ cannot be separated in the time-scale plane. Thus we reach our well-separated conditions of CWTs with a time-varying $\gs=\gs(b)$.

\begin{theo} Let
$x(t)=\sum_{k=1}^K x_k(t)$, where each $x_k(t)=A_k(t)e^{i2\pi\phi_k(t)}$ is a linear chirp signal or its adaptive CWT
$\wt W_{x_k}(b, a)$
with $\psi_{\gs(b)}$ can be well approximated by \eqref{linear_chirp_model}, and $\phi'_{k-1}(t)<\phi'_k(t)$. 
If for any $b$ with  $|\phi''_k(b)|+|\phi''_{k-1}(b)|\not=0$,
\begin{eqnarray}
\label{sep_cond1}
&& 4\ga \sqrt{\pi} \sqrt {|\phi''_k(b)|+|\phi''_{k-1}(b)|}\le \phi'_k(b)-\phi'_{k-1}(b), \quad k=2, \cdots, K, \; \hbox{and}\\
 \label{sep_cond2}
 && \max\Big\{\frac \ga\mu, \frac{\gb_k(b)-\sqrt{\Upsilon_k(b)}}{2\ga_k(b)}: 2\le k\le K\Big\}\le
\min_{2\le k\le K}\Big\{\frac{\gb_k(b)+\sqrt{\Upsilon_k(b)}}{2\ga_k(b)}\Big\},
 \end{eqnarray}
then the components of $x(t)$ are well-separable in time-scale plane in the sense that $\wt W_{x_k}(b, a), 1\le k\le K$ with $\gs(b)$ chosen to satisfy \eqref{gs_ineq} lie in non-overlapping regions in the time-scale plane.
 \end{theo}

 Considering again the fact that a smaller $\gs$ results in a sharper synchrosqueezed representation, we choose the smallest $\gs(b)$ such that \eqref{gs_ineq} holds. Hence,
we propose the linear chirp signal-based choice for $\gs$, denoted by $\gs_2(b)$, to be
\begin{equation}
\label{def_gs2}
\gs_2(b)=\left\{
\begin{array}{ll}
\max\Big\{\frac\ga\mu, \frac{\gb_k(b)-\sqrt{\Upsilon_k(b)}}{2\ga_k(b)}: \; 2\le k\le K\Big\}, &\hbox{if $|\phi''_k(b)|+|\phi''_{k-1}(b)|\not=0$}, \\
& \\
 \max\Big\{\frac \ga\mu \frac{\phi'_k(b)+\phi'_{k-1}(b)}{\phi'_k(b)-\phi'_{k-1}(b)}: \; 2\le k\le K\Big\}, &\hbox{if $\phi''_k(b)=\phi''_{k-1}(b)=0$.}
\end{array}
\right.
\end{equation}

\begin{figure}[th]
\centering
\begin{tabular}{ccc}
\resizebox{2.0in}{1.5in}{\includegraphics{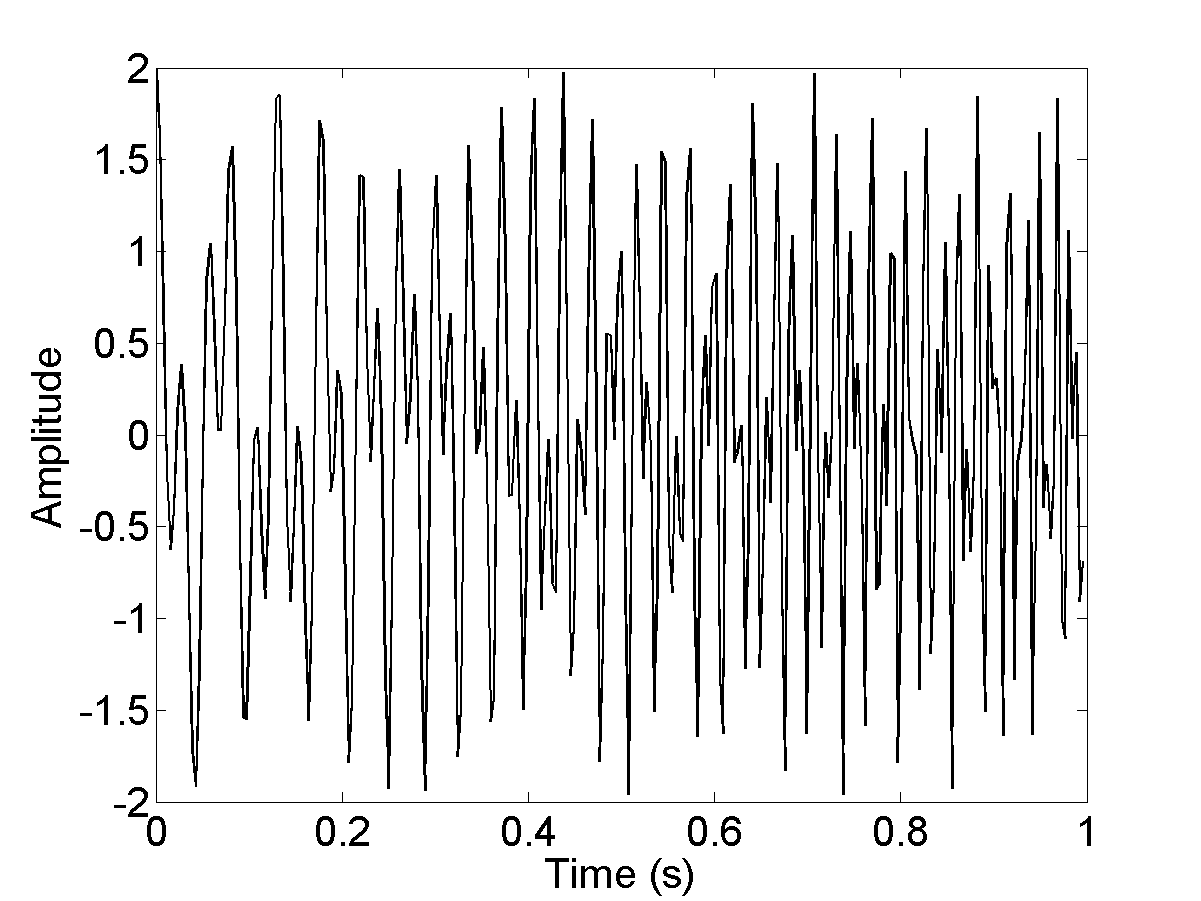}}
&\resizebox{2.0in}{1.5in}{\includegraphics{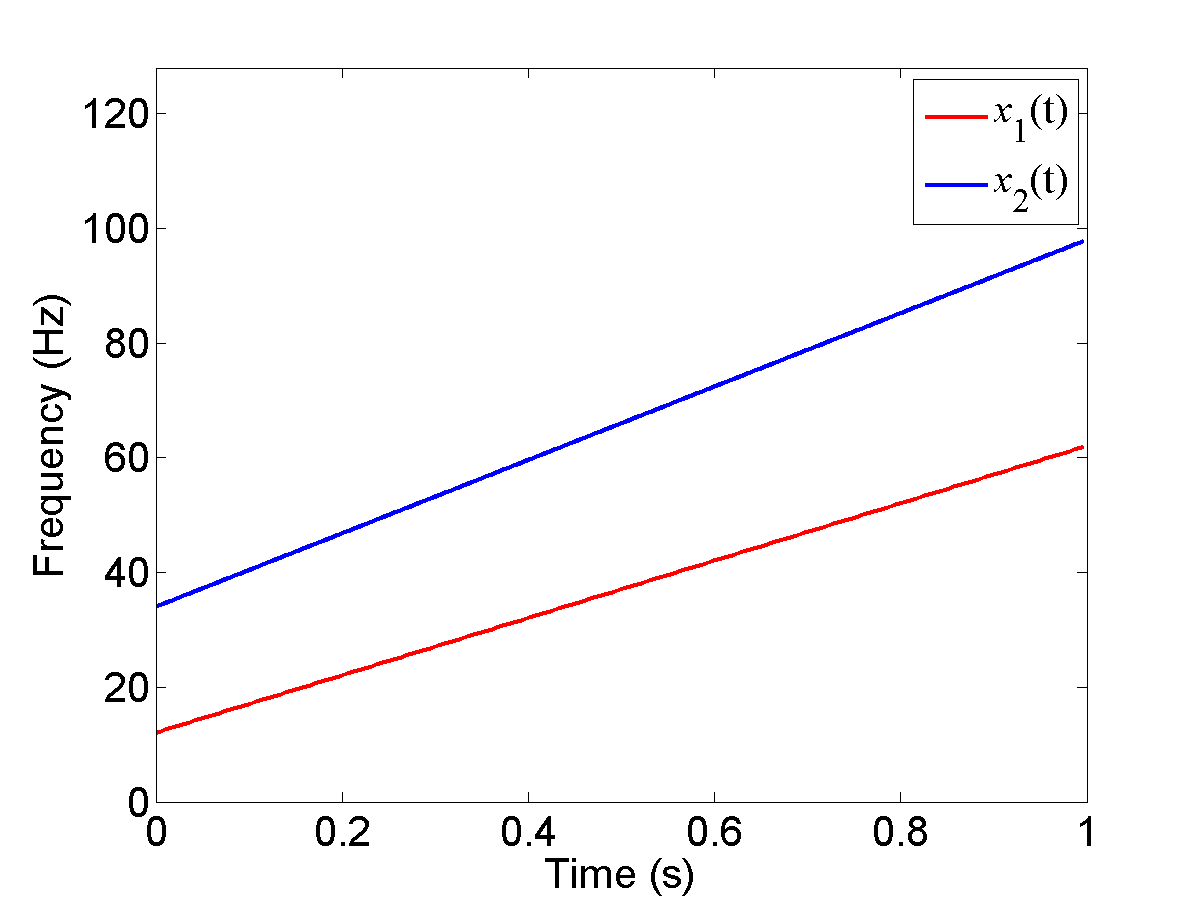}}
&\resizebox{2.0in}{1.5in}{\includegraphics{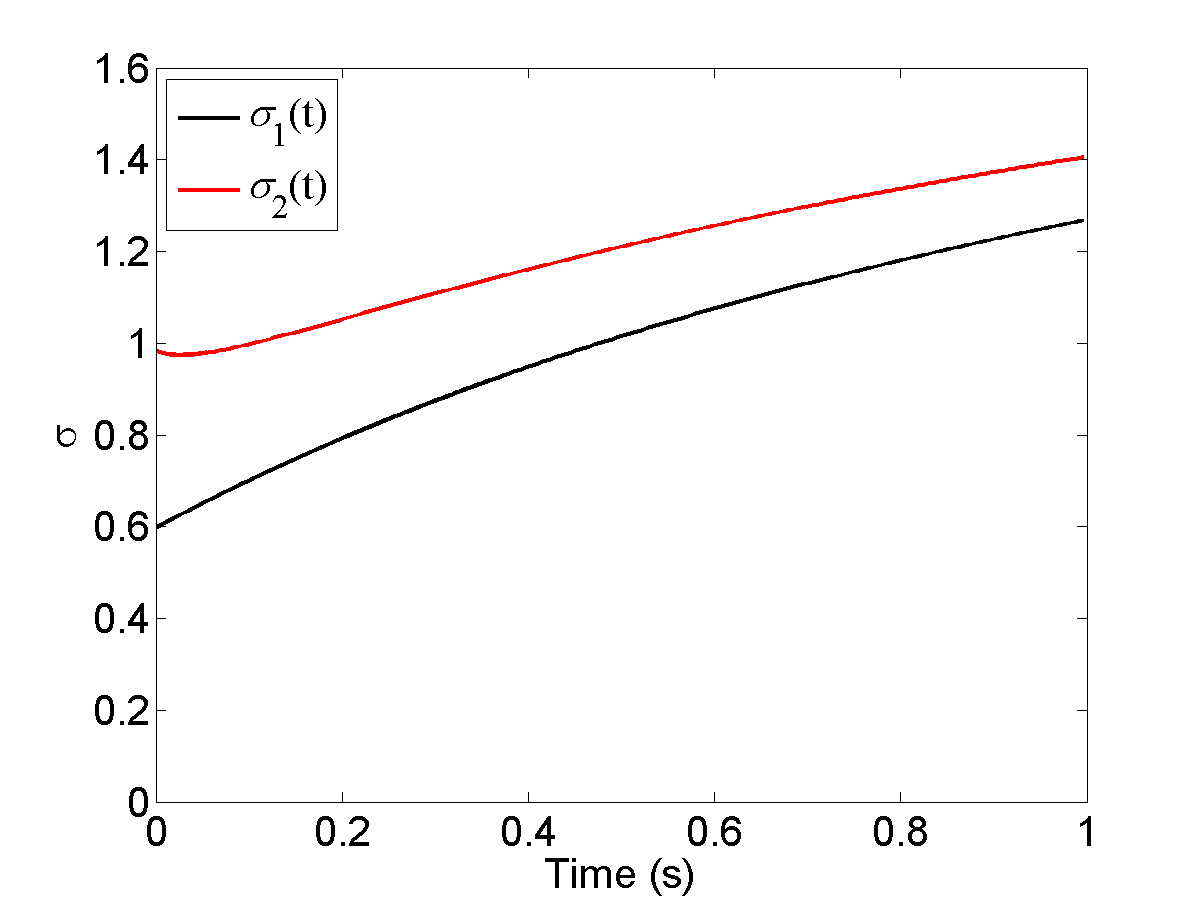}}\\
\resizebox{2.0in}{1.5in}{\includegraphics{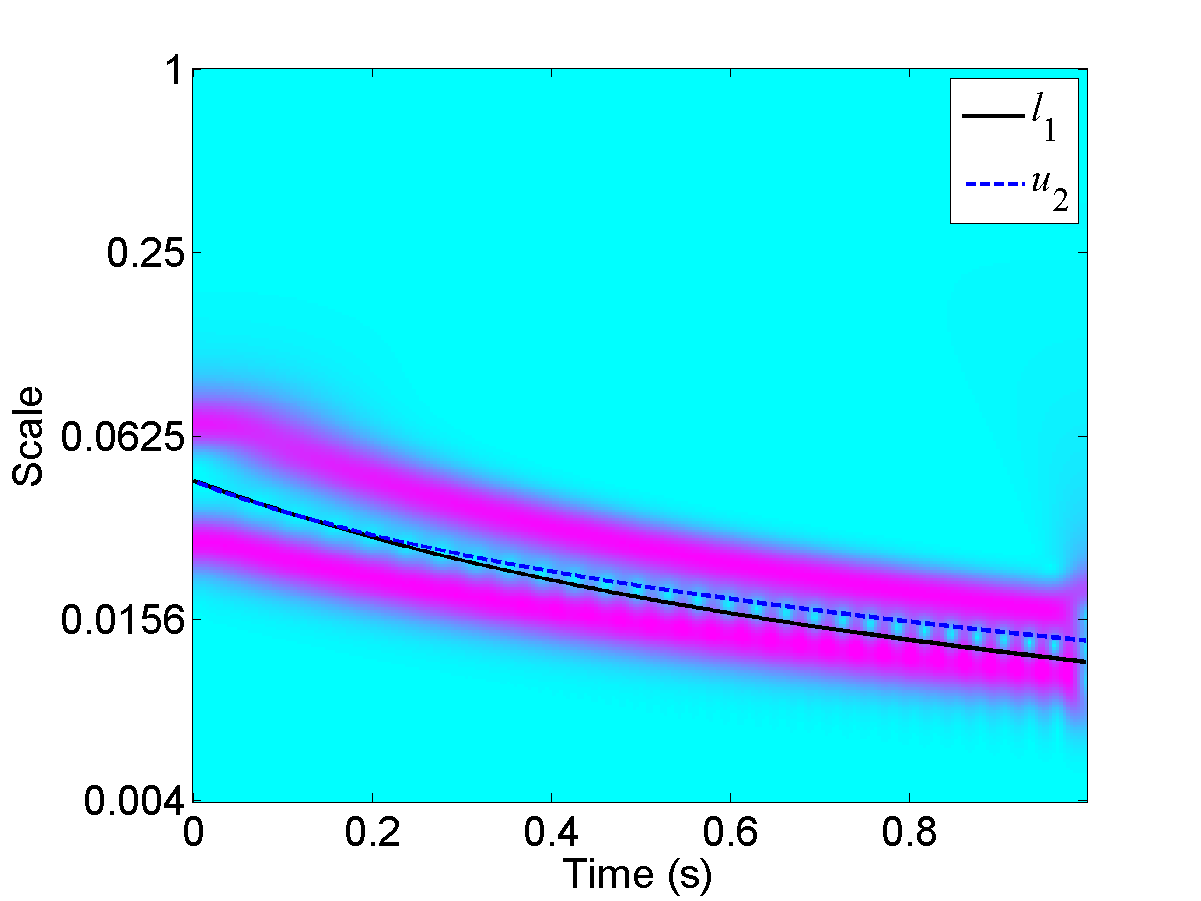}}
&\resizebox{2.0in}{1.5in}{\includegraphics{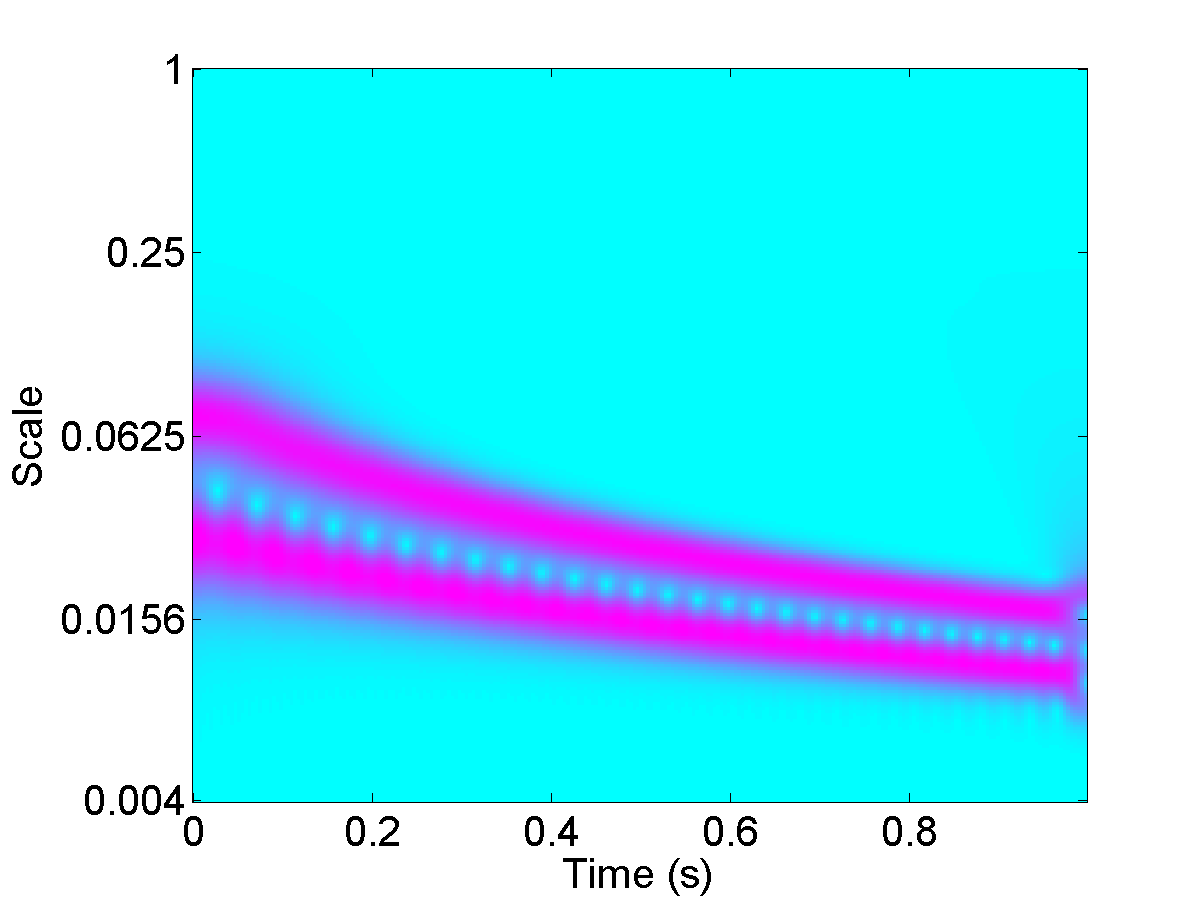}}
&\resizebox{2.0in}{1.5in}{\includegraphics{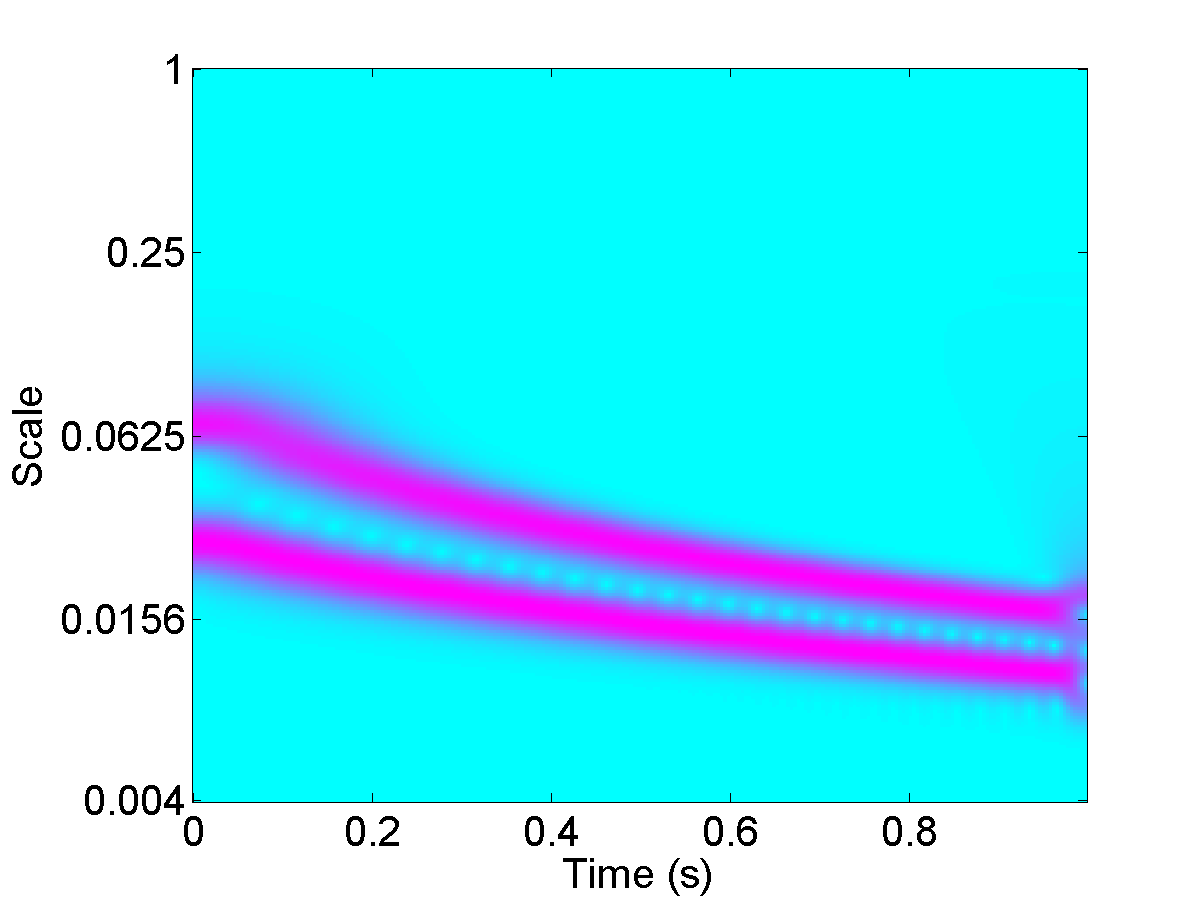}}\\
\resizebox{2.0in}{1.5in}{\includegraphics{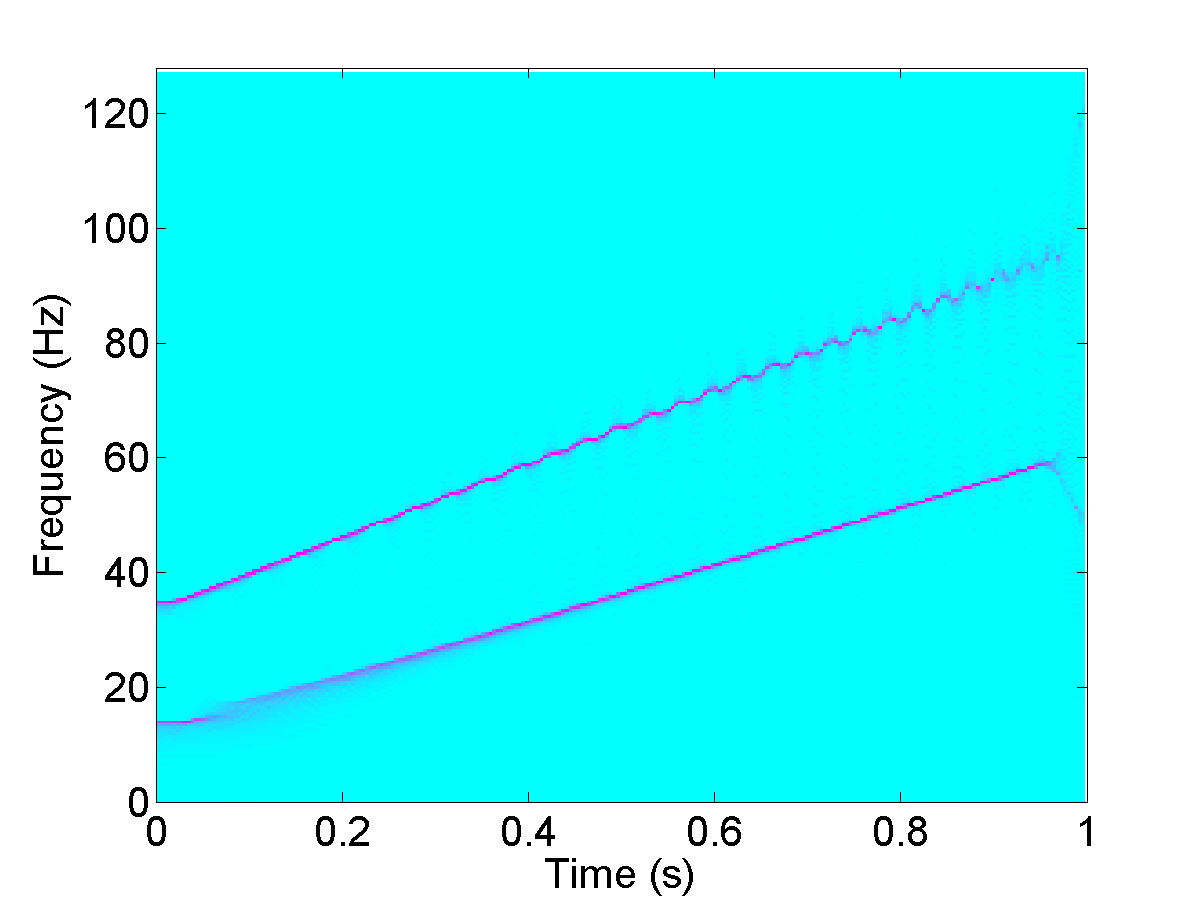}}
&\resizebox{2.0in}{1.5in}{\includegraphics{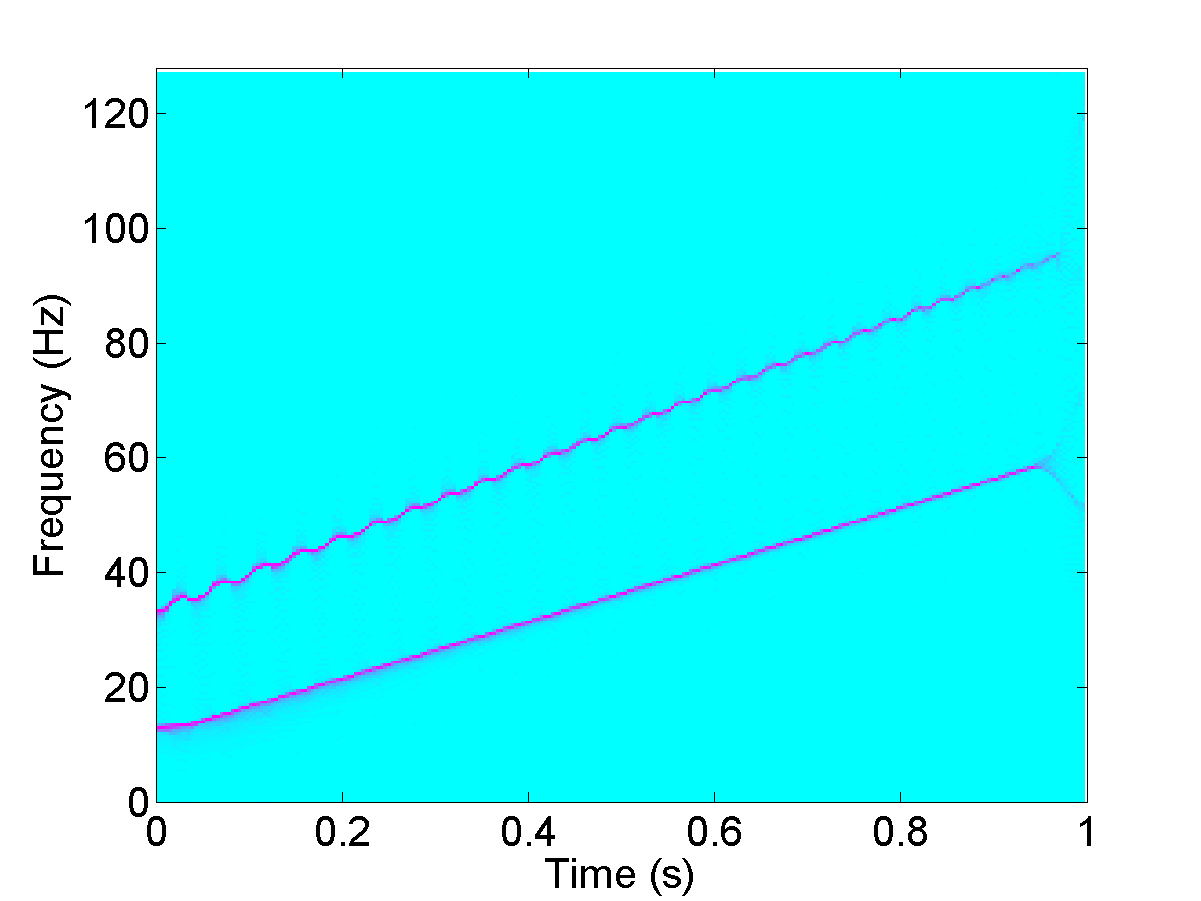}}
&\resizebox{2.0in}{1.5in}{\includegraphics{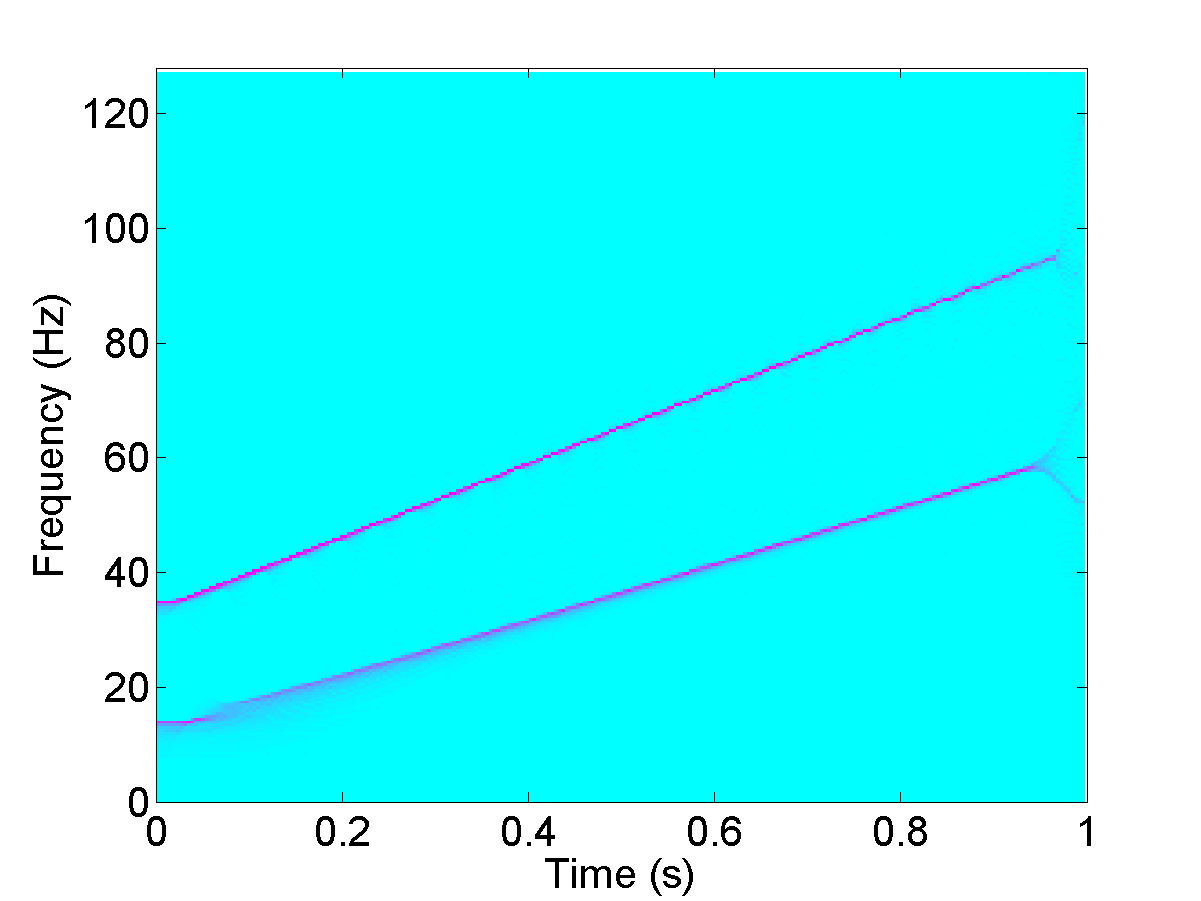}}\\
\resizebox{2.0in}{1.5in}{\includegraphics{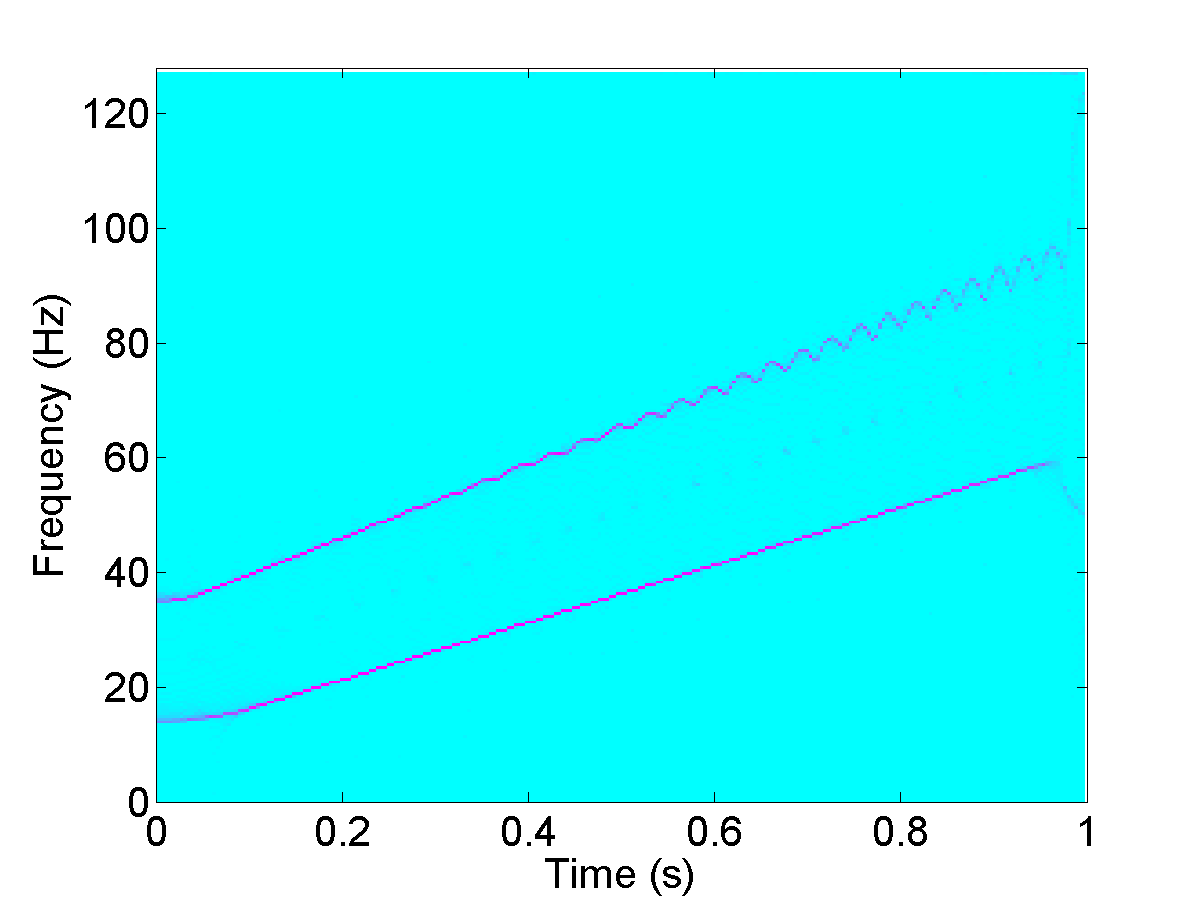}}
&\resizebox{2.0in}{1.5in}{\includegraphics{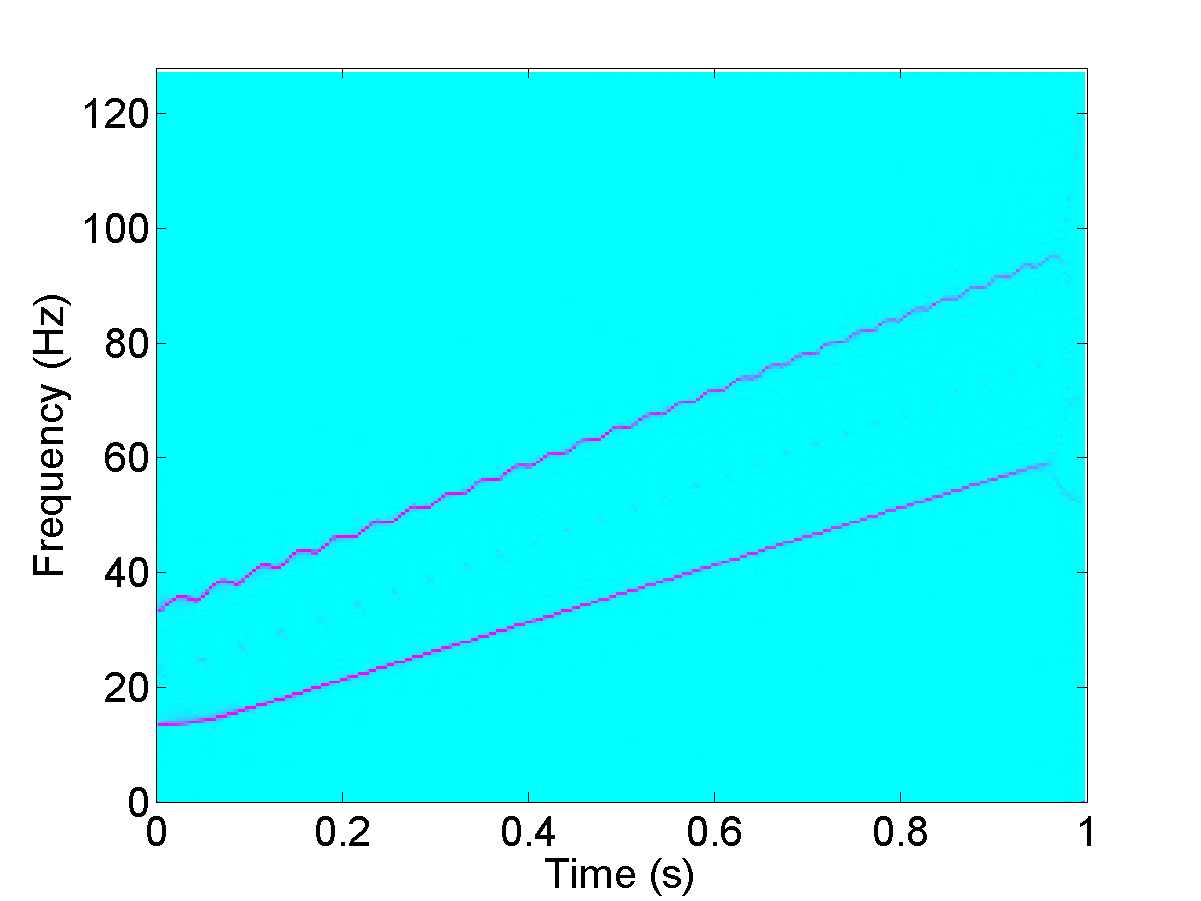}}
&\resizebox{2.0in}{1.5in}{\includegraphics{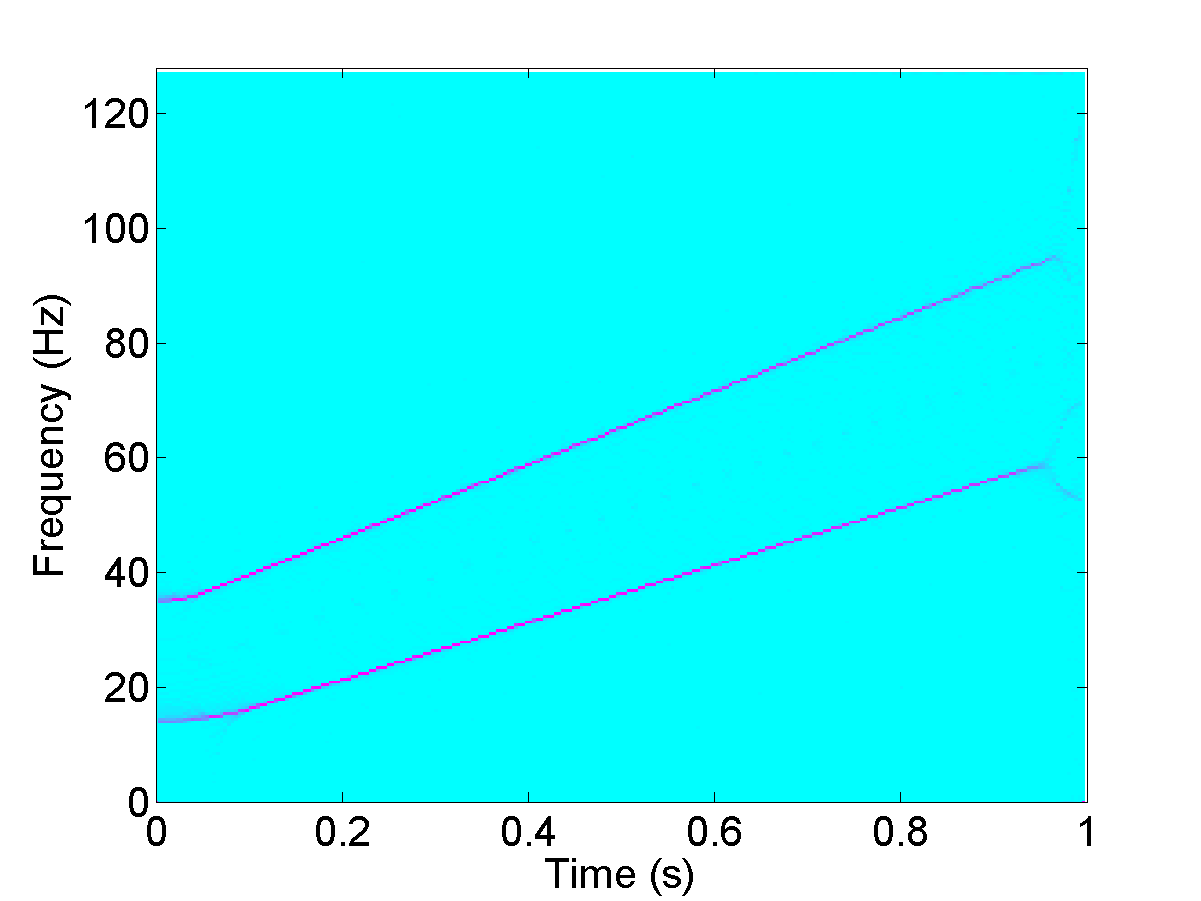}}
\end{tabular}
\caption{\small Example of the two-component LFM signal $x(t)$ in \eqref{two_linear_chirps}. Top row: waveform of $x(t)$ (left panel), instantaneous frequencies of the $x_1(t)$ and  $x_2(t)$ (middle panel), time-varying parameters $\gs_1(t)$ and $\gs_2(t)$ (right panel); Second row:  conventional CWT when $\gs=1$ (left panel), adaptive CWT with $\gs_1(t)$ (middle panel), adaptive CWT with $\gs_2(t)$ (right panel); Third row: conventional SST when $\gs=1$ (left panel), adaptive SST with $\gs_1(t)$ (middle panel), adaptive SST with $\gs_2(t)$ (right panel); Bottom row:  conventional 2nd-order SST when $\gs=1$ (left panel), 2nd-order adaptive SST with $\gs_1(t)$ (middle panel), 2nd-order adaptive SST with $\gs_2(t)$ (right panel). }
\label{fig:adaptiveSST_true}
\end{figure}

\clearpage

Next we show some experiment results. Let $x(t)$ be a signal with two linear chirps:
\begin{equation}
\label{two_linear_chirps}
x(t)=x_1(t)+x_2(t)=\cos\big(2\pi (c_1+\frac 12 r_1 t)t\big)+\cos\big(2\pi (c_2+\frac 12 r_2 t)t\big), t \in [0,1],
\end{equation}
where  the starting frequencies: $c_1=12$, $c_2=34$,  and the chirp rates: $r_1=50$, $r_2=64$. Here $x(t)$
is sampled uniformly with $ N=256$ sample points.
We let $\mu$ in Morlet's wavelet $\psi_\ga$ be 1. The scale variable $a$ is discretized as $a_j=2^{j/n_\nu} \Delta t$ with $n_\nu=32$, $j=1, 2, \cdots, n_\nu \log_2 N$.  We choose $\tau_0$ in \eqref{def_gs2} to be $1/5$. Note that we set the same values of $\mu$, $n_\nu$ and $\tau_0$ for all experiments in this and next sections.

Fig.\ref{fig:adaptiveSST_true} (top-right panel)  shows the waveform of  $x(t)$ in \eqref{two_linear_chirps}. The left panel in the second row also shows the boundaries $l_1$ and $u_2$, namely the low boundary of $x_1(t)$ and upper boundary of $x_2(t)$ by \eqref{zone_k_eq_big_low} and \eqref{zone_k_eq_big_upper} when $\gs=1$. So when $\gs=1$, $x_1(t)$ and $x_2(t)$ are not separable in the time-scale plane of CWT except for the part with the time $t$ near $0$. By comparing with the actual instantaneous frequencies $\phi'_1(t)=12+50 t$ and $\phi'_2(t)=34+64 t$
 of $x_1(t)$ and $x_2(t)$ resp., the adaptive SST defined by \eqref{def_SST_para_simple}  and 2nd-order adaptive SST defined by \eqref{def_2ndSST_para_simple}  with the proposed time-varying parameters $\gs_1(t)$ and $\gs_2(t)$
give sharper and more correct representations of IFs than the conventional SST and conventional 2nd-order SST with a constant parameter.

\section{An algorithm to select the time-varying parameter $\gs(b)$ automatically}

Suppose $x(t)$ given by \eqref{AHM} is separable, meaning \eqref{sep_cond1} and \eqref{sep_cond2} hold
when an LFM signal is used to approximate the signal during any local time. If we know $\phi'_k(b)$ and $\phi''_k(b)$, then we can choose a $\gs(b)$ such that it satisfies \eqref{gs_ineq} to define  the adaptive CWT and adaptive SST for sharp representations of $x_k(t)$ in the time-frequency plane and accurate recovery of $x_k(t)$.  However in practice, we in general have no prior knowledge of $\phi'_k(b)$ and $\phi''_k(b)$. Hence, we need a method to find a suitable $\gs(b)$. In this section, we propose an algorithm for obtaining a $\gs(b)$ based on the LFM-model.

First, for $\psi _\gs(t)$ defined by
by \eqref{def_Morlet0_time},  we have that  the amplitude of wavelet $\psi _{(a,b)}=\frac 1a \psi _\gs(\frac {t-b}a)$ is
\[\left| \psi _{(a,b)}(t) \right| = \frac{1}{{a\sigma \sqrt {2\pi } }}e^{ - \frac{1}{{2\sigma ^2 a^2 }}(t - b)^2 }. \]
Following \eqref{def_ga},  the duration of $\psi _{(a,b)}$  is defined as,
\[    L_{\psi _{(a,b)} }  = 4\pi \alpha \sigma a.   \]

Next we describe our idea of selecting $\gs(b)$ for a multicomponent signal.
For a fixed pair $(b, \gs)$, denote $W_{(b, \gs)}(a)=W_x(a, b, \gs)$,  $x(t)$'s CWT with a time-varying parameter defined by \eqref{def_CWT_para}.
First of all, for temporarily fixed $b$ and $\gs$, we extract the peaks (local maxima) of $|W_{(b, \gs)}(a)|$ above certain height. More precisely, let $\Gamma_3>0$ is a given threshold. We find local maximum points $a_1, a_2, \cdots, a_m$ of $|W_{(b, \gs)}(a)|$ at which $|W_{(b, \gs)}(a)|$ attains local maxima with
$$|W_{(b, \gs)}(a_k)|> \Gamma_3, \; k=1, \cdots, m.
$$
Observe that $m$ may depend on $b$ and $\gs$. We assume $a_1<a_2< \cdots <a_m$.
For each local maximum point $a_k$, we treat $(a_k, b)$ as the local maximum of the adaptive CWT $W_{x_k}(b, a, \gs)$ of a potential component, denoted by $x_k$, of $x(t)$.  To check whether $x_k$ is indeed a component of $x(t)$ or not, we consider the support interval $[g_k, h_k]$ for $W_{x_k}(a, b, \gs)$ for fixed $b$ and $\gs$ with $W_{x_k}(a, b, \gs)\approx 0$ for $a\not\in [g_k, h_k]$. If there is no overlap among  $[g_k, h_k]$, $[g_{k-1}, h_{k-1}]$, $[g_{k+1}, h_{k+1}]$, then we decide that  $x_k(t)$ is indeed a component of $x(t)$, where $[g_{k-1}, h_{k-1}]$, $[g_{k+1}, h_{k+1}]$ are the support intervals for CWTs of $x_{k-1}$ and $x_{k+1}$ defined similarly.
Next we provide a method to estimate $g_k, h_k$.

With our LMF model, if the estimated IF $\phi'_k(t)$ of $x_k(t)$ is
 $\wh c_k+ \wh r_k (t-b)$, then by \eqref{zone_k_eq_big_upper} and  \eqref{zone_k_eq_big_low} with $\phi'_k(b)=\wh c_k, \phi''_k(b)=\wh r_k$,
\begin{eqnarray}
\label{estimate_hk}
&&h_k=\frac {2(\mu+\frac \ga\gs)}{\wh c_k+\sqrt{\wh c_k^2-8\pi \ga (\ga+\mu \gs)|\wh r_k|}},\\
&&g_k=\frac {2(\mu-\frac \ga\gs)}{\wh c_k+\sqrt{\wh c_k^2+8\pi \ga (\mu \gs-\ga)|\wh r_k|}}.
\label{estimate_gk}
\end{eqnarray}
Thus  to obtain $g_k, h_k$, we need to estimate $\wh c_k$ and the chirp rate $\wh r_k$ of $x_k(t)$.
To this regard, we extract a small piece of curve in the time-scale plane passing through $(a_k, b)$ which corresponds to the local ridge on $|W_{(t, \gs)}(a)|$. More precisely, letting
$$
t_{k1}=b-\frac{1}{2}L_{\psi _{(a_k,b)} }=b-2\pi \ga \gs a_k, \quad
t_{k2}=b-\frac{1}{2}L_{\psi _{(a_k,b)} }=b+2\pi \ga \gs a_k,
$$
we define
$$
d_k(t)=\underset{ \hbox{$a$: $a$ {\small is near $a_k$}}}{\rm argmax}   |W_{(t, \gs)}(a)|, \quad t\in [t_{k1},t_{k2}].
$$
Note that $d_k(b)=a_k$ and $(a_k, b)$ is a point lying on the curve  in the time-scale of $(a, t)$ given by
$$
L=\{(d_k(t), t): t\in [t_{k1},t_{k2}]\}=\{(a, t): a=d_k(t),  t\in [t_{k1},t_{k2}]\}.
$$
Most importantly, $\{|W_{(t, \gs)}(a)|: (a, t)\in L\}$ is the local ridge on $|W_{(t, \gs)}(a)|$  near $(a_k, b)$, and thus, it is also the local ridge on $|W_{x_k}(a, t, \gs)|$. Observe that from the CWT of an LFM signal given by \eqref{CWT_LinearChip}, the local ridge on $|W_{x_k}(a, t, \gs)|$ occurs when $\frac \mu a=\phi'_k(t)=c_k+r_k t$, namely
 the local ridge on $|W_{x_k}(a, t, \gs)|$ is given by
  $\{|W_{x_k} (a, t, \gs)|: \frac {\mu}a=c_k+r_k t\}$.
 Thus the curve $L$ given by $a=d_k(t)$ can be used to estimate $c_k$ and $r_k$:
 $$
 c_k+r_k t \approx \wh f_k(t) =\mu/d_k(t).
$$
With the LFM model, we use the linear function
\begin{equation*}
\label{local_L_functions}
f_k(t) =\wh r_k(t-b)+  \wh c_k , t \in [t_{k1},t_{k2}]
\end{equation*}
 to fit $\wh f_k(t)$. 
With these $\wh c_k$ and $\wh r_k$, we have $h_k, g_k$ given in \eqref{estimate_hk} and \eqref{estimate_gk}.  Especially when $\wh r_k=0$, recalling the support zone of a sinusoidal signal mode in \S5, we have
$$
h_k=\frac{\mu+\ga/\gs}{\wh c_k},  \; g_k=\frac{\mu-\ga/\gs}{\wh c_k}.
$$

In this way we obtain the collection of support intervals for $W_{x}(a, b, \gs)$ for fixed $b$ and $\gs$:
\begin{equation}
\label{def_s_intervals}
{\bf s}=\{[g_1, h_1], \cdots, [g_m, h_m] \}.
\end{equation}
If adjacent intervals of ${\bf s}$ do not overlap, namely,
\begin{equation}
\label{nonoverlap_hkgk}
h_k \le g_{k+1}, \; \hbox{ for all $k=1, 2, \cdots, m-1$}
\end{equation}
holds, then this $\gs$  is a right parameter to separate the components and such a $\gs$  is a good candidate which we should consider to select. Otherwise, if a pair of adjacent intervals of ${\bf s}$ overlap, namely, \eqref{nonoverlap_hkgk} does not hold, then this is not the parameter we shall choose and we need to consider a different  $\gs$.

In the above description of our idea for the algorithm, we start with a $\gs$ and (fixed) $b$, then we decide whether this $\gs$ is a good candidate to select based on the criterion \eqref{nonoverlap_hkgk}. The choice of the initial $\gs$ plays a critical role for the success of our algorithm due to the fact that on one hand, as we have mentioned above, a smaller $\gs$ will in general result in a sharper representation of SST, and hence, we should find $\gs$  as small as possible such that \eqref{nonoverlap_hkgk} holds; and
on the other hand,  different $\gs$ with which \eqref{nonoverlap_hkgk} holds may result in different number of intervals $m$ in
\eqref {def_s_intervals} even for the same time instance $b$.  To keep the number $m$ (the number of components) unchanged
when we search for different $\gs$ with a fixed $b$,
the initial $\gs$ is required to provide a good estimation on the number of  the components of a multicomponent signal $x(t)$. To this end, in this paper we propose to use the R${\rm \acute e}$nyi entropy to determine the initial $\gs(b)$.
The R${\rm \acute e}$nyi entropy approach provides a sharp representation of the CWTs of 
the components of $x(t)$ and hence, it facilities us to determine the number of intervals $m$ in \eqref {def_s_intervals} when we search for smaller $\gs(b)$ for a fixed $b$.

The R${\rm \acute e}$nyi entropy  is a method to evaluate the concentration of a time-frequency representation  \cite{Baraniuk01, Stankovic01}. For a time-frequency representation $D(\upsilon,\xi )$ of a signal $x(t)$, such as CWT, STFT, SST, etc. of $x(t)$,   the R${\rm \acute e}$nyi entropy $R_{\ell ,\zeta} (t)$ is defined by
 \begin{equation}
\label{def_renyi_entropy}
R_{\ell ,\zeta} (t) = \frac{1}{{1 - \ell  }}\log _2 \frac{{\int_{t - \zeta }^{t + \zeta }{\int_{-\infty}^\infty   {\left| {D (\xi, b)} \right|^{2\ell  }d\xi db} } }}{{\left( {\int_{t - \zeta }^{t + \zeta }{\int_{-\infty}^\infty  {\left| {D (\xi, b )} \right|^2  d\xi db} } } \right)^\ell   }} \\,
\end{equation}
where $\ell$  is a constant and usually $\ell \ge 2$ (see \cite{Stankovic01}), $\zeta$ is another constant and $[t - \zeta, t +\zeta]$ is a local range around $t$ to be integrated. Taking the CWT $W(a,b)$ of a signal $x(t)$ as an example, and assuming $\ell=2.5$ (which is also used for the experiments in our paper), we have
  \begin{equation}
 \label{def_renyi_entropy_spec}
R_{\zeta} (t) = -\frac{2}{3}\log _2 \frac{{\int_{t - \zeta }^{t + \zeta }{\int_{ 0 }^\infty   {\left| {W (a, b)} \right|^5 da db} } }}{{\left({\int_{t - \zeta }^{t + \zeta } {\int_{0 }^\infty   {\left| {W (a, b)} \right|^2 da db} } } \right)^{2.5}   }}.
\end{equation}
Observe that $R_{\zeta} (t)<0$. Note that the smaller the R${\rm \acute e}$nyi entropy, the better the time-frequency resolution. So for a fixed time $t$, we can use \eqref{def_renyi_entropy_spec} to find a $\gs$ (denoted as $\gs_u(t)$) with the best time-frequency concentration of  $W(a, b, \gs)$, where $W(a, b, \gs)$ is the CWT of $x(t)$ with $\psi_\gs$ with a parameter $\gs$.
More precisely, replacing $W (a,b)$ in \eqref{def_renyi_entropy_spec} by
$W(a, b, \gs)$, we define the R${\rm \acute e}$nyi entropy $R_{\varsigma} (t, \sigma)$ of $W(a, b, \gs)$, and then, obtain
  \begin{equation}
 \label{def_renyi_entropy_best}
\sigma _u (t) = \mathop {\rm argmin }\limits_{\sigma\ge \frac \ga\mu}  \left\{ {R_{\varsigma  } (t, \sigma)} \right\}.
\end{equation}
We set $\gs_u(t)$ as the upper bound of $\gs(t)$ for a fixed $t$.

With these discussions, we propose an algorithm to estimate $\gs(t)$  as follows.

{\bf Algorithm 1.} (Separability parameter estimation)
Let  $\{\gs_j, j=1, 2, \cdots, n\}$ be an uniform discretization of $\gs$  with $\gs_1>\gs_2>\cdots>\gs_n>\frac {\ga}\mu$ and sampling step $\gD\gs = \gs_{j-1}-\gs_j$. The discrete sequence $s(t),$ $t=t_1, t_2, \cdots , t_N $  (or $t=0, 1, \cdots, N-1$) is the signal to be analyzed.
 \begin{itemize}
\item[] {\bf Step 1.} Let $t$ be a given time. Find $\gs_u$ in \eqref{def_renyi_entropy_best} with $\gs\in \{\gs_j, j=1, 2, \cdots, n\}$. 
\item[] {\bf Step 2.} Let ${\bf s}$ be the set of the intervals given by \eqref{def_s_intervals} with  $\gs=\gs_u$. Let $z=\gs_u$.
If \eqref{nonoverlap_hkgk} holds, then go to Step 3. Otherwise, go to Step 5.
\item[]{\bf Step 3.} Let $\gs=z-\gD\gs$. If the number of intervals $m$ in \eqref{def_s_intervals} with this new $\gs$ remains unchanged, $\gs\ge\gs_n$ and \eqref{nonoverlap_hkgk} holds, then go to Step 4. Otherwise, go to Step 5.
\item[]{\bf Step 4.} Repeat Step 3 with $z=\gs$.
\item[]{\bf Step 5.} Let $C(t)=z$, and repeat Step 1 to Step 4 for the next value of $t$.
\item[]{\bf Step 6.} Smooth  $C(t)$ with a low-pass filter $B(t)$:
\begin{equation}
\label{smooth_C}
\gs_{est}(t)=(C*B)(t).
\end{equation}
\end{itemize}

We call $\gs_{est}(t)$ the estimation of the separability time-varying parameter $\gs_2(t)$ in \eqref{def_gs2}.
We repeat Step 1 through Step 5 with $t=t_1$, then $t= t_2$, $\cdots$, and finally $t=t_N$. In Step 6, we use a low-pass filter
$B(t)$ to smooth $C(t)$. This is because of the assumption of the continuity condition for $A_k(t)$ and $\phi_k(t)$.
With the estimated $\gs_{est}(t)$, we can define the adaptive CWT, the adaptive SST and the 2nd-order adaptive  SST with a time-varying parameter $\gs(t)=\gs_{est}(t)$.

In \cite{Wu17}, the time-varying window was proposed for the sharp representation of SST.
More precisely, denote the R${\rm \acute e}$nyi entropies of SST and the 2nd-order SST by $R^{SST}_{\ell,\zeta, \gs} (t)$  and $R^{SST2}_{\ell ,\zeta, \gs} (t)$ respectively, which are defined by  \eqref{def_renyi_entropy} with $D(\xi, b)$ to be the regular SST $T_x(\xi, b)$ and the regular 2nd-order $T_x^{2nd}(\xi, b)$ of $x(t)$ (with the phase transformation $\go_x^{2nd}(a, b)$ given in \cite{OM17}) with a continuous wavelet $\psi_\gs$ defined by \eqref{def_SST_simple} and \eqref{def_2ndSST_simple} respectively. The time-varying parameter $\gs$ is obtained by minimizing $R^{SST}_{\ell ,\zeta, \gs} (t)$  and $R^{SST2}_{\ell ,\zeta, \gs} (t)$:
\begin{equation}
\label{def_Wu_optimal_SST_para}
\gs_{Re}(t)=\mathop {\rm argmin }\limits_{\gs>0}R^{SST}_{\ell ,\; \zeta, \; \gs} (t),  \;
\gs_{Re2}(t)=\mathop {\rm argmin }\limits_{\gs>0}R^{SST2}_{\ell ,\; \zeta, \; \gs} (t).
\end{equation}
 
With $\gs_{Re}(t)$ and $\gs_{Re2}(t)$ obtained by \eqref{def_Wu_optimal_SST_para}, the time-varying-window SST   with $\gs=\gs_{Re}(t)$ in \cite{Wu17} is defined by
\eqref{def_SST_para_simple} but with the phase transformation $\go^{adp}_x(a, b)$ in \eqref{def_SST_para_simple} replaced by the regular phase transformation \eqref{def_phase} for the conventional SST. Similarly, the 2nd-order time-varying-window SST  with $\gs=\gs_{Re2}(t)$ in  \cite{Wu17} is defined by
\eqref{def_2ndSST_para_simple} but with the phase transformation $\go^{2adp}_x(a, b)$ in \eqref{def_2ndSST_para_simple} replaced by the regular phase transformation defined in  \cite{OM17} for the conventional 2nd-order SST. With PT representing phase transformation,
we call them the regular-PT adaptive SST and the 2nd-order regular-PT adaptive SST, respectively.

\begin{figure}[th]
\centering
\begin{tabular}{cc}
\resizebox{2.4in}{1.8in}{\includegraphics{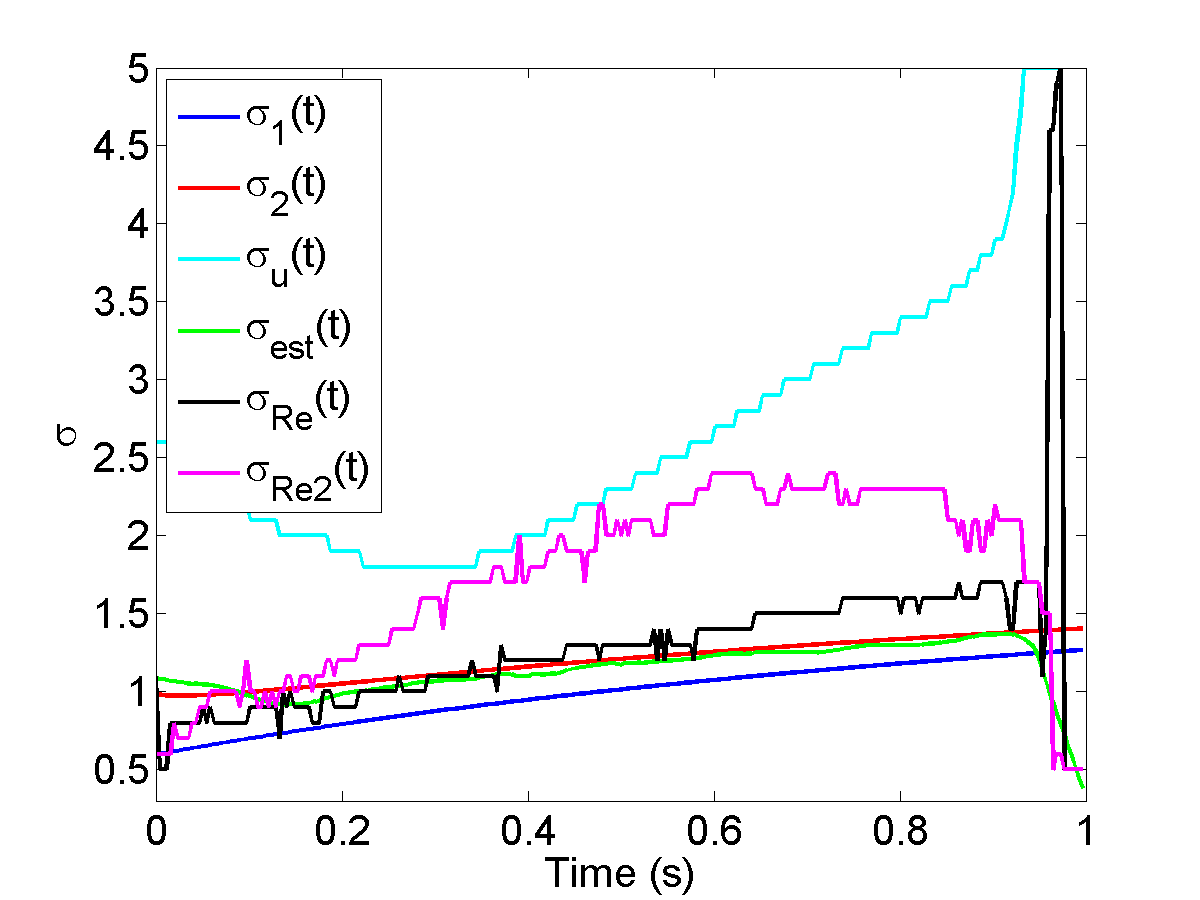}}&\\
\resizebox{2.4in}{1.8in}{\includegraphics{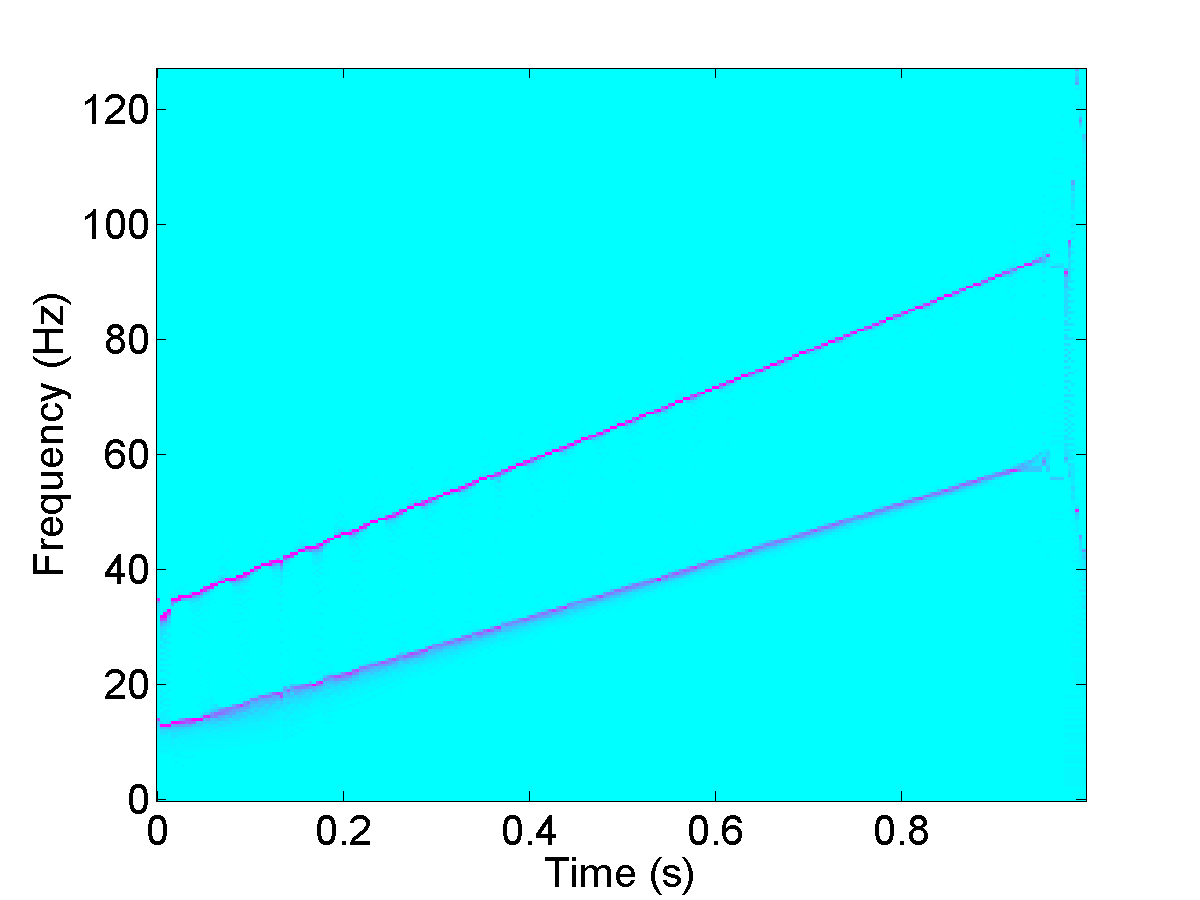}}
&\resizebox{2.4in}{1.8in}{\includegraphics{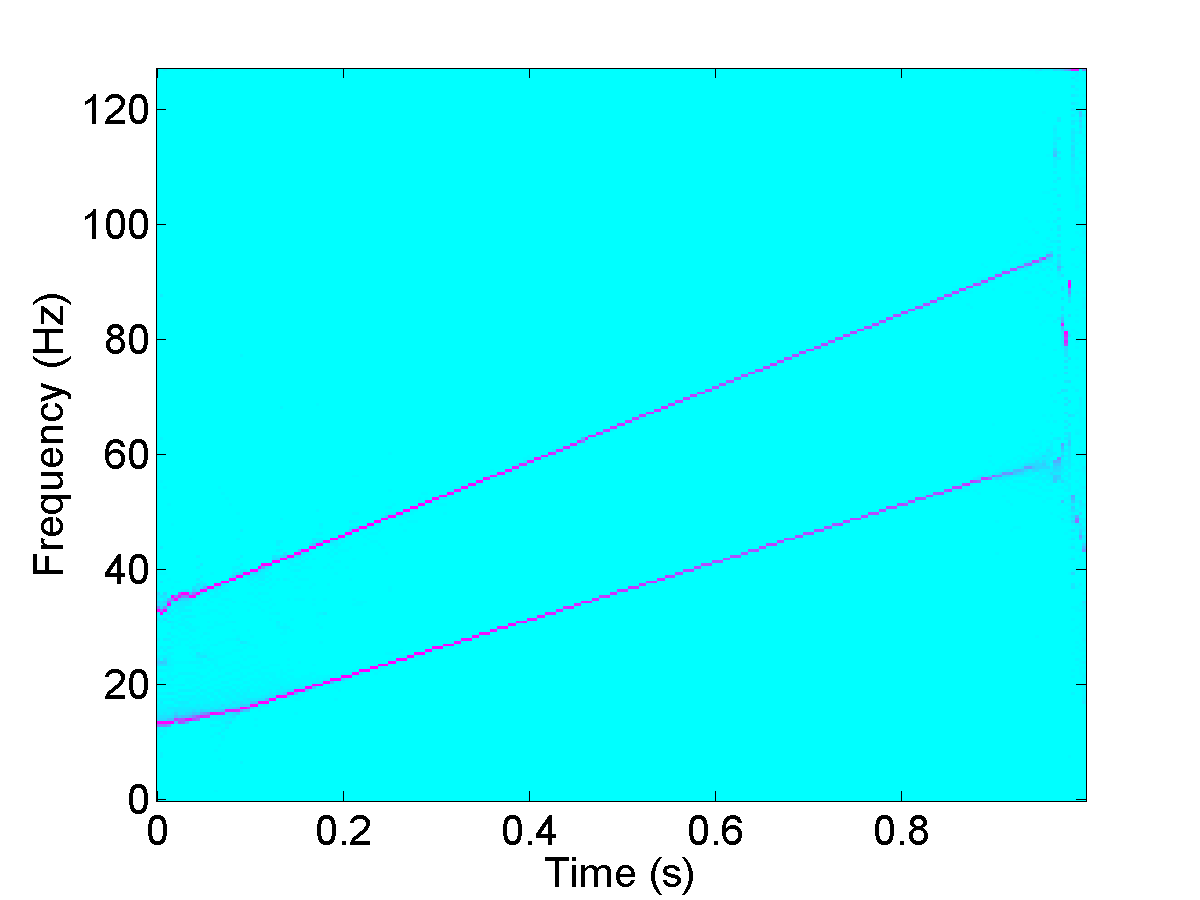}}\\
\resizebox{2.4in}{1.8in}{\includegraphics{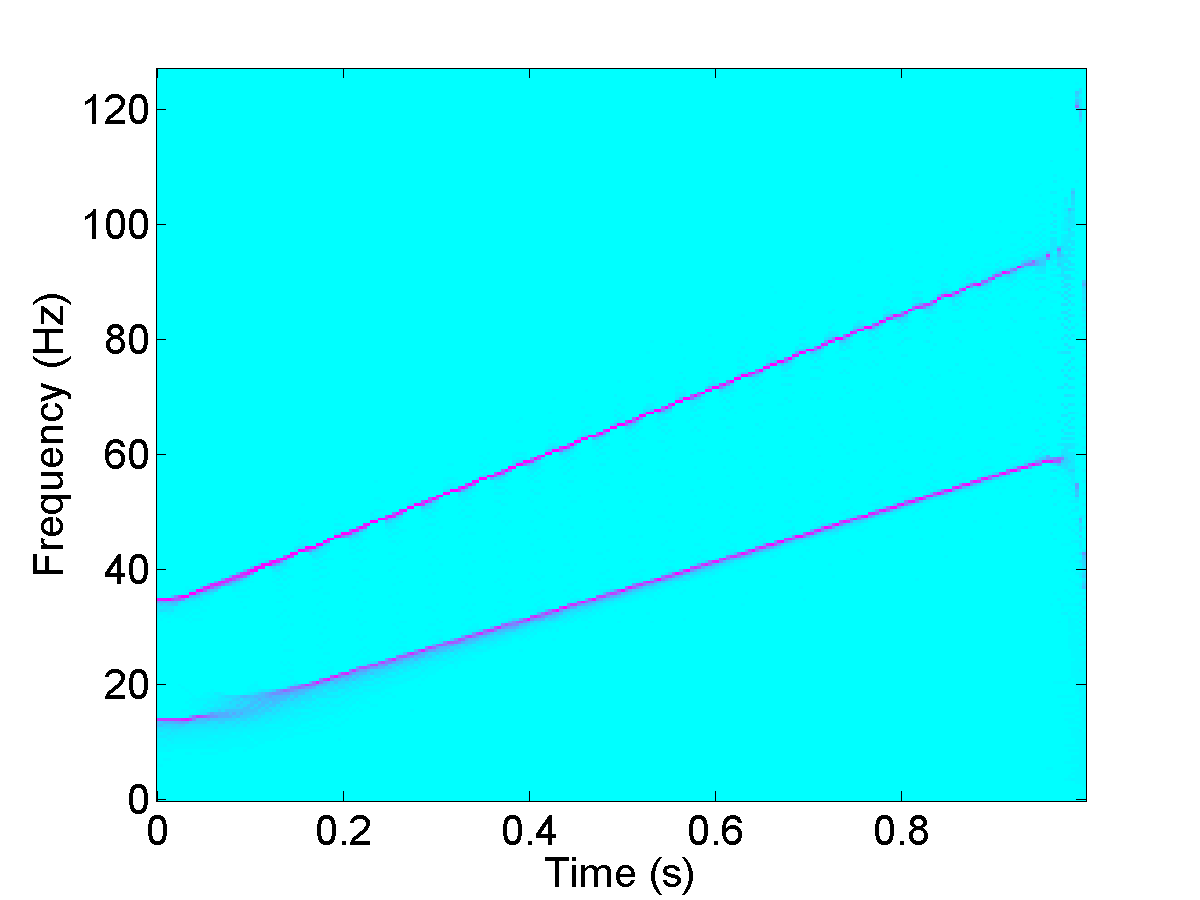}}
&\resizebox{2.4in}{1.8in}{\includegraphics{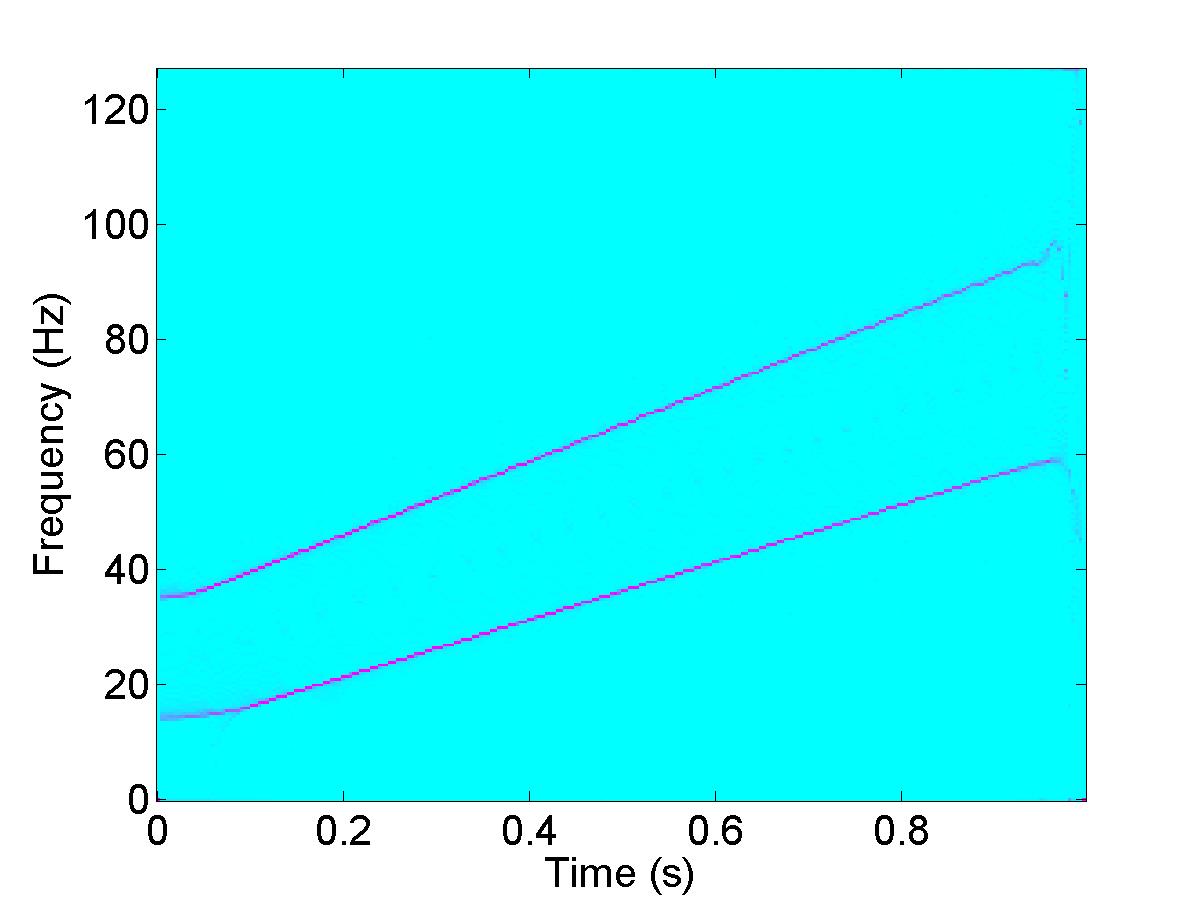}}\\
\end{tabular}
\caption{\small  Example of the two-component LFM signal $x(t)$ in \eqref{two_linear_chirps}. Top: various time-varying parameters;  Middle-left: regular-phase-transformation adaptive SST with $\gs_{Re}(t)$;
Middle-right: 2nd-order regular-phase-transformation adaptive SST with $\gs_{Re2}(t)$;
Bottom-left: adaptive SST with $\gs_{est}(t)$; Bottom-right: 2nd-order adaptive SST with $\gs_{est}(t)$.}
\label{fig:adaptiveSST_estimate_rev}
\end{figure}

We use the proposed algorithm to process the two-component linear chirp signal $x(t)$ in \eqref{two_linear_chirps} and compare the performance
of this algorithm with those of regular SST and  regular-PT adaptive SST in \cite{Wu17}. The different time-varying parameters are shown in the top row of Fig.\ref{fig:adaptiveSST_estimate_rev}, where  $\gs_1(t)$, $\gs_2(t)$, $\gs_u(t)$, $\gs_{est}(t)$, $\gs_{Re}(t)$ and  $\gs_{Re2}(t)$ are defined by \eqref{def_gs1}, \eqref{def_gs2},  \eqref{def_renyi_entropy_best}, \eqref{smooth_C} and \eqref {def_Wu_optimal_SST_para},  respectively.
Here we let $\gs \in [0.5, 10]$ with $\gD\gs=0.05$, namely $\gs_1 = 10$ in {\bf Algorithm 1}.  We set $\ell=2.5$, $ \zeta=4$ (sampling points, for discrete signal) and $\Gamma_3=0.2$. Note that we set the same values of $\ell$, $ \zeta$, and $\Gamma_3$ for the other experiments in \S7.
We use a simple rectangular window $B=\{1/5, 1/5, 1/5, 1/5, 1/5\}$ as the low-pass filter.
Note that $\gs_1(t)$ and $\gs_2(t)$ are the same curves as those plotted in  Fig.\ref{fig:adaptiveSST_true}.
The estimation $\gs_{est}(t)$ by  {\bf Algorithm 1} is very close to $\gs_2(t)$ except for at the boundary near $t=1$. So the estimation algorithm is an efficient method to estimate the well-separated time-varying parameter $\gs_2(t)$.
From Fig.\ref{fig:adaptiveSST_estimate_rev}, we observe that the proposed adaptive SST with $\gs=\gs_{est}(t)$ is similar to the regular-PT adaptive SST with $\gs=\gs_{Re}(t)$, and both of them are much better than
the conventional SST which is shown in Fig.\ref{fig:adaptiveSST_true}.
The 2nd-order adaptive SST with the estimated parameter $\gs_{est}(t)$ is as sharp as the 2nd-order adaptive SST with parameter $\gs_2(t)$ in Fig.\ref{fig:adaptiveSST_true}. In addition, we observe that the regular-PT adaptive FSST also performs well in  the time-frequency energy concentration of this two-component signal.

The Matlab routines for Algorithm 1, the adaptive SST and regular-PT adaptive SST can be downloaded at the website of one of the authors \cite{Jiang_Web}.

\clearpage

\section{Experiments on multicomponent signals}

In this section we consider signals with more than 2 components. As demonstrated by
Figs.\ref{fig:adaptiveSST_true} and \ref{fig:adaptiveSST_estimate_rev},
the conventional 2nd-order SST and 2nd-order adaptive SST perform better than the first-order SST.
In this section we just show some results of the conventional 2nd-order  SST and the 2nd-order adaptive SST.

First, we consider a three-component signal,
\begin{equation}
\label{three_component_signal}
 s(t) = s_1 (t) + s_2 (t) + s_3 (t) \\
  = \cos \big( 16\pi t \big) + \cos \big ( 96\pi t + 30\cos (4\pi t) \big) + \cos \big( 180\pi t + 30\cos (4\pi t)  \big), \\
\end{equation}
where $t\in[0,1]$, $s_1 (t)$ is a single-tone mode, $s_2 (t)$ and $s_3 (t)$ are sinusoidal frequency modulation modes. $s(t)$ is sampled uniformly with $ N=512$ sample points. Hence the sampling rate is $F_s=512$ Hz.
We let $\mu$ in Morlet's wavelet $\psi_\gs$ be 1, and $\tau_0$ in \eqref{def_ga} be $1/5$.

\begin{figure}[th]
\centering
\begin{tabular}{cc}
\resizebox{2.4in}{1.8in}{\includegraphics{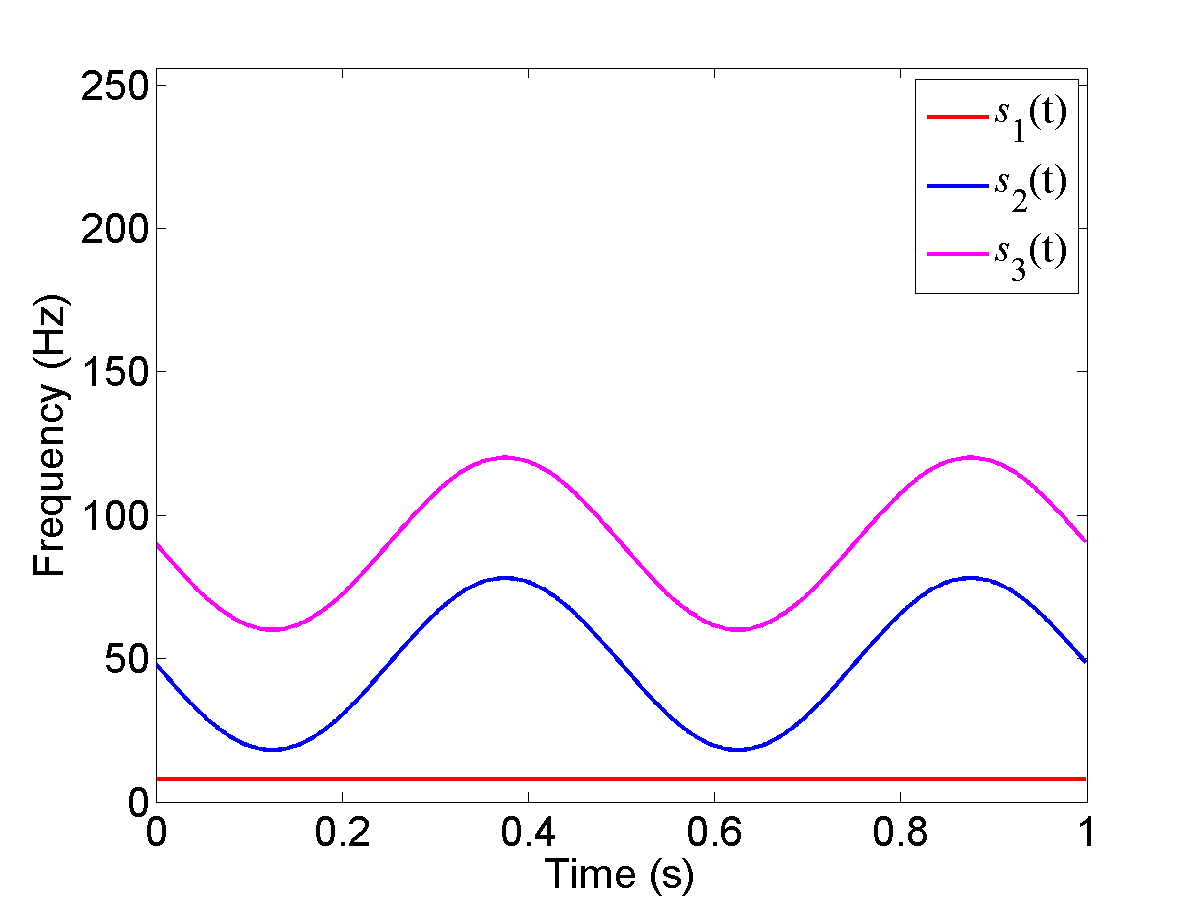}}
&\resizebox{2.4in}{1.8in}{\includegraphics{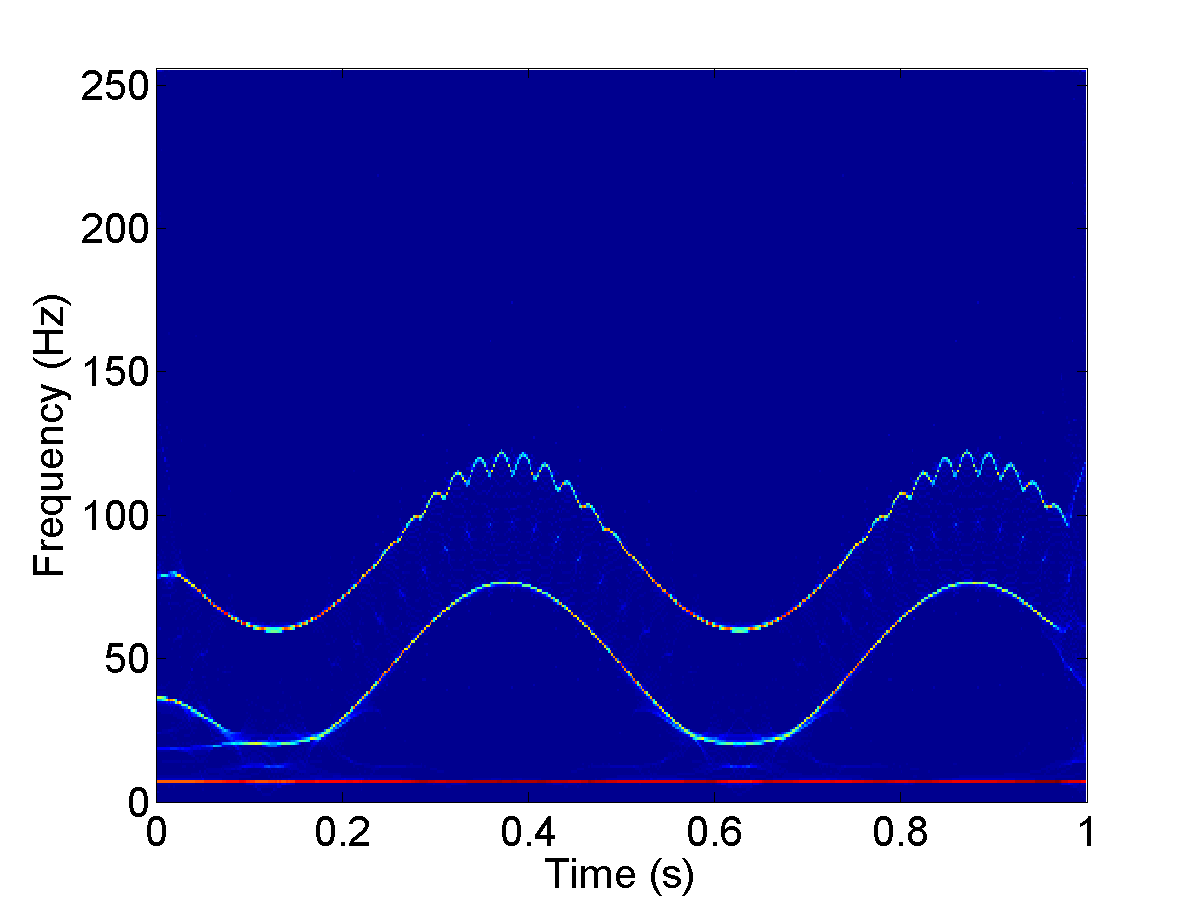}}\\
\resizebox{2.4in}{1.8in}{\includegraphics{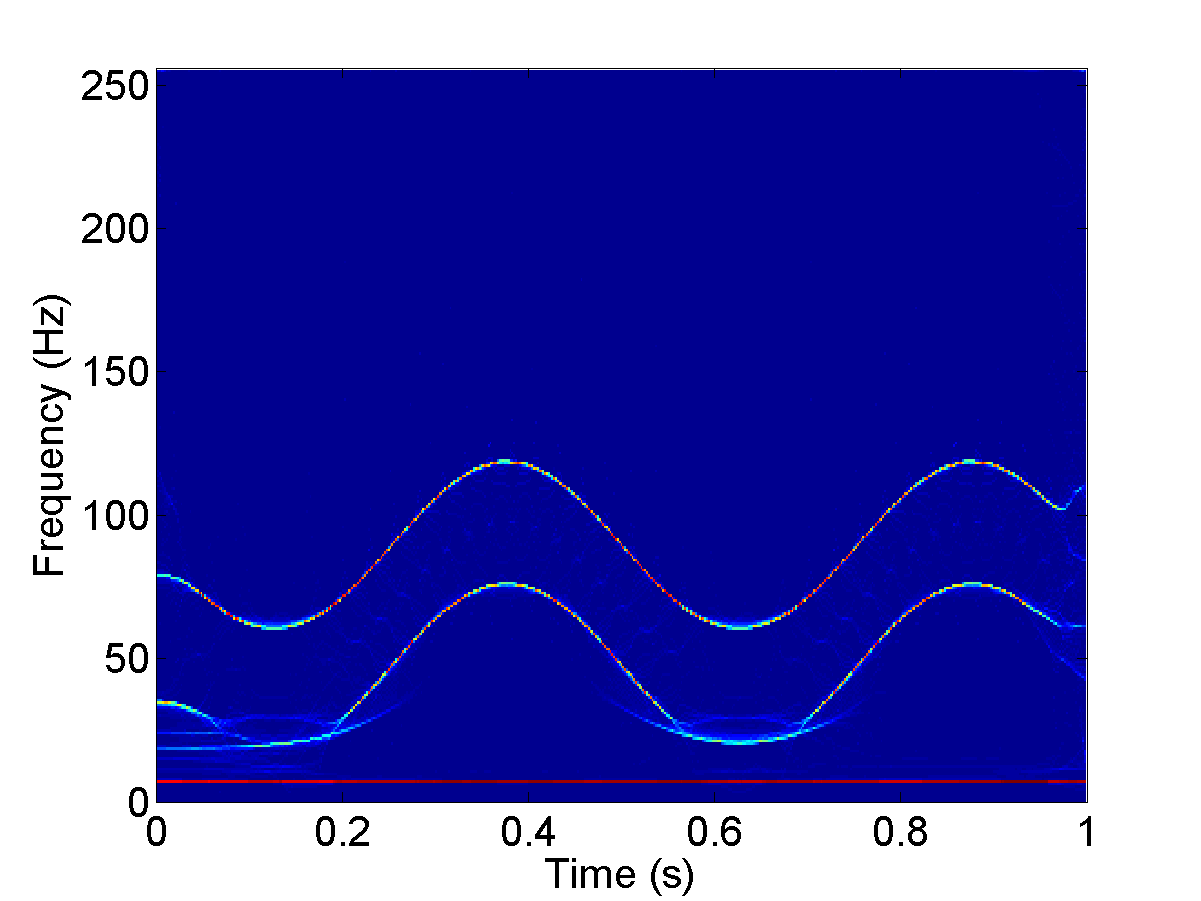}}
&\resizebox{2.4in}{1.8in}{\includegraphics{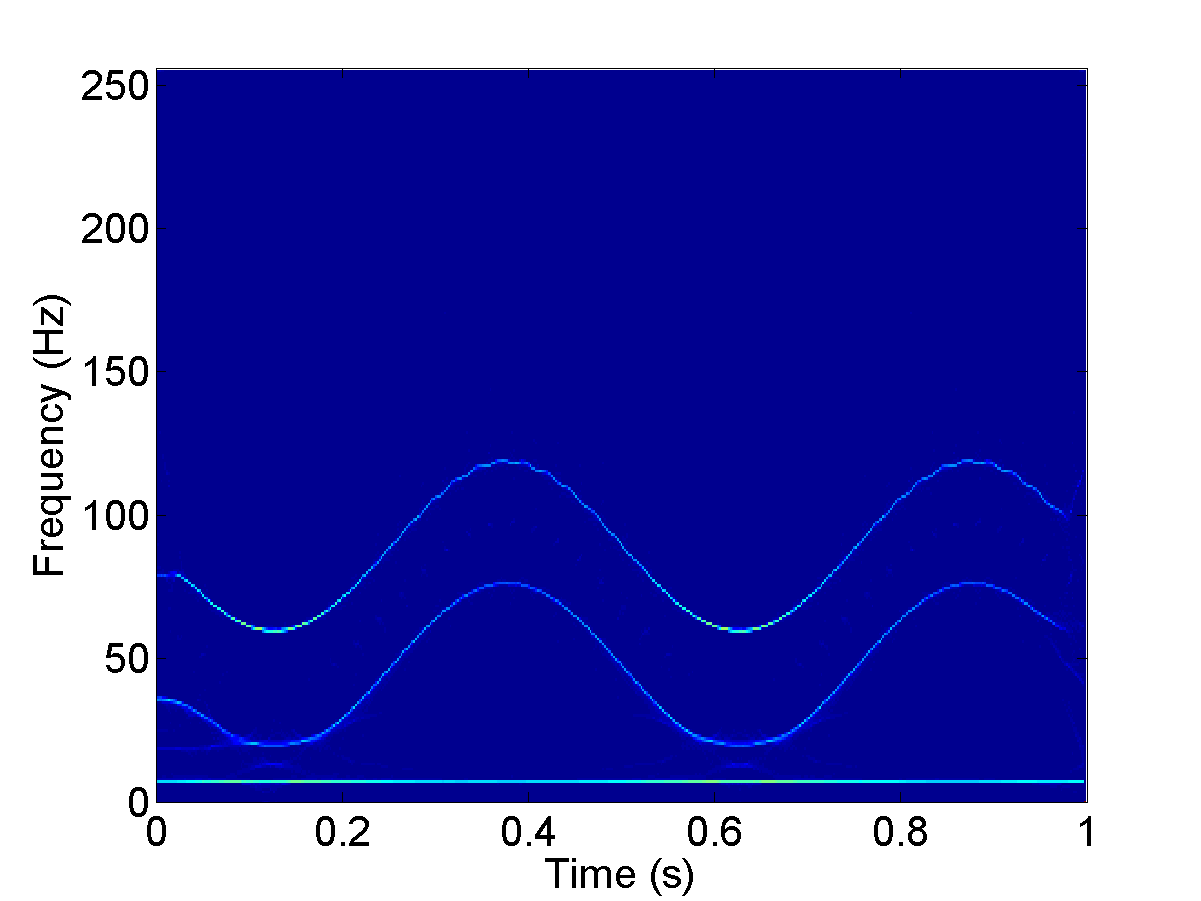}}\\
\end{tabular}
\caption{\small Example of the three-component signal $s(t)$ in \eqref{three_component_signal}.
Top-left: instantaneous frequencies of $s_1 (t)$, $s_2 (t)$ and  $s_3 (t)$; Top-right:  conventional 2nd-order SST with $\gs=1$;  Bottom-left: conventional 2nd-order SST with $\gs=1.5$; Bottom-right:
2nd-order adaptive SST with  time-varying $\gs(t)=\gs_{est}(t)$.}
\label{fig:SST_3_comp}
\end{figure}

Fig.\ref{fig:SST_3_comp} shows the experimental results of the  three-component signal $s(t)$.
Observe that the 2nd-order SSTs represent well for the single-tone mode $s_1 (t)$. For the conventional 2nd-order SST, it is difficult to find a $\gs$ to represent well for both of the sinusoidal frequency modulation modes $s_2 (t)$ and $s_3 (t)$. As shown in Fig.\ref{fig:SST_3_comp}, $\gs=1$ is suitable for $s_2 (t)$, while $\gs=1.5$ is suitable for $s_3 (t)$.
Setting same parameters $\{\gs_j\}$,  $\gD\gs$, $\ell$, $ \zeta$, $\Gamma_3$ and $B$ as those in Fig.\ref{fig:adaptiveSST_estimate_rev}, we estimate the time-varying parameter  $\gs_{est}(t)$. Note that the sinusoidal frequency modulation modes $s_2 (t)$ and $s_3 (t)$ are approximated by LFM modes during any local time when using {\bf Algorithm 1} to estimate the time-varying parameter $\gs=\gs_{est}(t)$. The bottom-right of Fig.\ref{fig:SST_3_comp} shows the 2nd-order adaptive SST with $\gs=\gs_{est}(t)$. 
Obviously, the 2nd-order adaptive SST can represent signal $s(t)$ separately and sharply, and gives the highest energy concentration. 

\begin{figure}[th]
\centering
\begin{tabular}{ccc}
\resizebox{2.0in}{1.5in}{\includegraphics{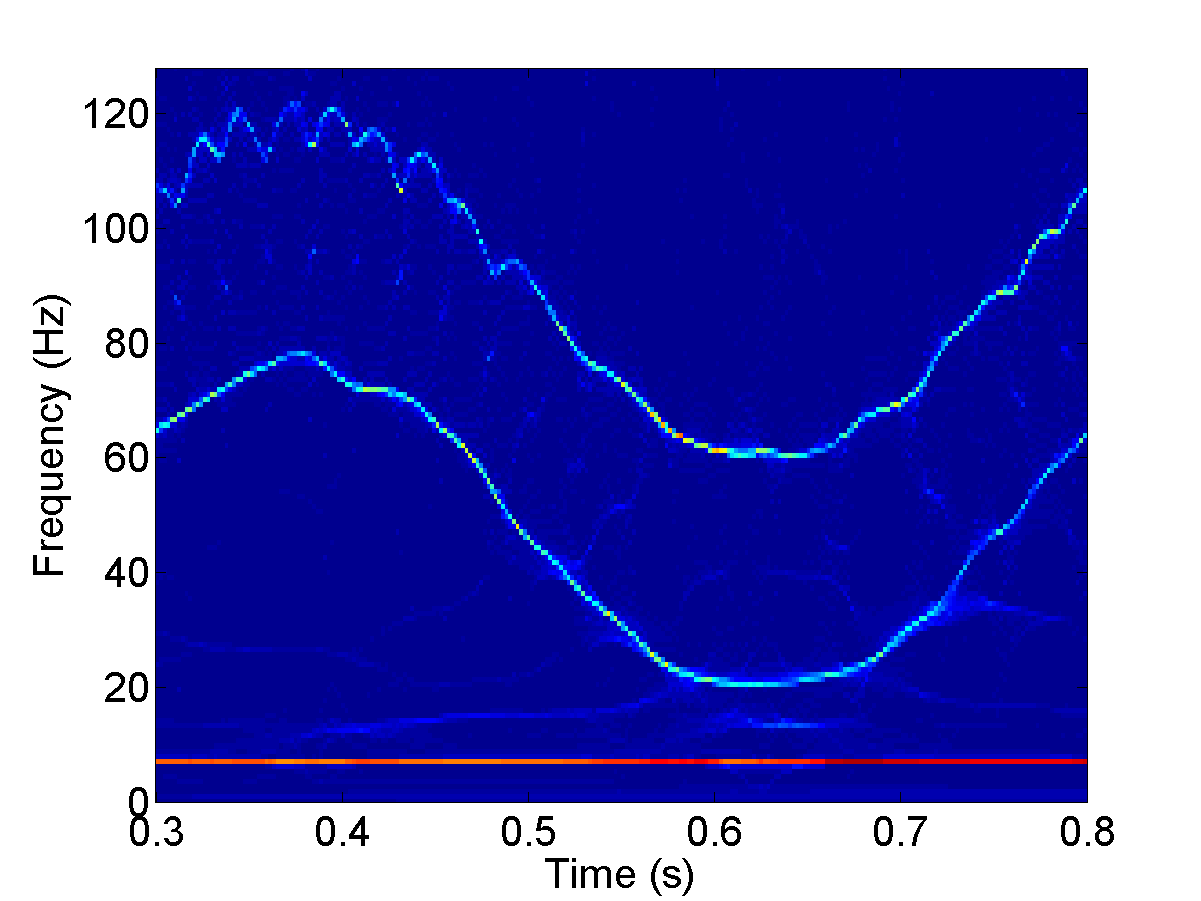}}
&\resizebox{2.0in}{1.5in}{\includegraphics{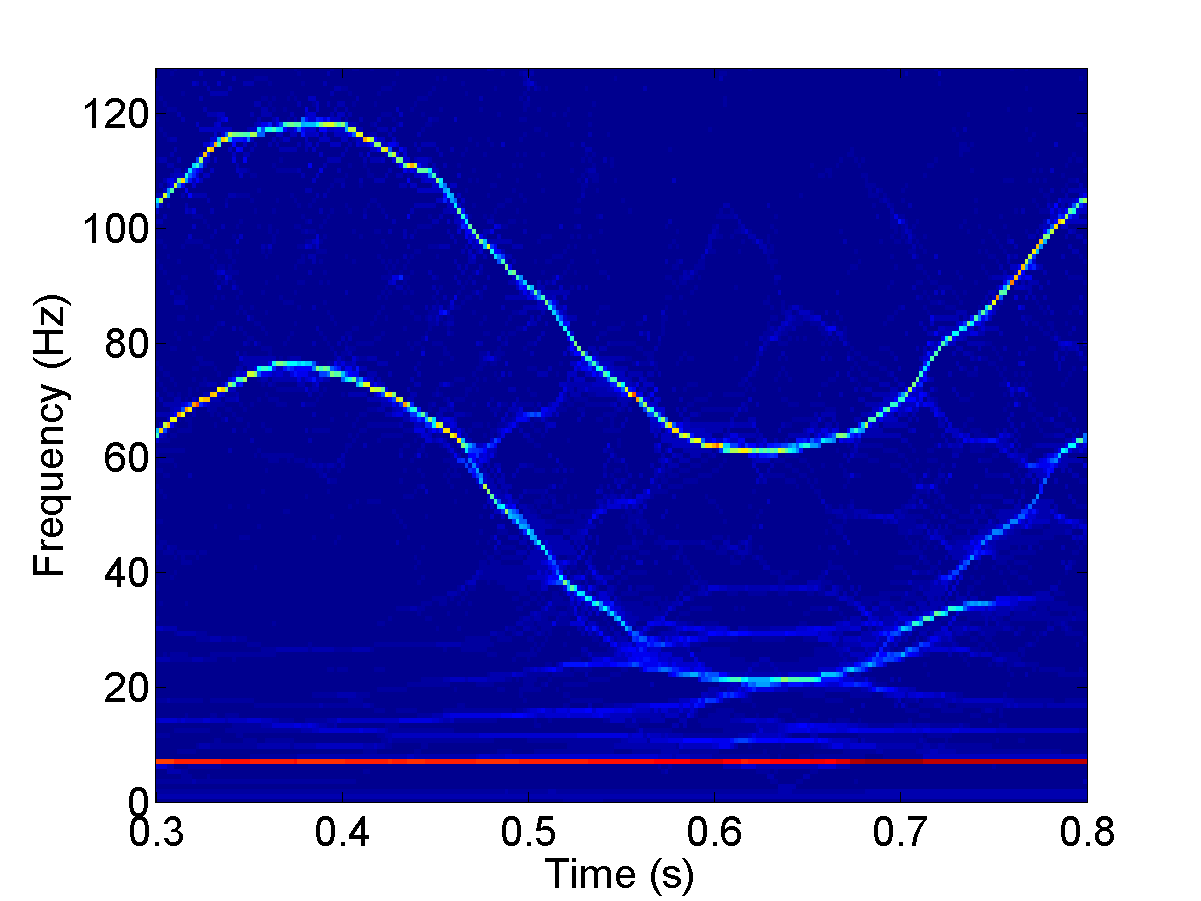}}
&\resizebox{2.0in}{1.5in}{\includegraphics{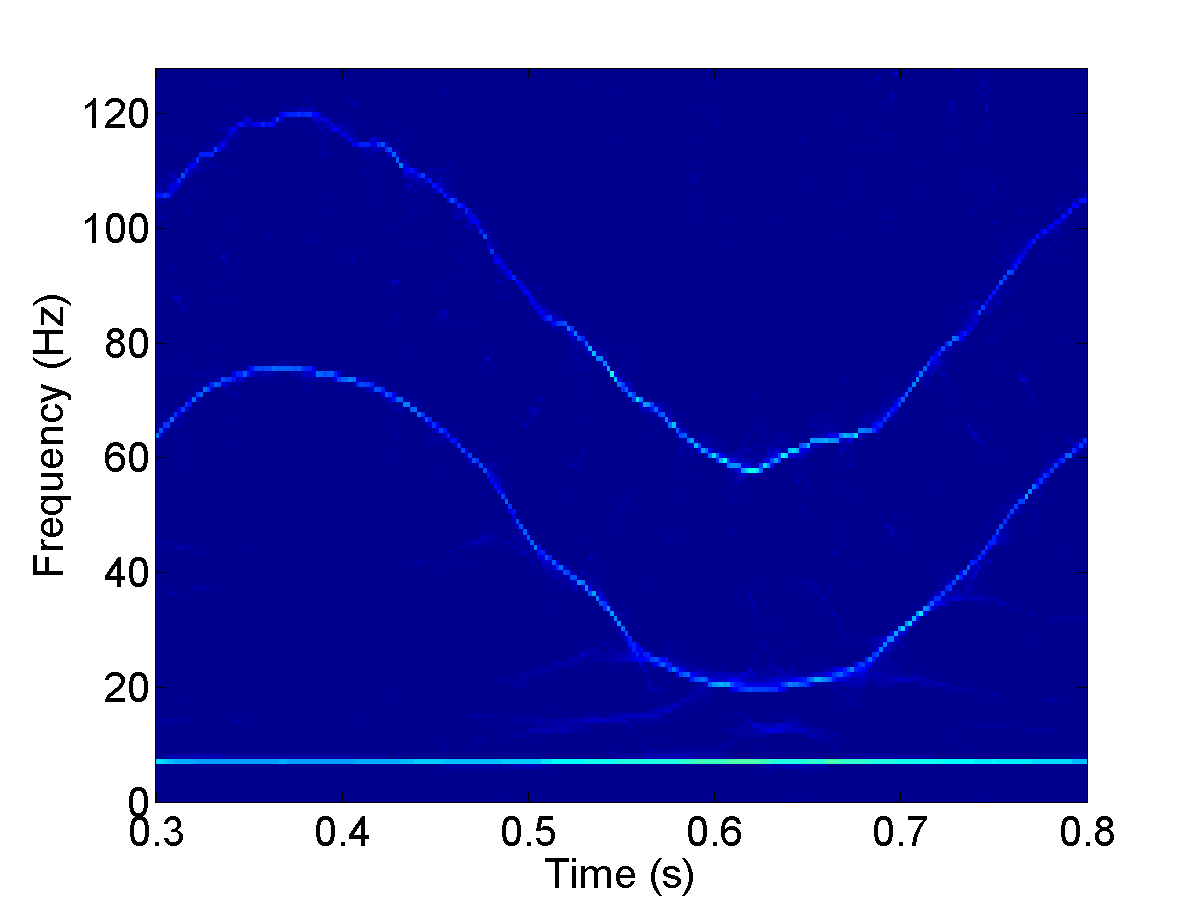}}
\end{tabular}
\caption{\small Example of the three-component signal $s(t)$ in \eqref{three_component_signal} with noise  SNR=10dB and enlarged scale. Left: conventional 2nd-order SST with $\gs=1$; Middle: conventional 2nd-order SST with $\gs=1.5$;
Right: 2nd-order adaptive SST with time-varying parameter $\gs(t)=\gs_{est}(t)$.
}
\label{fig:adaptiveSST_3_comp_noi}
\end{figure}

In real applications, signals are usually accompanied by noises and interferences. We add Gaussian noises to the three-component signal $s(t)$ in \eqref{three_component_signal} with signal-to-noise ratio (SNR) 10dB. Fig.\ref{fig:adaptiveSST_3_comp_noi} shows the experimental results with enlarged scale, namely $b \in [0.3,08] $ and $\xi \in [0,F_s/4]$ for time-frequency diagrams.
Although noises will affect the time-frequency distributions of SSTs and 2nd-order SSTs and decrease their energy concentration, the 2nd-order adaptive SST proposed in this paper is much clearer and sharper than other time-frequency distributions. This is because the 2nd-order adaptive SST has higher energy concentration as shown in Fig.\ref{fig:SST_3_comp}, and therefore is more efficient in noise suppression.

\begin{figure}[tbh]
\centering
\begin{tabular}{cc}
\resizebox{2.4in}{1.8in}{\includegraphics{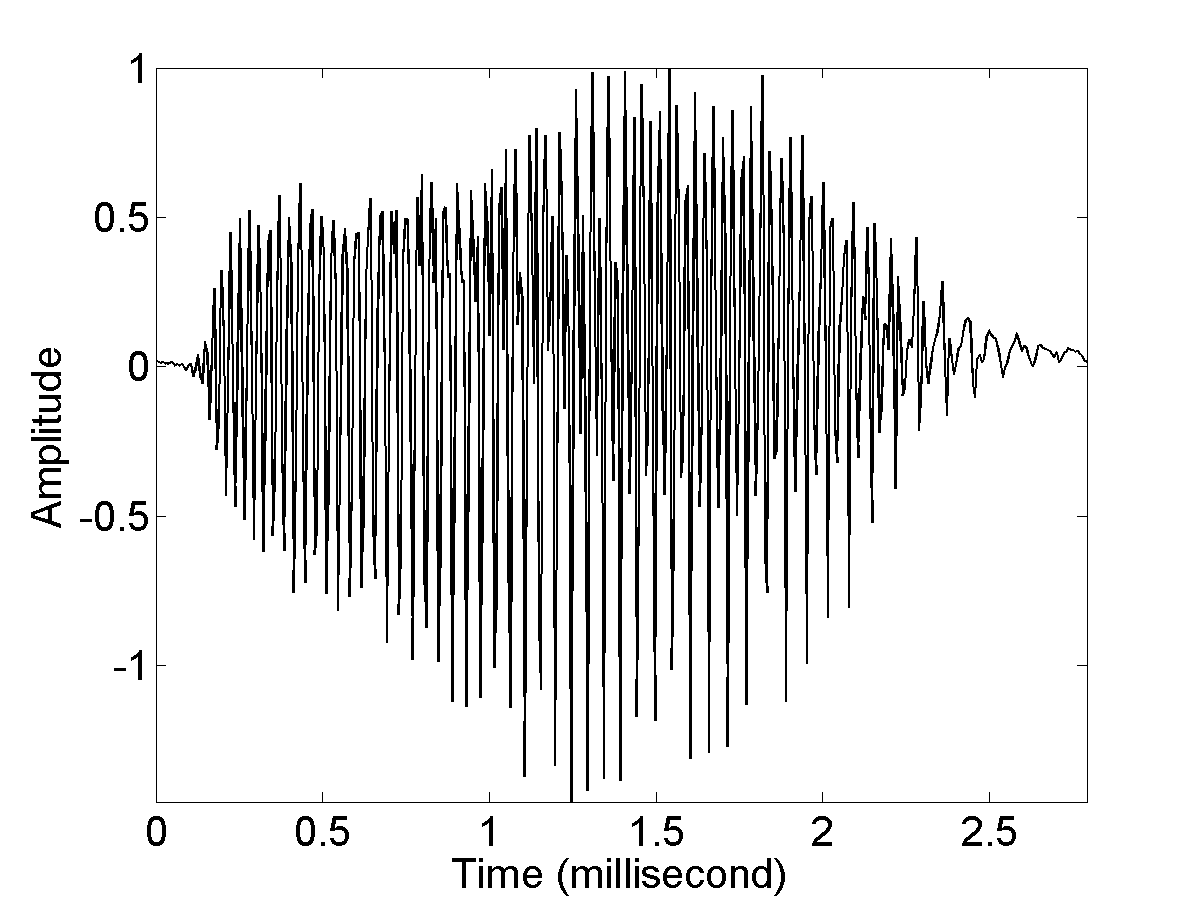}}
&\resizebox{2.4in}{1.8in}{\includegraphics{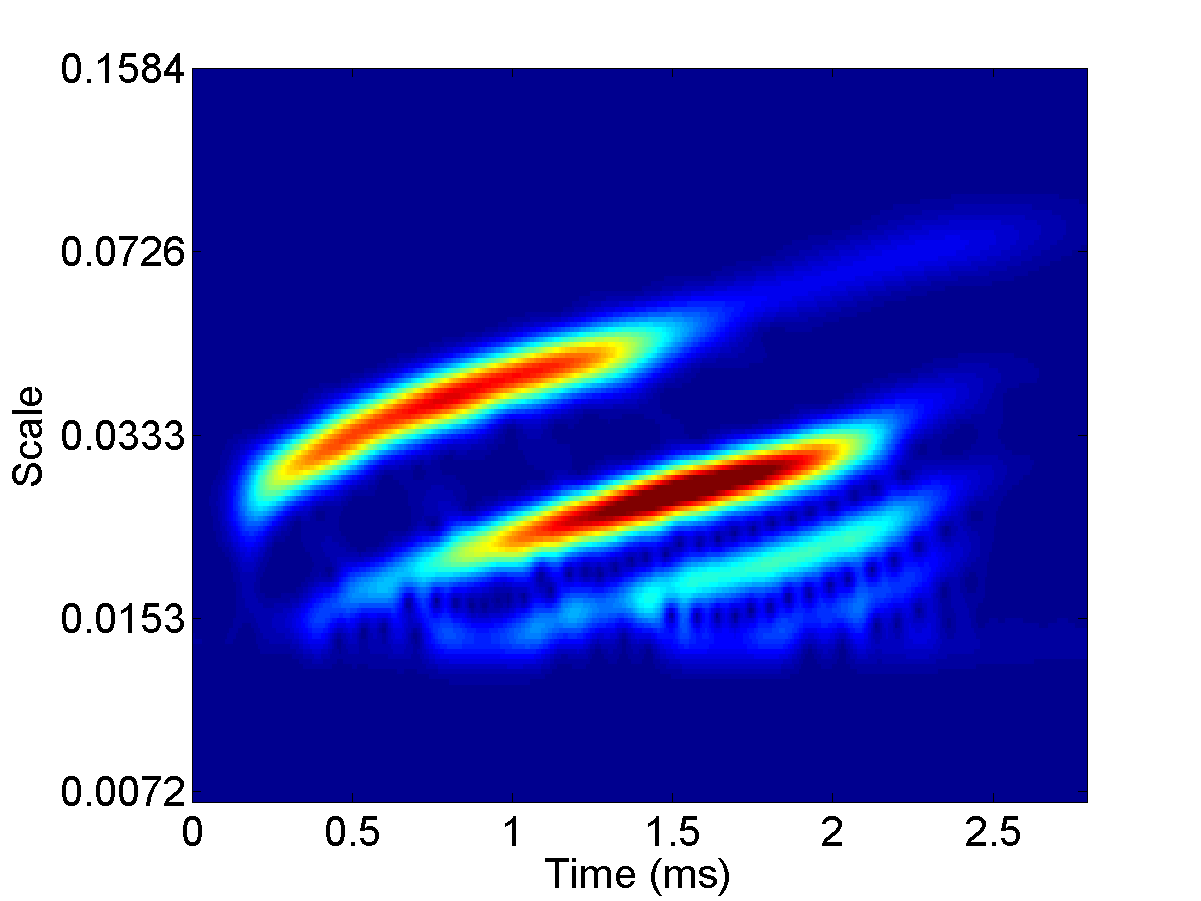}}\\
\resizebox{2.4in}{1.8in}{\includegraphics{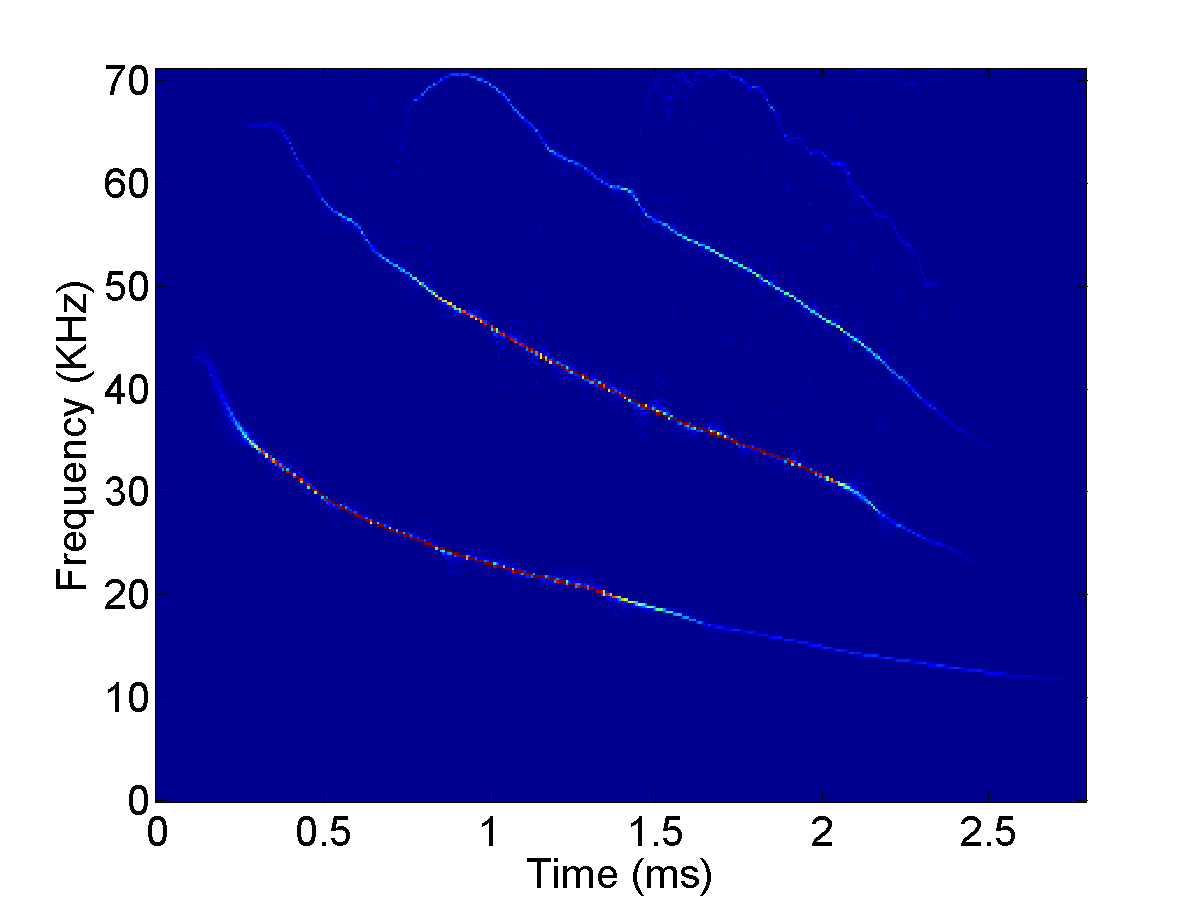}}
&\resizebox{2.4in}{1.8in}{\includegraphics{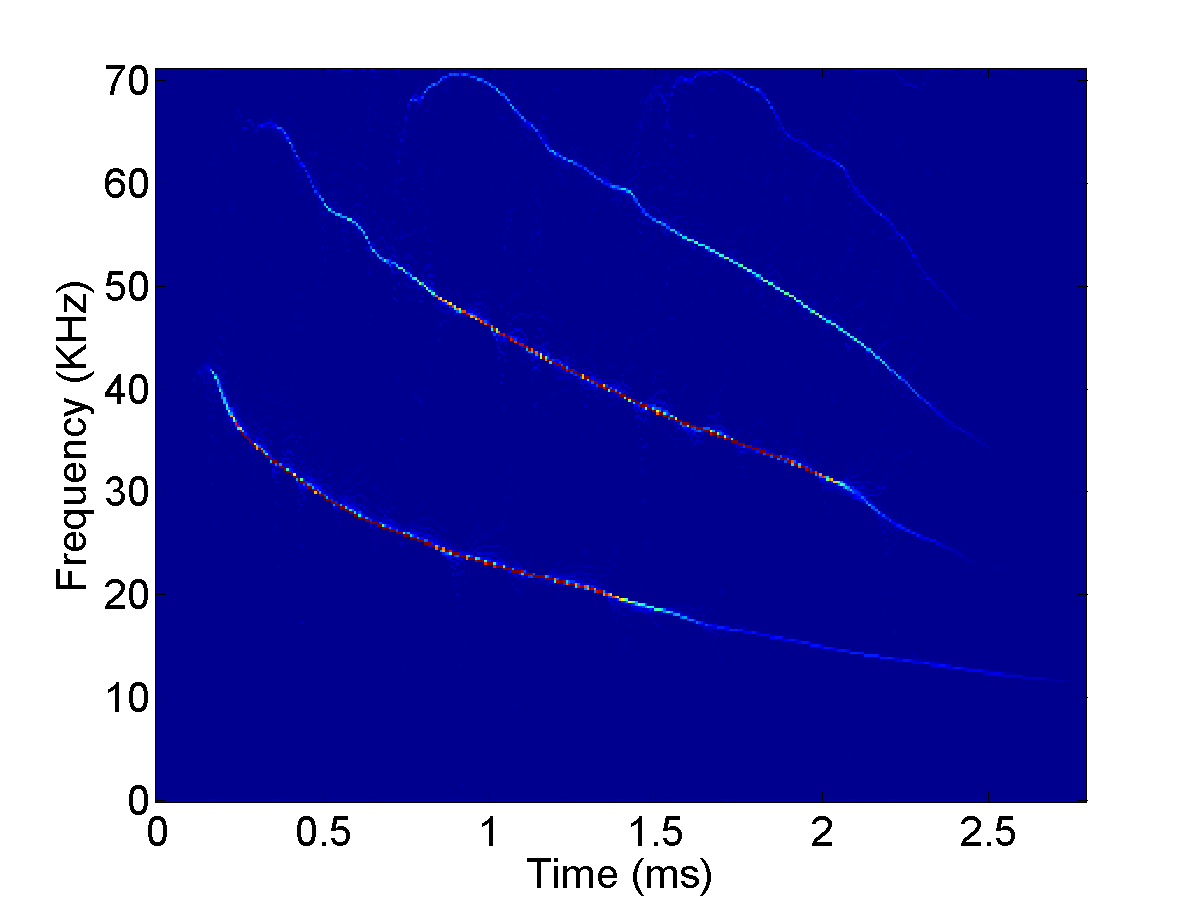}}
\end{tabular}
\caption{\small Example of the bat echolocation signal. Top-left:  waveform; Top-right: conventional CWT with $\gs=2$;
Bottom-left: conventional 2nd-order SST with $\gs=2$;
Bottom-right: 2nd-order  adaptive SST with time-varying parameter $\gs(t)=\gs_{est}(t)$ obtained by our proposed Algorithm 1.}
\label{fig:adaptiveSST_bat}
\end{figure}

In order to further verify the reliability of the proposed algorithm, we test our method on a real dataset containing a bat echolocation signal emitted by a large brown bat \cite{bat}. There are 400 samples with the sampling period 7 microseconds (sampling rate $F_s\approx142.86$ kHz). From its CWT presented in Fig.\ref{fig:adaptiveSST_bat}, the echolocation signal is a multicomponent signal, which consists of nonlinear FM components.
Fig.\ref{fig:adaptiveSST_bat} shows the time-frequency representations of the echolocation signal: 
the conventional 2nd-order SST with $\gs=2$ 
and the proposed 2nd-order adaptive SST  with the estimated time-varying parameter $\gs_{est}(t)$.
Unlike the three-component signal $s(t)$ in \eqref{three_component_signal},  
the four components in the bat signal are much well separated.
Thus, both the conventional 2nd-order SST and the 2nd-order adaptive SST
can separate well the components of the signal.
In addition, they both give sharp representations in the time-frequency plane.
Comparing with the conventional 2nd-order SST, the 2nd-order adaptive SST with $\gs=\gs_{est}(t)$ 
gives a better representation for the fourth component (the highest frequency component) and the two ends of the signal.
One may tempt to try other choices of $\gs$ for the conventional SST.
For example, one may increase the value of $\gs$ to obtain a sharper representation of the fourth component with the conventional 2nd-order SST, but this will affect the concentration of the first component (the lowest frequency component).

\bigskip

\begin{figure}[th]
\centering
\begin{tabular}{cc}
\resizebox{3.2in}{2.4in}{\includegraphics{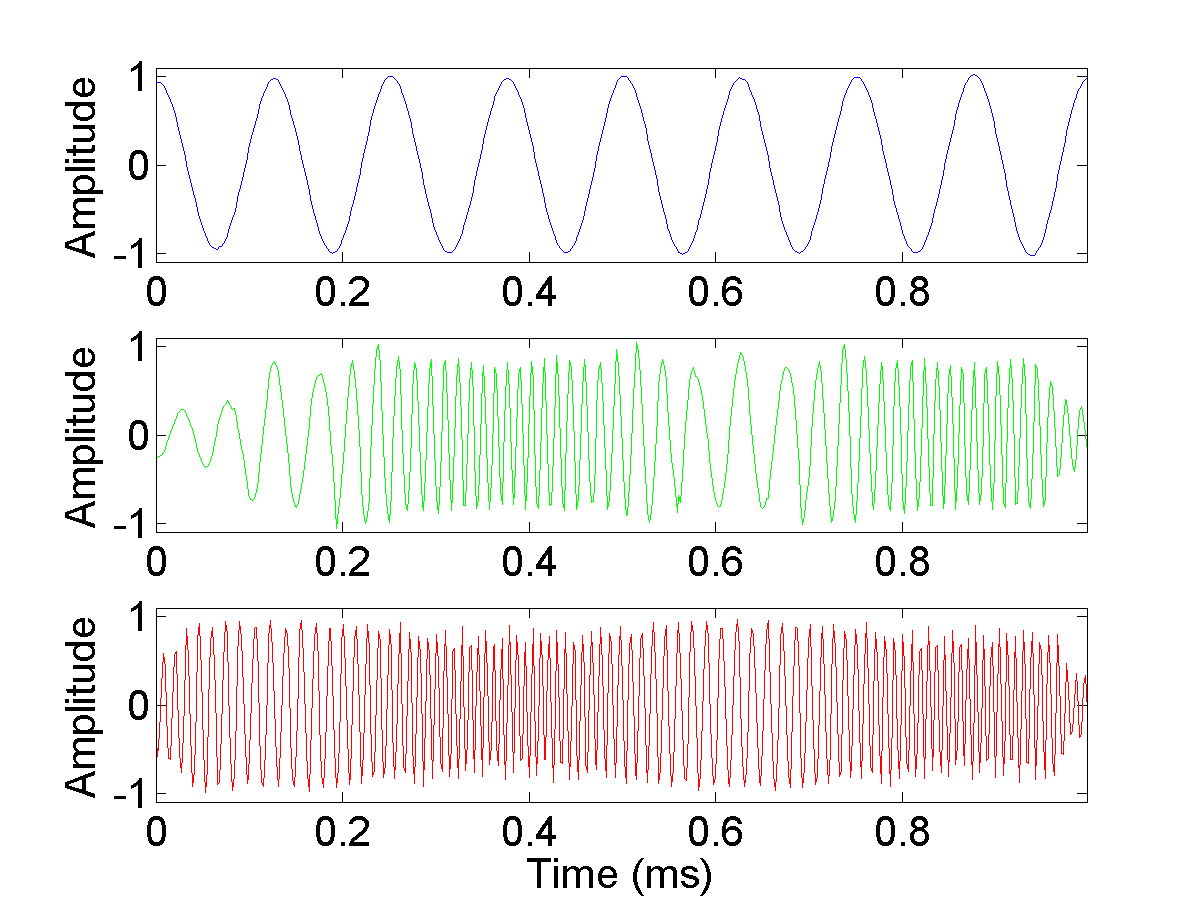}} &
\resizebox{3.2in}{2.4in}{\includegraphics{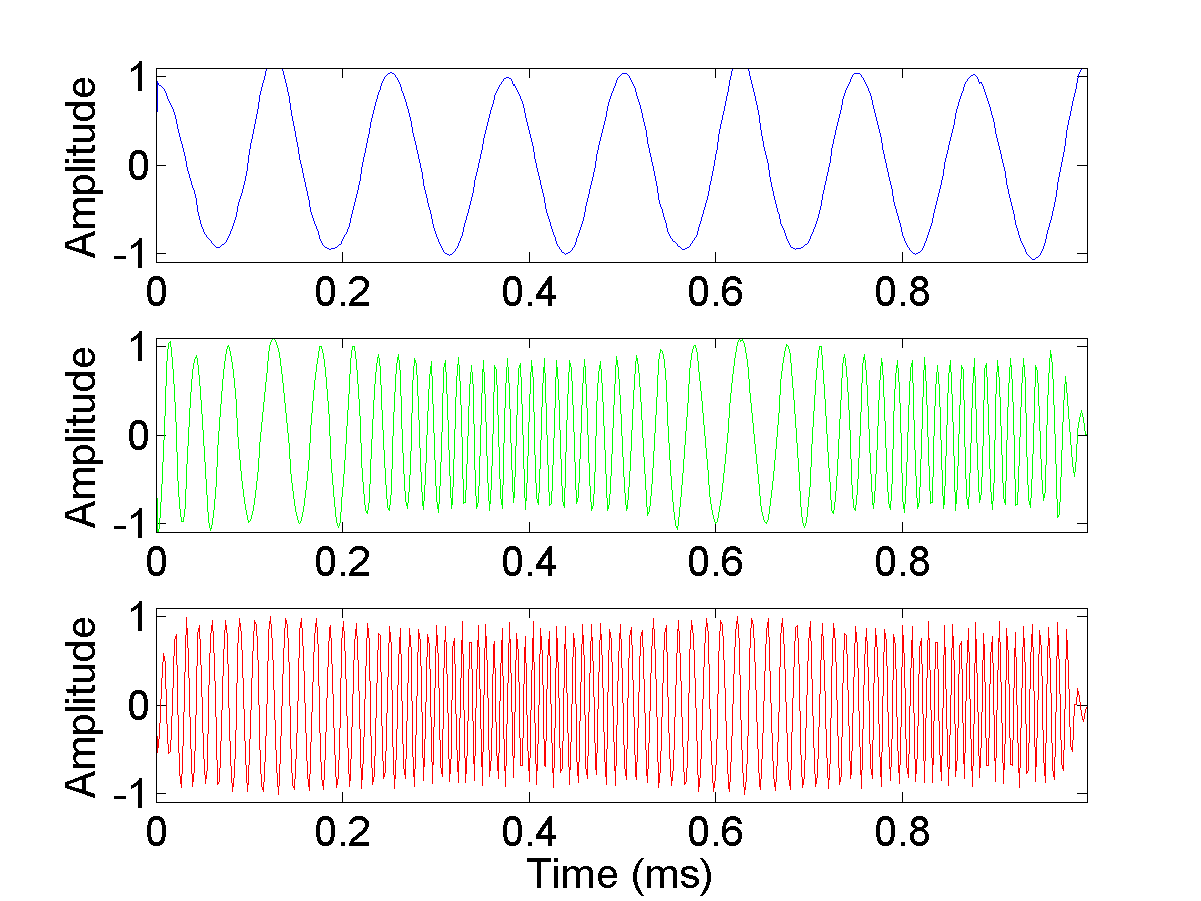}}
\end{tabular}
\caption{\small Reconstruction results of  the components of $s(t)$ given in \eqref{three_component_signal}. Reconstructed $s_1(t), s_2(t)$ and $s_3(t)$ by
regular 2nd-order SST (left column) and by  2nd-order adaptive SST (right column)}
\label{fig:reconst_SST_three_component}
\end{figure}

\begin{figure}[th]
\centering
\begin{tabular}{cc}
\resizebox{3.2in}{2.6in}{\includegraphics{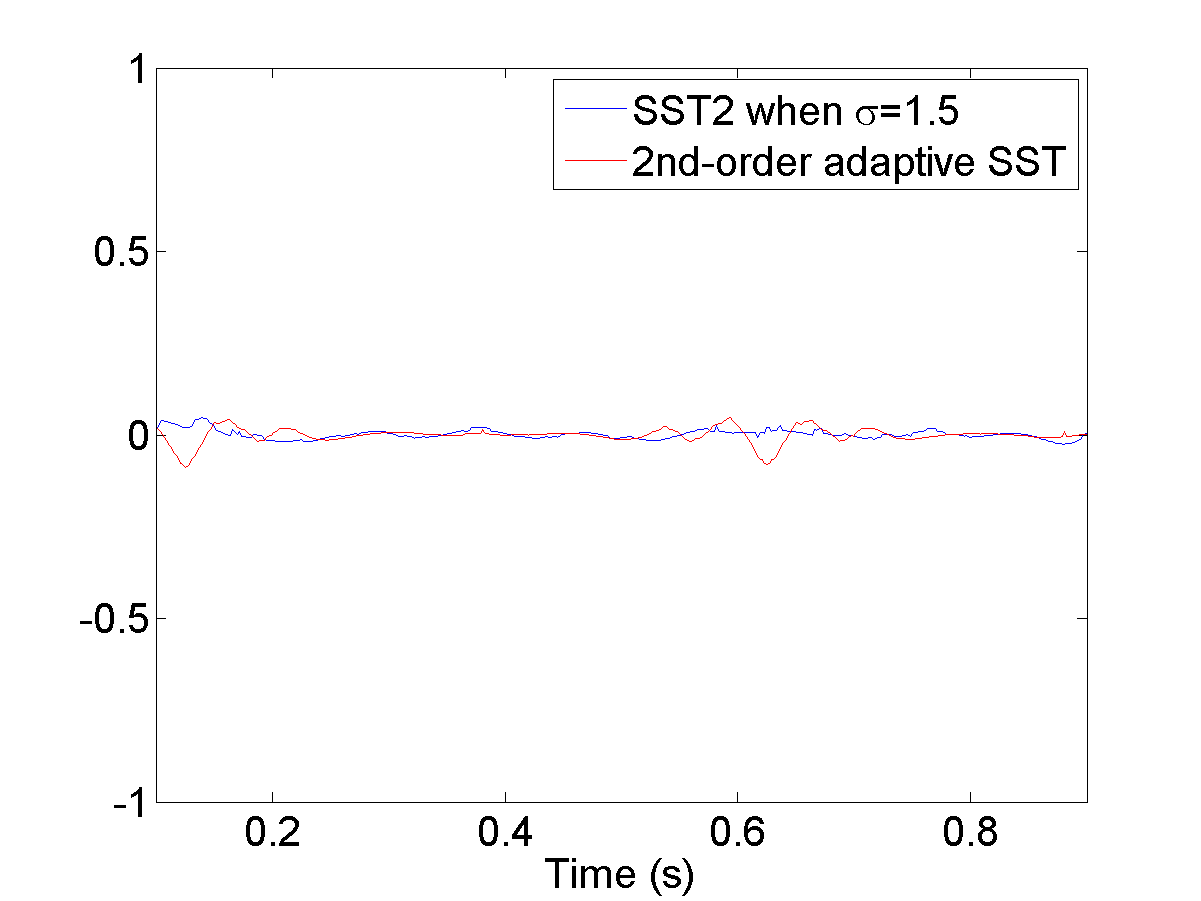}} &
\resizebox{3.2in}{2.6in}{\includegraphics{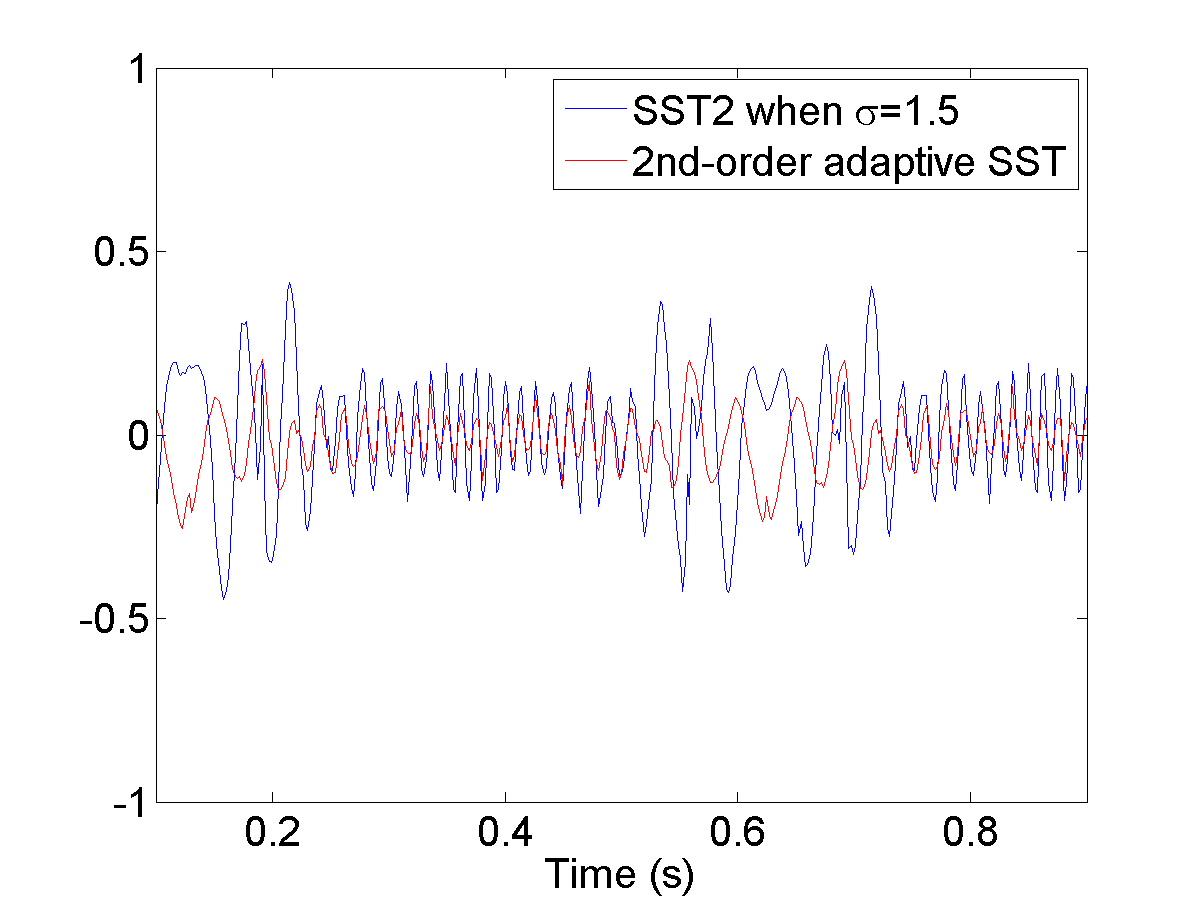}}\\
\resizebox{3.6in}{2.6in}{\includegraphics{difference_compoent_2.png}}&
\end{tabular}
\caption{\small Reconstruction results of  the components of $s(t)$ given in \eqref{three_component_signal}. Difference of reconstructed $s_1(t)$ (top-left panel),
$s_2(t)$ (top-right panel), $s_3(t)$ (bottom panel)  with original $s_1(t), s_2(t)$, $3_2(t)$ by
regular 2nd-order SST and 2nd-order adaptive SST  }
\label{fig:diff_reconst_SST_three_component}
\end{figure}

\begin{figure}[th]
\centering
\begin{tabular}{cc}
\resizebox{3.4in}{3.0in}{\includegraphics{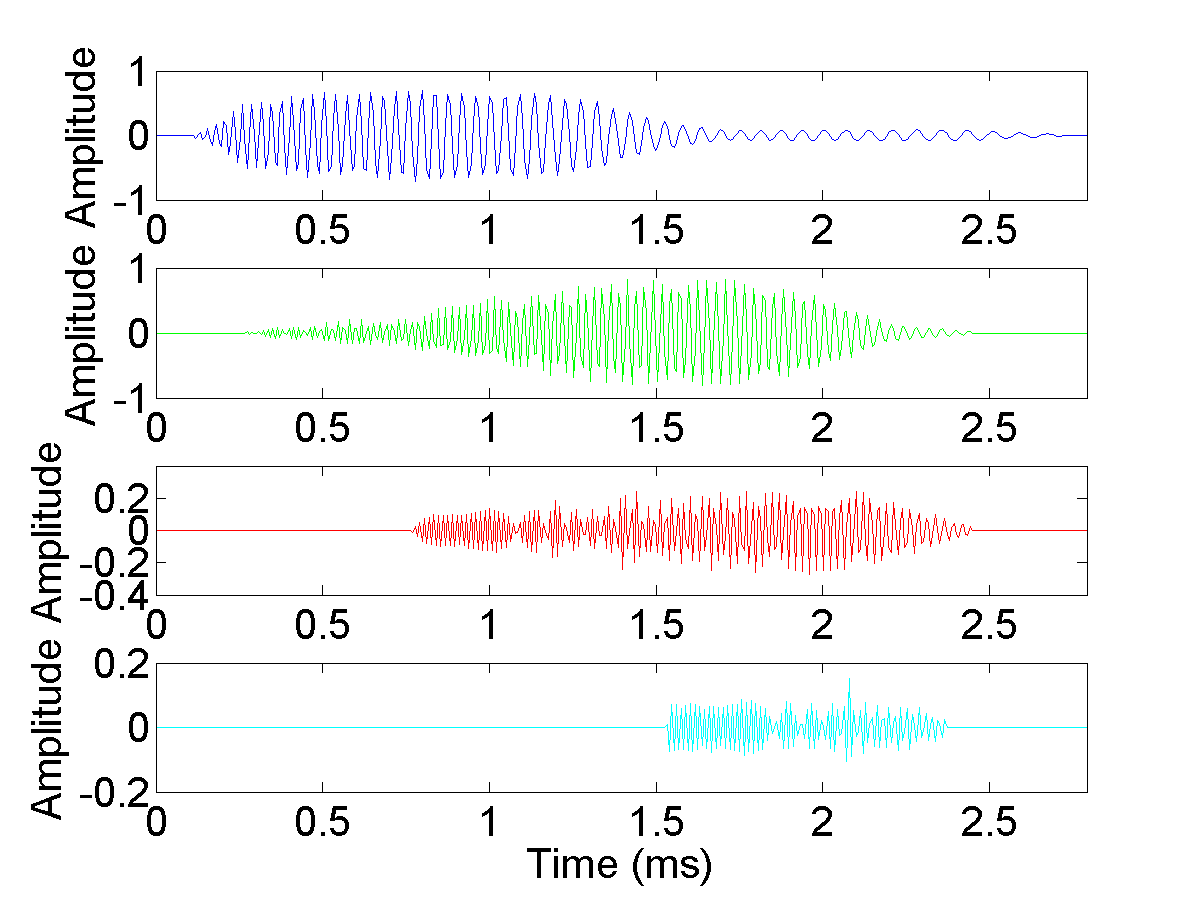}} &
\resizebox{3.4in}{3.0in}{\includegraphics{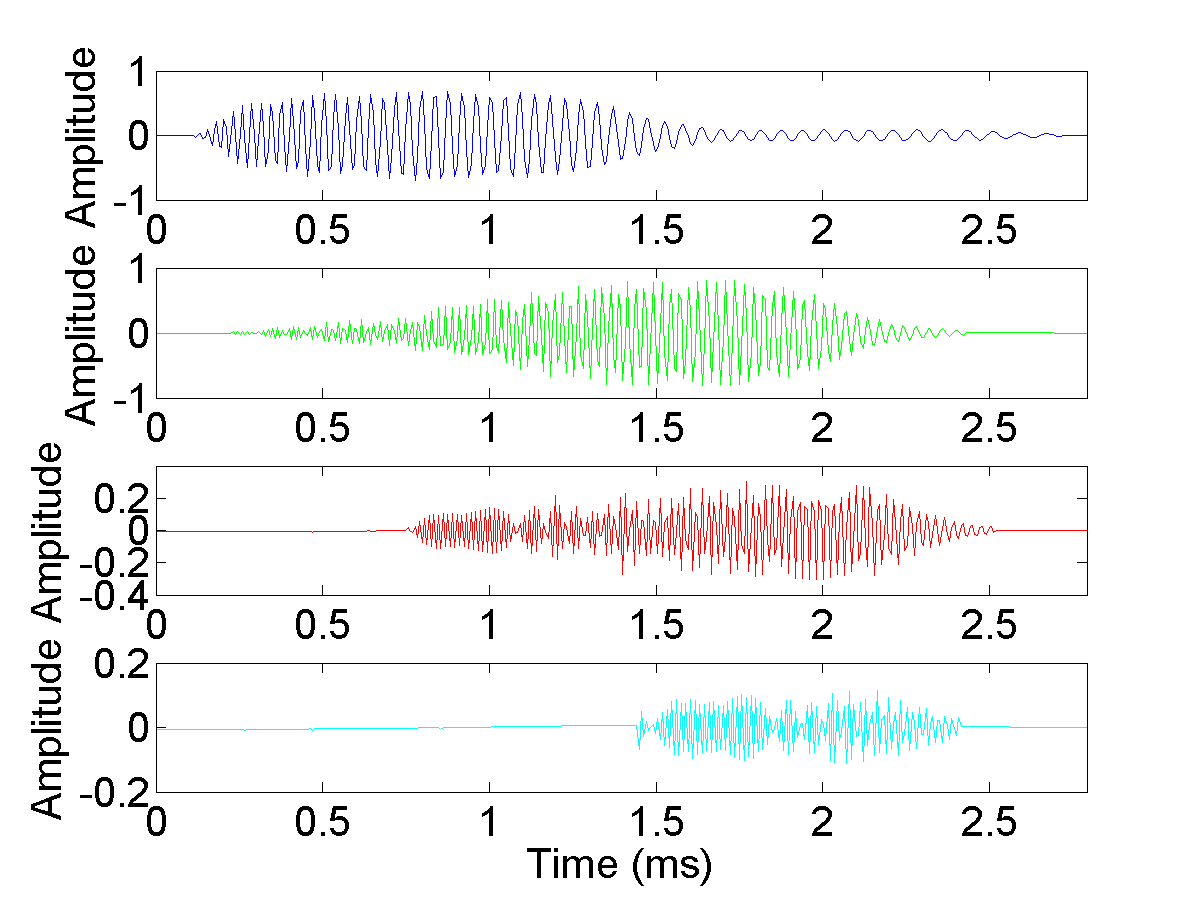}}
\end{tabular}
\caption{\small  Reconstructed components of the bat signal by
regular 2nd-order SST (left column) and by  2nd-order adaptive SST (right column)}
\label{fig:reconst_bat}
\end{figure}

 Next we show some results on component recovery/separation of multicomponent signals.
 We consider the three-component signal $s(t)$ in \eqref{three_component_signal} and the bat signal
 with the 2nd-order SST.
We use \eqref{reconst_SST_component} and \eqref{reconst_SST_para_component} with $T_x$ and $T_x^{adp}$  replaced by $T_x^{2nd}$ and $T_x^{2adp}$ respectively to recover the signal components for conventional 2nd-order SST and 2nd-order adaptive SST, respectively. We use the maximum values on the SST plane to search for the IF ridges $\phi'_k(t)$ one by one. Then integrate around the ridges with $\Gamma_1=\Gamma_2=2$ (discrete value, unitless).
 In Fig.\ref{fig:reconst_SST_three_component} we show the reconstructed components. The reconstructed components with either the regular or the adaptive 2nd-order SST are close to the original components.
 We show the differences between the reconstructed components and the original components by these two methods in
 Fig.\ref{fig:diff_reconst_SST_three_component}. Our method outperforms the regular 2nd-order SST. Finally,
 in  Fig.\ref{fig:reconst_bat} we show the reconstructed components of the bat signal.

\section{Conclusion}
In this paper, we propose the adaptive CWT, the adaptive SST and the 2nd-order adaptive SST, all with a time-varying parameter, for the multicomponent signal separation. We define a bandwidth of Gaussian window to describe the supported zones of the CWT of a multicomponent signal in the time-scale plane efficiently. We derive the well-separated conditions of a multicomponent signal. Both the sinusoidal signal model and
the linear frequency modulation (linear chirp) signal model are proposed.
We propose a new algorithm which selects automatically the time-varying parameter for multicomponent signal separation. The simulation experiments on multicomponent signals demonstrate the validity of the proposed method in noisy environment.  In this paper, we consider the CWT-based SST.
The method and algorithm proposed in this paper can be applied to the case of the STFT-based SST. We will report our results on the adaptive STFT and adaptive STFT-based SST in an accompanying paper. In addition, we will carry out the study of the adaptive SST with the quadratic chirp and other higher order chirp models.

\bigskip
\n {\bf Acknowledgments:} The authors would like to thank Professor Hongbing Ji for helpful discussions. 
The authors also wish to thank Curtis Condon, Ken White, and Al Feng of the Beckman Institute of the University of Illinois for the bat data in  Fig.\ref{fig:adaptiveSST_bat} and for permission to use it in this paper.

\subsection*{Appendix}
In this appendix, we provide the proof of Propositions \ref{pro:recover_CWT_para} and \ref{pro:CWT_linear_chirp} and Theorem \ref{theo:2nd_phase_para}.

{\bf Proof of Proposition \ref{pro:recover_CWT_para}.} \quad
From \eqref{CWT_xreqdomain_analytic}, we have
\begin{eqnarray*}
&&
\int_0^\infty \wt W_x(a, b)\;  \frac {da}{a}=\int_0^\infty
 \int_0^\infty \wh x(\xi) \overline{\wh \psi _{\gs(b)}\big(a \xi\big)} e^{i2\pi b \xi} d\xi\;   \frac {da}{a}\\
&& =\int_0^\infty
\wh x(\xi) e^{i2\pi b \xi}  \int_{0}^\infty  \overline{\wh \psi_{\gs(b)} \big(a \xi\big)} \frac {da}{a}  \; d\xi\\
&& =\int_0^\infty \wh x(\xi) e^{i2\pi b \xi}d\xi  \int_{0}^\infty  \overline{\wh \psi_{\gs(b)} \big(a \xi\big)} \; \frac {da}{a}  \\
&& =\int_0^\infty \wh x(\xi) e^{i2\pi b \xi}d\xi  \int_{0}^\infty  \overline{\wh \psi_{\gs(b)}(a)} \; \frac {da}{a}  \\
&& =c_\psi(b) \int_0^\infty \wh x(\xi) e^{i2\pi b \xi} \; d\xi=c_\psi(b) \; x(b).
\end{eqnarray*}
This completes the proof of \eqref{CWT_para_recover_analytic_a_only}.

If $x(t)$ is real, then we have $\overline{\wh x(\xi)}=\wh x(-\xi)$. Thus,
$$
\int_{-\infty}^0 \wh x(\xi) e^{i2\pi b\xi} d\xi=\int_0^\infty  \wh x(-\xi) e^{-i2\pi b\xi} d\xi=\overline{\int_0^\infty  \wh x(\xi) e^{i2\pi b\xi} d\xi},
$$
and hence
\begin{eqnarray*}
&&x(b)=\int_{-\infty}^\infty \wh x(\xi) e^{i2\pi b\xi} d\xi=\int_{-\infty}^0 \wh x(\xi) e^{i2\pi b\xi} d\xi
+\int_0^\infty  \wh x(\xi) e^{i2\pi b\xi} d\xi\\
&&=2 {\rm Re }\Big(\int_0^\infty  \wh x(\xi) e^{i2\pi b\xi} d\xi\Big).
\end{eqnarray*}
From the proof of  \eqref{CWT_para_recover_analytic_a_only}, we have
 $$
\int_0^\infty \wt W_x(a, b)\;  \frac {da}{a} =c_\psi(b) \int_0^\infty \wh x(\xi) e^{i2\pi b \xi} d \xi.
$$
Therefore,
$$
x(b)=2 {\rm Re }\Big(\int_0^\infty  \wh x(\xi) e^{i2\pi b\xi} d\xi\Big)=
{\rm Re }\Big(\frac 2{c_\psi(b)} \int_0^\infty \wt W_x(a, b) \frac {d a}a \Big).
$$
This proves \eqref{CWT_para_recover_real_a_only}.
\hfill$\blacksquare$

 \bigskip

{\bf Proof of Theorem  \ref{theo:2nd_phase_para}.}
For $s=s(t)$ given by \eqref{def_chip_At},  from $s'(t)=\big(p+qt+i2\pi (c+rt )\big) s(t)$ and \eqref{def_CWT_para},
we have
\begin{eqnarray*}
&&\frac {\partial} {\partial b} \wt W_s(a, b)=\int_{-\infty}^\infty s'(b+at)\;\frac 1{\gs(b)}{g(\frac t{\gs(b)})}e^{-i2\pi \mu t} dt + \int_{-\infty}^\infty s(b+at) (-\frac {\gs'(b)}{\gs(b)^2}){g(\frac t{\gs(b)})}e^{-i2\pi \mu t} dt\\
&&\qquad + \int_{-\infty}^\infty s(b+at) (-\frac {\gs'(b) t}{\gs(b)^3}){g'(\frac t{\gs(b)})}e^{-i2\pi \mu t} dt\\
&&=\big(p+qb+i2\pi (c+rb)\big) \wt W_s(a, b)+(q+i 2\pi r) a \;  \int_{-\infty}^\infty t s(b+at)\;\frac 1{\gs(b)}{g(\frac t{\gs(b)})}e^{-i2\pi \mu t} dt\\
&&\qquad - \frac {\gs'(b)}{\gs(b)}\wt W_s(a, b) -\frac {\gs'(b)}{\gs(b)}\wt W^{g_2}_s(a, b)\\
&&=\big(p+qb+i2\pi (c+rb)\big) \wt W_s(a, b)+(q+i 2\pi r) a \gs(b) \wt W^{g_1}_s(a, b)
- \frac {\gs'(b)}{\gs(b)}\wt W_s(a, b) -\frac {\gs'(b)}{\gs(b)}\wt W^{g_2}_s(a, b)
\end{eqnarray*}
Thus, if $\wt W_s(a, b)\not=0$, we have
\begin{equation}
\label{2nd_para_derivation}
\frac {\frac {\partial}{\partial b} \wt W_s(a, b)}{\wt W_s(a, b)}=
p+qb+i2\pi (c+rb) +(q+i 2\pi r) a \gs(b)\; \frac {\wt W^{g_1}_s(a, b)}{\wt W_s(a, b)}- \frac {\gs'(b)}{\gs(b)}- \frac {\gs'(b)}{\gs(b)} \frac {\wt W^{g_2}_s(a, b)}{\wt W_s(a, b)}.
\end{equation}
Taking partial derivative $\frac {\partial}{\partial a}$ to both sides of \eqref{2nd_para_derivation},
$$
\frac {\partial}{\partial a}\Big(\frac {\frac {\partial}{\partial b} \wt W_s(a, b)}{\wt W_s(a, b)}\Big)=
(q+i 2\pi r) \gs(b)\; \frac {\partial}{\partial a}\Big(a \frac {\wt W^{g_1}_s(a, b)}{\wt W_s(a, b)}\Big)- \frac {\gs'(b)}{\gs(b)}
\frac {\partial}{\partial a}\Big(\frac {\wt W^{g_2}_s(a, b)}{\wt W_s(a, b)}\Big).
$$
Therefore, if in addition, $\frac {\partial}{\partial a}\Big(a \frac {\wt W^{g_1}_s(a, b)}{\wt W_s(a, b)}\Big)\not=0$,   then $(q+i 2\pi r)\gs(b)=R_0(a, b)$, where $R_0(a, b)$ is defined by \eqref{def_R0}.

Back to \eqref{2nd_para_derivation} , we have
$$
\frac {\frac {\partial}{\partial b} \wt W_s(a, b)}{\wt W_s(a, b)}=
p+qb+i2\pi (c+rb) + R_0(a, b) \frac {a \wt W^{g_1}_s(a, b)}{\wt W_s(a, b)}- \frac {\gs'(b)}{\gs(b)}- \frac {\gs'(b)}{\gs(b)} \frac {\wt W^{g_2}_s(a, b)}{\wt W_s(a, b)}.
$$
Hence,
$$
\phi'(b)=c+rb =\frac {\frac{\partial}{\partial b} \wt W_s(a, b)}{i2\pi \wt W_s(a, b)}
-\frac {p+qb}{i2\pi}
- a \frac{\wt W^{g_1}_s(a, b)}{i2\pi \wt W_s(a, b)} R_0(a, b)+ \frac {\gs'(b)}{i2\pi \gs(b)}
+ \frac {\gs'(b)}{\gs(b)} \frac {\wt W^{g_2}_s(a, b)}{i2\pi \wt W_s(a, b)}.
$$
Since $\phi'(b)$ is real, we conclude that
$$
\phi'(b)=c+rb ={\rm Re}\Big\{\frac {\frac{\partial}{\partial b} \wt W_s(a, b)}{i2\pi \wt W_s(a, b)}\Big\}
- a \;{\rm Re}\Big\{ \frac{\wt W^{g_1}_s(a, b)}{i2\pi \wt W_s(a, b)} R_0(a, b)\Big\}
+ \frac {\gs'(b)}{\gs(b)} {\rm Re}\Big\{ \frac {\wt W^{g_2}_s(a, b)}{i2\pi \wt W_s(a, b)} \Big\}.
$$
Thus for an LFM signal $x(t)$ given by \eqref{def_chip_At}, at $(a, b)$ where $\frac {\partial}{\partial a}\Big(a \frac {\wt W^{g_1}_x(a, b)}{\wt W_x(a, b)}\Big)\not=0$ and $\wt W_x(a, b)\not=0$,
$\go^{2adp}_x(a, b)$ defined by \eqref{2nd_phase_para} is
$\phi'(b)=c+rb$, the IF of $x(t)$. This shows Theorem \ref{theo:2nd_phase_para}.
\hfill $\blacksquare$

\bigskip

{\bf Proof of Proposition \ref{pro:CWT_linear_chirp}.} \quad Let $s(t)$ be the linear chirp signal given by \eqref{def_chip}. Then the CWT of $s(t)$ with $\psi_\gs$ is given by
\begin{eqnarray*}
W_s(a, b)\hskip -0.6cm &&=\int_{-\infty}^\infty s(t)\overline{\psi_\gs(\frac {t-b}a)}\frac {dt}a=\int_{-\infty}^\infty s(b+a\tau)\overline{\psi_\gs(\tau)}d\tau\\
&&=\int_{-\infty}^\infty A e^{i2\pi \big(c(b+a\tau) +\frac 12 r (b+a\tau)^2\big)}\frac 1{\gs \sqrt {2\pi}}e^{-\frac{\tau^2}{2 \gs^2}}e^{-i2\pi \mu \tau }d\tau\\
&&=\frac A{\gs \sqrt {2\pi}}\int_{-\infty}^\infty
e^{i2\pi \big(c b+ca\tau +\frac r2 b^2+rb a\tau+\frac r2 a^2\tau^2\big)}e^{-\frac{\tau^2}{2 \gs^2}}e^{-i2\pi \mu \tau}d\tau\\
&&=\frac A{\gs \sqrt {2\pi}}e^{i2\pi \big(c b+\frac r2 b^2\big) }\int_{-\infty}^\infty
e^{-\frac{\tau^2}{2 \gs^2}+i\pi r a^2\tau^2+ i2\pi a \big(c +r b -\frac \mu a\big)\tau}d\tau\\
&&=\frac A{\gs \sqrt {2\pi}}\; e^{i2\pi \big(c b+\frac r2 b^2\big)} \; \frac{\sqrt \pi}{\sqrt{\frac 1{2\gs^2}-i\pi r a^2}}
\; e^{-2\pi^2 (a\gs)^2 (c +r b -\frac \mu a)^2\frac 1{1-i2\pi \gs^2 r a^2 }}\\
&&=\frac {A}{\sqrt{1-i2\pi \gs^2 a^2 r}}\;  e^{i2\pi \big(c b+\frac r2 b^2\big)}  e^{-\frac{2\pi^2(a \gs)^2}{1+(2\pi \gs^2 a^2 r)^2}(c +r b-\frac \mu a)^2(1+i2\pi \gs^2 a^2 r)},
\end{eqnarray*}
where the second last equality follows from Lemma \ref{lem:FT_for_LinearChirp}. Thus \eqref{CWT_LinearChip} holds.
\hfill $\blacksquare$


\end{document}